\documentclass[12pt]{iopart}

\usepackage{latexsym}
\usepackage{amssymb}
\usepackage{epsfig}

\begin{document}

\newcommand{\muB}{\hat{\mu}}

\newcommand{\connE}{\ensuremath{\tilde{\nabla}}}
\newcommand{\phib}{\ensuremath{\bar{\phi}}}
\newcommand{\connM}{\ensuremath{\nabla}}

\newcommand{\conngam}{\ensuremath{\;^{(\tilde{\gamma})}\tilde{\nabla}}}
\newcommand{\conneta}{\ensuremath{\bar{\nabla}}}
\newcommand{\conng}{\ensuremath{\tilde{\nabla}}}

\newcommand{\conngamt}{\ensuremath{\;^{(\gamma)}\nabla}}
\newcommand{\connetat}{\ensuremath{\hat{\nabla}}}
\newcommand{\conngt}{\ensuremath{\nabla}}

\newcommand{\nablaM}{\ensuremath{\nabla}}
\newcommand{\nablaE}{\ensuremath{\tilde{\nabla}}}
\newcommand{\nablaS}{\ensuremath{\hat{\nabla}}}

\newcommand{\metE}{\ensuremath{\tilde{g}}}
\newcommand{\metM}{\ensuremath{g}}
\newcommand{\RWmetE}{\ensuremath{\tilde{\gamma}}}
\newcommand{\RWmetM}{\ensuremath{\gamma}}
\newcommand{\etaE}{\ensuremath{\tilde{\eta}}}
\newcommand{\etaM}{\ensuremath{\eta}}
\newcommand{\permetE}{\ensuremath{\tilde{h}}}
\newcommand{\permetM}{\ensuremath{h}}

\newcommand{\volE}{\ensuremath{\sqrt{-\metE}}}
\newcommand{\volM}{\ensuremath{\sqrt{-\metM}}}
\newcommand{\RiemE}{\ensuremath{\tilde{R}}}
\newcommand{\RiemM}{\ensuremath{R}}
\newcommand{\EinE}{\ensuremath{\tilde{G}}}
\newcommand{\EinM}{\ensuremath{G}}

\newcommand{\Vp}{\ensuremath{\frac{dV}{d\mu}}}

\newcommand{\bim}{\ensuremath{B}}
\newcommand{\ibim}{\ensuremath{C}}

\newcommand{\grad}{\ensuremath{\vec{\nabla}}}

\newcommand{\metS}{\ensuremath{\hat{g}}}
\newcommand{\AS}{\ensuremath{\hat{A}}}
\newcommand{\GS}{\ensuremath{\hat{G}}}
\newcommand{\KS}{\ensuremath{\hat{K}}}
\newcommand{\SeS}{\ensuremath{\hat{S}}}
\newcommand{\JS}{\ensuremath{\hat{J}}}
\newcommand{\cS}{\ensuremath{\hat{c}}}

\newcommand{\KE}{\ensuremath{\tilde{K}}}
\newcommand{\aE}{\ensuremath{\tilde{A}}}
\newcommand{\cE}{\ensuremath{\tilde{c}}}

\paper{Cosmological Perturbation Theory With Background Anisotropic Curvature}
\author{Tom Zlosnik}
\ead{t.zlosnik@imperial.ac.uk}
\address{Imperial College Theoretical Physics, Huxley Building, London SW7}

\begin{abstract}
The theory of cosmological perturbations is extended to spacetimes displaying isotropic expansion but anisotropic
curvature. The perturbed Einstein equation and Boltzmann equations for massless and massive particles are derived in a general gauge
and a decomposition of perturbations into harmonic modes and moments is proposed.
Generalization to the case where anisotropic expansion is also present in the background is discussed.
\end{abstract}

\section{Introduction}
\label{intro}

The observed approximate homogeneity and isotropy of the cosmic microwave background (CMB) remains one of the most striking discoveries in cosmology.  
This approximation holds so well that deviations are adequately characterized by a small perturbation to the temperature $\Theta\equiv \Delta T/T $ as well a polarization of the photon field, expected to be of comparable or smaller order on various scales 
\cite{Dodelson:2003ft}.
Both of these perturbational quantities are time dependent, inhomogeneous (having a non-trivial spatial dependence) and anisotropic (having a non-trivial dependence upon the angle that an observer see an incoming photon).
The influx of data from precision cosmology has allowed for a more searching question to be asked - that of whether these quantities are \emph{statistically} isotropic. For the case of the temperature
perturbation $\Theta$, fluctuations are statistically isotropic if the following identity holds:

\begin{eqnarray}
\label{sts}
\left<\Theta(\textbf{n}_{(1)}^{i})\Theta(\textbf{n}_{(2)}^{i})\right>= f ( \textbf{n}_{(1)}^{i}\textbf{n}_{(2)i})
\end{eqnarray}
 i.e. the above expectation value for the product of temperature fluctuations at two different directions in the sky $\textbf{n}_{(1)}^{i}$ and $\textbf{n}_{(2)}^{i}$ is a 
 function of only their projection along one another, the implication being that there is no other spatial direction present in the physics of the generation and evolution
 of temperature perturbations. The degree to which statistical isotropy of CMB temperature fluctuations is true given the data has been the focus of considerable
 debate. There exist indications that observations are indeed implying that (\ref{sts}) is not exact in our universe on large scales, notably there appear to be a number of anomalies that 
 are relatable to the presence of  directions other than  $\textbf{n}_{(1)}^{i}$ and $\textbf{n}_{(2)}^{i}$ (see for instance \cite{Copi:2010na},\cite{Land:2005ad},\cite{Hoftuft:2009rq}).
 
Given these indications one may speculate as to whether the presence of an extra direction in a given CMB anomaly is as a consequence of the presence of fields 
which impose an actual preferred spatial direction
in the universe. This would not be unprecedented in cosmology. For instance, the existence of a field to generate
a primordial period of inflation is conventionally taken to explain the existence of a preferred 
set of clocks on large scales in cosmology \cite{Brout:1997wg}. 

If there does existed a preferred spatial direction, this is expected to be encoded in the gravitational
field either at the background or the perturbed level. Consider the following class of background metrics:

\begin{eqnarray}
\label{back}
ds^{2} &=&  a^{2}(t)\left(-dt^{2}+\left(\frac{b(t)}{a(t)}\right)^{2}dz^{2}+\gamma_{cd}dx^{c}dx^{d}\right)\\
\gamma_{cd} &=& \frac{1}{|K|}\tilde{\gamma}_{cd}
\end{eqnarray}
The interpretation of the 
tensor $\tilde{\gamma}_{cd}$ is as follows:

\begin{itemize}
\item For the case $K>0$, $\tilde{\gamma}_{ab}$ is the metric on a 2-sphere
with two dimensional Ricci scalar $^{2}R=2$ (i.e unit radius) 
\item For the case $K<0$, $\tilde{\gamma}_{ab}$ is the metric on the upper
sheet of a 2-hyperboloid of two sheets  with two dimensional Ricci scalar $^{2}R=-2$.
\item For the case $K=0$,  the tensor $(1/|K|)\tilde{\gamma}_{ab}$ is a two dimensional, flat Euclidean metric (written in cylindrical coordinates).
\end{itemize}

For non-zero values of $a(t)$ and $b(t)$, one may intuitively think of these spaces as an infinite spatial line (the `z direction') with a two-dimensional surface at each point
along this line- these surfaces may either a flat 2-space (the case $K=0$), a 2-sphere (the case $K>0$), or the upper sheet of a two dimensional hyperboloid 
of two sheets ($K<0$). 

When $\frac{db}{dt}\neq\frac{da}{dt}$ the expansion rate is anisotropic and hence there exists shear in the congruence of timelike geodesics labelled by proper-time
$ \tau = \int a dt$. The form of (\ref{back}) additionally implies that this congruence has vanishing vorticity.
Cosmological perturbation theory for the gravitational field coupled to a scalar field for the case with shear and $K=0$ has been developed in detail 
(\cite{Pereira:2007yy},\cite{Pitrou:2008gk},\cite{Gumrukcuoglu:2007bx}). The case where shear is present and $K\neq 0$ has also been considered at the level of the
background \cite{Graham:2010hh}. The authors explicitly consider a scenario of primordial anisotropic inflation, followed by isotropic, flat FRW expansion (such expansion corresponds to the choice $K=0$, $a=b$ in (\ref{back})). The anisotropy
is then encoded in the perturbations and not the background. However, we will see in Section \ref{secstatanis} that this scenario is also straightforwardly addressed in choosing 
a `2+1' decomposition of the background space as a result of how this allowance for spatial anisotropy is encoded into the perturbation theory.

Alternatively, deviations from Friedmann-Robertson-Walker (FRW) geometry may occur in the absence of shear but in the presence of anisotropic curvature (i.e. $|K| \neq 0$).
This case was discussed \cite{Mimoso:1993ym} and further developed in \cite{Coley:1994yt},\cite{McManus:1994ys}. 
Crucially, in the absence of shear the metric (\ref{back}) is conformally static and therefore admits a homogeneous and isotropic solution for the CMB at the background level \cite{Ehlers:1966ad}.
Anisotropic cosmological models which lack the conformally static property typically have fine tuning problems as they they induce anisotropies in the CMB radiation which 
are known to be consistently regarded as small perturbations around a homogeneous, isotropic solution. Clearly the models are also immune to constraints on directionality of the expansion
rate so in the shearless case departures for the predictions of
the FRW model begin instead via the anisotropy of the luminosity distance relation which can used to constrain $K$ by comparing with data from supernovae. It was found \cite{Koivisto:2010dr}
that constraints on the quantity $\Omega_{K}$ (i.e the contribution to the critical density of components with energy density scaling as $a^{-2}$) from the Union SnE dataset \cite{Kowalski:2008ez} are comparable to 
those obtained in the FRW model.

The intention of this paper is to develop the theory of cosmological perturbations in the case where the cosmological background belongs to the class (\ref{back}), is shear free but may posses
anisotropic curvature. In Section \ref{cosper}  we discuss the harmonic decomposition of perturbations and present the perturbed Einstein tensor in a general gauge. Following this we consider perturbations
in the matter sectors. Firstly in Section \ref{sf} we consider perturbations to a putative field supporting the background shear-free condition. Then in Section \ref{boltmassless} we consider the Boltzmann equation for massless particles,
and in doing so propose a decomposition of the angular dependence of the particle distribution function in terms of moments. In Section \ref{moment} we compare the proposed moment
expansion to that conventionally employed in the FRW limit, and in Section \ref{freest} we discuss the consequences that the mere presence of anisotropic curvature at the background level may have on the evolution of 
of the photon perturbation.
Then, in Section \ref{boltmassive}, we consider the Boltzmann equation in the case of massive particles. In Section \ref{polcorr} we present expressions for the angular correlation function for polar temperature perturbations and in Section \ref{secstatanis} uses these results for comparison to previous results in the literature. The extension of the formalism to the case where shear is present in the background is 
discussed in Section \ref{inclshear} and conclusions are presented in Section \ref{conc}.

Additionally there are a number of appendices: some necessary mathematical background to the calculations is presented in \ref{appa}; the explicit calculation of the polar angular correlation function is given in \ref{calpolar}; the form of spacetime gauge transformations and the construction of gauge invariant perturbations is given in \ref{gaugetran}; comparison to scalar-vector-tensor perturbations in the FRW limit is given in 
\ref{compfrw}; and a table of notation is given in \ref{tableof}.

\section{Cosmological perturbations}
\label{cosper}

For the shearless case, as in the case of linear perturbations around an FRW spacetime,
the problem of deducing the spacetime metric, if sufficiently `close' to the background metric, is equivalent to a 
problem of examining the evolution of a set of fields on a `co-moving spacetime', this spacetime defined by the metric $g^{(C)}_{\mu\nu}$ :

\begin{eqnarray}
\label{backshearless}
g^{(C)}_{\mu\nu}dx^{\mu}dx^{\nu} &=& -dt^{2}+dz^{2}+\gamma_{cd}dx^{c}dx^{d}
\end{eqnarray}
 In the case of linear perturbations around an FRW background, surfaces of constant conformal time on the comoving spacetime are homogeneous and isotropic. This isotropy 
allows for a simple decomposition of fields into scalar, vector, and tensor perturbations, which evolve independently of one another \cite{Straumann:2008cj}. 
This decomposition is not appropriate in the case (\ref{backshearless}) where $K\neq 0$, due to the inherent anisotropy of surfaces of constant conformal time.  If attempting to do so, one would
find coupling between scalar, vector, and tensor modes, so considerably complicating the analysis \cite{Buniy:2008xs}.

Our approach is to follow (\cite{Gerlach:1979rw},\cite{Clarkson:2009sc}) and take advantage of a more limited isotropy present in the spacetime - i.e. the isotropy of surfaces of constant $(t,z)$. 
Due to this isotropy, we may decompose a function $y(t,z,x^{a})$ existing on the co-moving spacetime as follows:

\begin{eqnarray}
\label{scalard}
y(t,z,x^{a}) = \sum_{k,m}\tilde{y}(t,z){\mathcal E}^{k}_{m}(x^{a})
\end{eqnarray}
Unless otherwise stated, coordinate indices $a,b,c..$ refer exclusively to components in the 2-surface. 
The functions ${\mathcal E}^{k}_{m}(x^{a})$ form a complete set of eigenfunctions of the Laplace-Beltrami operator on surfaces of constant $(t,z)$, and indeed for a general value of the curvature $K$, we define the complex label $k$ via the following equation:

\begin{eqnarray}
\label{eig1}
\nabla_{a}\nabla^{a}{\mathcal E}^{k}_{m}(x^{a})=- kk^{*}{\mathcal E}^{k}_{m}(x^{a})
\end{eqnarray}
where $\nabla_{a}$ is the derivative operator compatible with the metric $\gamma_{cd}$.

Due to our choice of coordinates, the label $m$ is always discrete, though with a range dependent upon the curvature of the surface, whilst $k$ may be continuous or discrete depending upon the curvature of the surface In the continuous case the summation over $k$ must be replaced by an integration with appropriate measure. For instance, for $K>0$, the functions are the spherical harmonics $Y^{l}_{m}$, and the eigenvalues $kk^{*}$ are equal to $|K|l(l+1)$. See Appendix...for discussion of the corresponding functions in the open and flat case.

Similarly, a vector field $y_{a}(t,z,x^{b})$ (i.e. a two component vector field existing on surfaces of constant $(t,z)$) may be decomposed as follows:

\begin{eqnarray}
\label{vectord}
y_{a}(t,z,x^{b}) &=& \sum_{k,m}\left(\tilde{y}^{(VP)}(t,z)\nabla_{a}{\mathcal E}^{k}_{m}(x^{b})+\tilde{y}^{(VA)}(t,z)\bar{\nabla}_{a}{\mathcal E}^{k}_{m}(x^{b})\right)
\end{eqnarray}
Where we have now introduced the differential operator
operator $\bar{\nabla}_{a}\equiv \epsilon_{a}^{\phantom{a}b}\nabla_{b}$ where $\epsilon_{ab}$ is the 
volume form on the co-moving 2-surface. The labels $P$ and $A$ refer to polar and axial terms respectively. It may be checked that $\nabla_{a}{\mathcal E}^{k}_{m}(x^{b})$ and $\bar{\nabla}_{a}{\mathcal E}^{k}_{m}(x^{b})$ are \emph{eigenvectors} of the Laplace-Beltrami operator on the comoving 2-surface.

We shall find that, much as scalar, vector, and tensor modes evolve independently of one another in the FRW case, the polar and axial modes evolve independently for small perturbations around the metric (\ref{back}). Finally a symmetric tensor field $y_{ab}(t,z,x^{c})$ existing on the 2-surface may be decomposed as follows:

\begin{eqnarray}
y_{ab}(t,z,x^{c}) &=& \sum_{k,m}(\tilde{y}^{(TP1)}(z,t)\gamma_{ab}{\mathcal E}^{k}_{m}+\tilde{y}^{(TP2)}(z,t)\nabla_{a}\nabla_{b}{\mathcal E}^{k}_{m} \nonumber \\
&&+\tilde{y}^{(TA)}(z,t)\nabla_{(a}\bar{\nabla}_{b)}{\mathcal E}^{k}_{m}) \label{tensord} 
\end{eqnarray}
where $P$ and $A$ again denote polar and axial modes, and $\nabla_{a}\nabla_{b}{\mathcal E}^{k}_{m}$ and $\nabla_{a}\bar{\nabla}_{b}{\mathcal E}^{k}_{m}$ are \emph{eigentensors} of the Laplace-Beltrami operator on the co-moving 2-surface.
Note that the appropriate range of $k$ need not be the same in the respective decomposition of functions, vectors, and tensors on
the 2-surface. The decompositions (\ref{scalard}),(\ref{vectord}), and (\ref{tensord}) then allow for the decomposition of a general perturbation $\delta g_{\mu\nu}$  to the metric
(\ref{back}) into polar and axial components as follows:

\begin{eqnarray}
\delta g_{\mu\nu}= \delta g^{(P)}_{\mu\nu}+\delta g^{(A)}_{\mu\nu}
\end{eqnarray}
where

\begin{eqnarray}
\delta g^{(P)}_{\mu\nu}dx^{\mu}dx^{\nu}&=& a^{2}\sum_{k,m}(V(z,t){\mathcal E}dt^{2}+E(z,t){\mathcal E}dzdt+F(z,t){\mathcal E}dz^{2}\nonumber\\
                                        && +B(z,t)\nabla_{d}{\mathcal E} dzdx^{d}+ C(z,t)\nabla_{d}{\mathcal E} dtdx^{d}\nonumber \\
     \label{polmet}                                   && +U(z,t)\gamma_{ab}{\mathcal E}dx^{a}dx^{b}+X(z,t)\nabla_{a}\nabla_{b}{\mathcal E} dx^{a}dx^{b})
\end{eqnarray}
and 

\begin{eqnarray}
\nonumber
\delta g^{(A)}_{\mu\nu}dx^{\mu}dx^{\nu}  &=&  a^{2}\sum_{k,m}(R(z,t) \bar{\nabla}_{a}{\mathcal E} dtdx^{a}+S(z,t)\bar{\nabla}_{a}{\mathcal E}dzdx^{a}\nonumber\\
&&+2Q(z,t)
\label{axmet}\nabla_{(a}\bar{\nabla}_{b)}{\mathcal E}dx^{a}dx^{b})
\end{eqnarray}
where the labels $k$ and $m$ on
${\mathcal E}^{k}_{m}$ have been suppressed for notational compactness.

Using the expansions (\ref{polmet}) and (\ref{axmet}) along with the eigenvalue equation, we will find that the Einstein equations
may be expanded in terms of sums of eigenfunctions, eigenvectors, and eigentensors, with coefficients of which being differential
equations for the harmonic modes of the expansion. These differential equations will contain partial derivatives with respect to time
and the z coordinate. It is desirable to transform this system into a set of ordinary derivatives with respect to time by instead
looking at the evolution of the Fourier transform of the function $\tilde{y}_{z,t}$, defined as follows:

\begin{eqnarray}
\label{fourz}
\tilde{y}(z,t) & =& \frac{1}{2\pi} \int \tilde{y}_{w}(t)e^{iwz} dw
\end{eqnarray}

\subsection{Polar perturbed Einstein tensor}

We now detail the components of the polar components of the perturbed Einstein tensor  $\delta G^{(P)\mu}_{\phantom{(P)\mu}\nu}$ in terms of these Fourier modes. We 
shall use a prime to denote ordinary derivatives with respect to conformal time $t$ and in our expressions the the appropriate multiplication by $e^{iwz}$ and sum/integral over labels $k,m,w$ is implicit.
For notational compactness we denote the real number $kk^{*}$ by $k^{2}$.

\begin{eqnarray}
\nonumber\delta G^{(P)t}_{\phantom{(P)t}t} &=& \frac{1}{a^{2}}(2{\mathcal H}iw E+3{\mathcal H}^{2}V-2{\mathcal H}k^{2} C  
 +iwk^{2}B\\
\nonumber &&   -\frac{1}{2}k^{2}F-{\mathcal H}F'   
-w^{2}U-2{\mathcal H}U'-\frac{1}{2}k^{2}U+KU  \\
&&
 +\frac{1}{2}k^{2}w^{2}X+{\mathcal H}k^{2}X'){\mathcal E}
\end{eqnarray}

\begin{eqnarray}
\delta G^{(P)t}_{\phantom{(P)t}z} &=& \frac{1}{a^{2}}(-\frac{k^{2}}{2}(E-iwC-B'+iwX') -{\mathcal H}iw V+iw U'){\mathcal E} 
\end{eqnarray}

\begin{eqnarray}
\nonumber\delta G^{(P)z}_{\phantom{(P)z}z} &=& \frac{1}{a^{2}}({\mathcal H}V'+\frac{1}{2}(-2{\mathcal H}^{2}+4\frac{a''}{a}-k^{2})V{\mathcal E}  \\
   \nonumber       &&    -k^{2}(C'+2{\mathcal H}C)\epsilon - (U'' +2{\mathcal H}U' -KU+\frac{1}{2}k^{2}U){\mathcal E}  \\
           &&   +k^{2}(\frac{1}{2}X''+{\mathcal H}X')){\mathcal E}
\end{eqnarray}

\begin{eqnarray}
\nonumber\delta G^{(P)t}_{\phantom{(P)t}a} &=&\frac{1}{2a^{2}}(-iwE-2{\mathcal H}V+(-w^{2}C+2K C) \\
&&-iwB'  +F'+ U'-KX')\nabla_{a}{\mathcal E}
 \end{eqnarray}
\begin{eqnarray}
\nonumber\delta G^{(P)z}_{\phantom{(P)z}a} &=& \frac{1}{2a^{2}}(-(E'+2{\mathcal H}E)-iwV-iw(C'+2{\mathcal H}C) \\
   && +(B''+2{\mathcal H}B'-2KB)-iwU+iwK X)\nabla_{a}{\mathcal E}
\end{eqnarray}
\begin{eqnarray}
\nonumber\delta G^{(P)a}_{\phantom{(P)a}b} &=& \frac{-1}{2a^{2}}\left((F''+2{\mathcal H}F')\gamma^{a}_{\phantom{a}b}{\mathcal E}+F(\nabla^{a}\nabla_{b}{\mathcal E}+k^{2} \gamma^{a}_{\phantom{a}b}{\mathcal E})\right)\\
\nonumber &&+ \frac{1}{2a^{2}}((-2{\mathcal H}^{2}V+4\frac{a''}{a}V-w^{2}V+2{\mathcal H}V')\gamma^{a}_{\phantom{a}b}{\mathcal E} \\
\nonumber &&-V(k^{2}\gamma^{a}_{\phantom{a}b}{\mathcal E}+\nabla^{a}\nabla_{b}{\mathcal E})) 
 + \frac{1}{a^{2}}iw B\left(\nabla^{a}\nabla_{b}{\mathcal E}+k^{2}\gamma^{a}_{\phantom{a}b}{\mathcal E}\right)\\
\nonumber && -\frac{1}{a^{2}}(C'+2{\mathcal H}C)\left(\nabla^{a}\nabla_{b}{\mathcal E}+k^{2}\gamma^{a}_{\phantom{a}b}{\mathcal E}\right)  +\frac{1}{a^{2}}iw\left(2{\mathcal H}E+E'\right)\gamma^{a}_{\phantom{a}b}{\mathcal E}\\
\nonumber &&+\frac{1}{2a^{2}}\left(-w^{2}U-U''-2{\mathcal H}U'\right)\gamma^{a}_{\phantom{a}b}{\mathcal E}\\
 && +\frac{1}{2a^{2}}(X''+w^{2}X+2{\mathcal H}X)\left(\nabla^{a}\nabla_{b}{\mathcal E}+k^{2}\gamma^{a}_{\phantom{a}b}{\mathcal E}\right) \label{space3}
\end{eqnarray}
The perturbation (\ref{space3}) has two tensor components: a sum of perturbations proportional to $\gamma^{a}_{\phantom{a}b}{\mathcal E}$
and a sum proportional to $\nabla^{a}\nabla_{b}{\mathcal E}$. One may isolate the latter components by operating on (\ref{space3}) with
$\nabla_{a}\nabla^{b}- \frac{1}{2}\gamma^{a}_{\phantom{a}b}\nabla^{2}$ which is nonzero only after contraction with traceless components.

 
 \subsection{Axial Perturbed Einstein Tensor}

We now consider the axial perturbations to the Einstein tensor:  

\begin{eqnarray}
\delta G^{(A)t}_{\phantom{(A)t}a} &=& \frac{1}{a^{2}}(\frac{(w^{2}+k^{2})}{2}R+K\frac{R}{2}-\frac{iw}{2}S'+\frac{k^{2}}{2}Q'-\frac{1}{2}KQ')\bar{\nabla}_{a}{\mathcal E}
\end{eqnarray}
\begin{eqnarray}
\nonumber\delta G^{(A)z}_{\phantom{(A)z}a}  &=& \frac{1}{a^{2}}(-\frac{iw}{2}(R'+2{\mathcal H}R)+\frac{1}{2}S''+{\mathcal H}S'+\frac{k^{2}}{2}S-\frac{1}{2}KS \\
&&-\frac{iwk^{2}}{2}Q+\frac{iw}{2}Q K)\bar{\nabla}_{a}{\mathcal E}
\end{eqnarray}
Observe that there are axial and polar contributions to $\delta G^{z}_{\phantom{z}a}$. However, their decoupling is guaranteed by orthogonality of vectors $\bar{\nabla}_{a}{\mathcal E}$ and
$\nabla_{a}{\mathcal E}$.

 \begin{eqnarray}
 \nonumber\delta G^{(A)a}_{\phantom{(A)a}b} &=&  \frac{1}{a^{2}}(-\frac{1}{2}(R'+2{\mathcal H}R)+\frac{iw}{2}S\\
    &&+(Q''+w^{2}Q+2{\mathcal H}^{2}Q-4\frac{a''}{a}Q+2{\mathcal H}Q'))\gamma^{ac}\nabla_{(c}\bar{\nabla}_{b)}
 \end{eqnarray}
 Again, orthogonality of axial and polar contributions to $\delta G^{a}_{\phantom{a}b}$ follows from the $\gamma^{ac}\nabla_{(c}\bar{\nabla}_{b)}$ having vanishing contraction
 with $\gamma_{a}^{\phantom{a}b}{\mathcal E}$ and $\nabla_{a}\nabla^{b}{\mathcal E}$.
 
 Given the perturbed Einstein tensor it is necessary then to obtain the perturbed stress energy tensors of matter fields in the universe, as well as their owned perturbed evolutions. We will obtain both
 of these things by a treatment of the Boltzmann equation for massless and massive particles in Sections \ref{boltmassless} and \ref{boltmassive}. 
 In the following section we discuss the perturbations of an explicit example of a field whose presence can allow for a shearless example of the spacetime (\ref{back}) to exist as a consistent
 solution to the Einstein and matter field equations.

\section{Scalar field}
\label{sf}

As noted in \cite{Mimoso:1993ym}, a shearless example of (\ref{back}) with $K\neq 0$ cannot be be a solution of the Einstein equations sourced by a perfect fluid. Rather, it is necessary
for there to exist a matter source in the Einstein equations with an anisotropic stress at the background level. Indeed it may be shown that if the matter source may be described by the following fluid stress energy tensor:

\begin{eqnarray}
T_{\mu\nu} = \rho U_{\mu}U_{\nu} +P h_{\mu\nu} + L V_{\mu}V_{\nu}
\end{eqnarray}
where $U_{\mu}$ is the fluid 4-velocity, $h_{\mu\nu}$ is the metric on surfaces of constant cosmic time, and $V_{\mu}$ is a unit vector (with respect to (\ref{back})) along the $(\partial_{z})^{\mu}$ direction. In turn, the anisotropic pressure $L$ must satisfy the following relation in order for  (\ref{back}) to be a solution:

\begin{eqnarray}
\label{shearco1}
 L = -\frac{K}{a^{2}}
 \end{eqnarray}
One would not expect there to be a unique field/particle origin of a matter candidate that could satisfy (\ref{shearco1}), but it is instructive to consider an explicit realization of the above condition.
This was provided in \cite{Koivisto:2010dr} via a massless canonical two-form field. Equivalently, such a theory is equivalent to a massless scalar field with the following action \cite{Carneiro:2001fz}:

\begin{eqnarray}
S= -\frac{{\mathcal C}}{16\pi G} \int d^{4}x\sqrt{-g} \nabla^{\mu}\phi\nabla_{\mu}\phi
\end{eqnarray}
We consider the following ansatz for the scalar field

\begin{eqnarray}
\phi(z,x^{a},t)= \frac{z}{z_{0}}+W(z,x^{a},t)
\end{eqnarray}
where the field $W$ is a small perturbation. The ansatz for the background field configuration guarantees that the background metric takes the shearless, curved form if $z_{0}$ is related to ${\mathcal C}$ and $K$ as follows:

\begin{eqnarray}
\label{restriction}
\frac{{\mathcal C}}{z_{0}^{2}}= -K
\end{eqnarray}
Up to first order in perturbations, the scalar field stress energy tensor $T^{\mu}_{\phantom{\mu}\nu}$ has the following nonvanishing components, up to first order
in perturbations:

\begin{eqnarray}
T^{t}_{\phantom{t}t} &=& -\frac{{\mathcal C}}{2a^{2}z_{0}^{2}}+\frac{{\mathcal C}}{2a^{2}z_{0}}\sum \left(\frac{F}{z_{0}}-2iw W \right){\mathcal E} e^{iwz}
\end{eqnarray}
where the summation is understood here to denote a sum/integral over the appropriate harmonic labels $k,m,w$

\begin{eqnarray}
T^{z}_{\phantom{z}z} &=& +\frac{{\mathcal C}}{2a^{2}z_{0}^{2}}-\frac{{\mathcal C}}{2a^{2}z_{0}}\sum \left(\frac{F}{z_{0}}-2iw W \right){\mathcal E} e^{iwz}
\end{eqnarray}
\begin{eqnarray}
T^{t}_{\phantom{t}z} &=&\frac{{\mathcal C}}{a^{2}z_{0}} \sum \left(\frac{E}{z_{0}}-W'\right){\mathcal E} e^{iwz}
\end{eqnarray}
\begin{eqnarray}
T^{z}_{\phantom{z}a} &=& \frac{{\mathcal C}}{a^{2}z_{0}}\sum W\nabla_{a}{\mathcal E}e^{iwz}
\end{eqnarray}
\begin{eqnarray}
T^{a}_{\phantom{z}b} &=& -\frac{{\mathcal C}}{2a^{2}z_{0}^{2}}\delta^{a}_{\phantom{a}b}+\frac{{\mathcal C}}{2a^{2}z_{0}}\sum \left(\frac{F}{z_{0}}-2iw W\right){\mathcal E} e^{iwz}\delta^{a}_{\phantom{a}b}
\end{eqnarray}
Additionally, the scalar field equation of motion is given by

\begin{eqnarray}
W''+2{\mathcal H}W'+(w^{2}+k^{2})W &=& 
    \frac{iwU}{z_{0}}-\frac{iw F}{2z_{0}}-\frac{iwk^{2} X}{2z_{0}}+\frac{E'}{z_{0}}
\end{eqnarray}
Note that the scalar field is coupled solely to polar perturbations in its own field equation and in its contribution to the Einstein equations.
The background energy density scales as a curvature component would. The restrictions on allowable values of $K$ (and thus ${\mathcal C}$ via
(\ref{restriction}) due to probes of the background spacetime imply that as a component of the cosmological energy density the scalar field will only become 
non-negligible at late times \cite{Koivisto:2010dr}. Note that the perturbed stress energy tensor also scales the density of a curvature component multiplied by a small
perturbation.

\section{Boltzmann Equation for massless particles}
\label{boltmassless}

We expect that in a realistic universe it is appropriate to describe a field $\Psi$ defined on the spacetime (\ref{backshearless}) in terms of its particle content via 
a quantity $f^{\Psi}(x^{\mu},P^{\mu})$ called the distribution function, which reflects particle number density in phase space at a time $t$. 
This function is expected to obey the \emph{Boltzmann} equation, which is used to calculate the function's total time derivative:

\begin{eqnarray}
\frac{df^{\Psi}}{dt} &=& \frac{\partial f^{\Psi} }{\partial t}+\frac{\partial f^{\Psi} }{\partial x^{i}}\frac{dx^{i}}{dt}
      +\frac{\partial f^{\Psi} }{\partial P^{i}}\frac{dP^{i}}{dt}  = {\mathcal C}
\end{eqnarray}
where $i$ denotes the collective labels $z$ and coordinates on the co-moving 2-surface, and the partial derivative is evaluated at constant phase-space point.
The term ${\mathcal C}$ is a collisional term and reflects couplings to other fields.

We initially consider the collisionless (${\mathcal C}=0$) limit of the Boltzmann equation for a massless field, described by a distribution function $f$ and restrict ourselves to the case where
the massless field is the photon field (though the results of course will apply to any massless field). A considerable simplification that follows from the condition ${\mathcal C}=0$ is that the particles
will follow geodesics of the perturbed spacetime. The dependence of a function
upon $z$ and $x^{a}$ is familiar from the preceding sections, but the dependence on 4-momentum $P^{\mu}$ is new.
For a massless particle, this momentum satisfies a null constraint:

\begin{eqnarray}
\label{nullcon}
g_{\mu\nu}P^{\mu}P^{\nu}=0
\end{eqnarray}
Therefore, only three components of the 4-vector $P^{\mu}$ may vary independently. We shall find it convenient throughout to use the constraint (\ref{nullcon}) to
eliminate the time-component $P^{t}$ of the photon 4-momentum. We write the 4-momentum as follows:

\begin{eqnarray}
P^{\mu}\partial_{\mu}&=& \frac{1}{a}(\bar{P}^{t}\partial_{t}+p\hat{P}^{i}\partial_{i}) \\
\bar{P}^{t} &\equiv& aP^{t}\\
\hat{P}^{i}\partial_{i} &\equiv & \alpha \partial_{z}+p\sqrt{1-\alpha^{2}}\hat{p}^{a}\partial_{a} 
\end{eqnarray}
What are the interpretations of $p$, $\alpha$, and $\hat{p}^{a}$?
Calculating the norm of the spatial part according to the background spatial metric $h_{ij}=a^{2}h^{(C)}_{ij}$, where $h^{(C)}_{ij}$ is the co-moving background spatial 3-metric,
we have that:

\begin{eqnarray}
h_{ij}P^{i}P^{j} = h^{(C)}_{ij}p^{2}\hat{P}^{i}\hat{P}^{j} &=& p^{2}\alpha^{2}+p^{2}(1-\alpha^{2})\gamma_{ab}\hat{p}^{a}\hat{p}^{b}
\end{eqnarray}
where $h^{(C)}_{ij}$ is the co-moving background spatial metric.
If we identify $\hat{p}^{a}$ as components of a unit vector living in the background co-moving 2-surface i.e satisfying $\gamma_{ab}\hat{p}^{a}\hat{p}^{b}=1$ then clearly $p^{2}\equiv h_{ij}P^{i}P^{j} $ 
i.e. it the norm-squared of the physical spatial momentum of the photon according to the background spatial metric.  The variable $\alpha$ is then interpreted 
as the projection of the co-moving spatial momentum alongside a vector in the z direction which is unit with
respect to the co-moving background 3-metric. The value of using the variables $\alpha$ and $\hat{p}^{a}$ is that they more clearly reflect the underlying symmetry of the background spacetime.

The total time derivative $df/dt$ is then decomposed as follows: 

\begin{eqnarray}
\frac{df}{dt} &=& \frac{\partial f}{\partial t}+ \frac{\partial f}{\partial x^{i}}\frac{dx^{i}}{dt}+\frac{\partial f}{\partial p}\frac{dp}{dt} +\frac{\partial f}{\partial \alpha}\frac{d\alpha}{dt}+\frac{\partial f}{\partial (\hat{p}^{a})}\frac{d (\hat{p}^{a})}{dt} \label{dfdt}
\end{eqnarray}
It is useful first to express the derivative $\frac{dx^{i}}{dt}$ in terms of the photon momentum. 
Firstly we note that we may define the particle four-momentum in terms of the affine parameter of null trajectories $\lambda$ as follows $P^{\mu}=dx^{\mu}/d\lambda$
Therefore we have that:

\begin{eqnarray}
\label{dxdt1}
dx^{j}/dt= (dx^{j}/d\lambda)/(dx^{t}/d\lambda)= P^{j}/P^{t}=p\hat{P}^{j}/\bar{P}^{t}
\end{eqnarray}
Next we eliminate $\bar{P}^{t}$ via the null-constraint (\ref{nullcon}). Up to first order in metric perturbations this equation reads:

\begin{eqnarray}
\nonumber 0 &=& -(1+V)(\bar{P}^{t})^{2}+2p(\alpha  E+(B_{a}+R_{a})\hat{p}^{a}\sqrt{1-\alpha^{2}})\bar{P}^{t} \\
   \nonumber && +p^{2}+U(1-\alpha^{2})p^{2}+F\alpha^{2}p^{2} +2(S_{a}+b_{a})\hat{p}^{a}\alpha\sqrt{1-\alpha^{2}}p^{2} \\
    && +(1-\alpha^{2})p^{2}(X_{ab}+2 Q_{ab})\hat{p}^{a}\hat{p}^{b}
  \end{eqnarray}
Where for notational compactness the metric perturbations here are the actual metric perturbations i.e. the appropriate sum, integral of all the harmonic components/Fourier modes
of the perturbation. Solving this equation for $\bar{P}^{t}$ we have up to linear order in perturbations that: 

\begin{eqnarray}
\bar{P}^{t} &=& p+p(\alpha E+(C_{a}+R_{a})\hat{p}^{a}\sqrt{1-\alpha^{2}}+\frac{U}{2}(1-\alpha^{2})\nonumber \\
 &&+\frac{F}{2}\alpha^{2} 
                  +(S_{a}+B_{a})\hat{p}^{a}\alpha\sqrt{1-\alpha^{2}}\nonumber\\
   \label{ptsolv}               &&+\frac{1}{2}(1-\alpha^{2})(X_{ab}+2Q_{ab})\hat{p}^{a}\hat{p}^{b}-\frac{V}{2}) \label{pt}
\end{eqnarray}
Consequently,  using (\ref{ptsolv}) in conjunction with (\ref{dxdt1}), we may express $dx^{i}/dt$ entirely in terms of the variables $p$,$\alpha$, $\hat{p}^{b}$, and metric perturbations.

Another term that we must calculate is $dp/dt$. From the time component of the geodesic equation we have that:

\begin{eqnarray}
\frac{dP^{t}}{d\lambda} =\frac{\bar{P}^{t}}{a}\frac{d}{dt}\left(\frac{\bar{P}^{t}}{a}\right)= -\Gamma^{t}_{\mu\nu}P^{\mu}P^{\nu}
\end{eqnarray}
We may use this equation in conjunction with the solution $ \label{ptsolv}$ and the observation that the geodesic equations (as may be checked) imply that
$d\alpha/dt$ and $d\hat{p}^{a}/dt$ are zero at the background level to find the following expression for $dp/dt$:

\begin{eqnarray}
\frac{dp}{dt} &=& -p{\mathcal H}-\sum\frac{p}{2}(\alpha^{3}\partial_{z}F+2\alpha E'+\alpha\partial_{z}V+2(1-\alpha^{2}) U' \nonumber\\
&&
   +2\alpha^{2}F'+\alpha(1-\alpha^{2}) \partial_{z}U){\mathcal E} \nonumber\\
         &&  +\sum\frac{\sqrt{1-\alpha^{2}}p}{2} ((\alpha^{2}-1)U -2\alpha^{2}\partial_{z}B
          \nonumber  \\
       &&     -2C'-4\alpha B'-V-\alpha^{2}F)\hat{p}^{a}\nabla_{a}{\mathcal E} \nonumber\\
       &&  +\sum(1-\alpha^{2})p(-B\alpha- X'-\frac{1}{2}\partial_{z}X\alpha)\hat{p}^{a}\hat{p}^{b}\nabla_{a}\nabla_{b}{\mathcal E}\nonumber\\
       && -\sum\frac{\sqrt{1-\alpha^{2}}}{2}(1-\alpha^{2})pX\hat{p}^{a}\hat{p}^{b}\hat{p}^{c}\nabla_{a}\nabla_{b}\nabla_{c}{\mathcal E}\nonumber\\
       && +\sum \sqrt{1-\alpha^{2}}p(-R'-2\alpha S'-\alpha^{2}\partial_{z}S)\hat{p}^{a}\bar{\nabla}_{a}{\mathcal E}\nonumber\\
       &&+\sum (1-\alpha^{2})p(-\alpha  \partial_{z}Q-2 Q'-\alpha S) \hat{p}^{a}\hat{p}^{b}\nabla_{a}\bar{\nabla}_{b}{\mathcal E}\nonumber\\
       && -\sum \sqrt{1-\alpha^{2}}(1-\alpha^{2})pQ \hat{p}^{a}\hat{p}^{b}\hat{p}^{c}\nabla_{a}\nabla_{b}\bar{\nabla}_{c}{\mathcal E} \label{dpdt}
\end{eqnarray}
where the sum denotes the appropriate sums/integrals over harmonic components and the plane wave $e^{iwz}$ is implicit.

We now decompose $f$ into an unperturbed component $f_{0}(p,t)$ and a perturbed component $\delta f(\hat{P}_{i},x^{i},t)$. Note that we have assumed that 
the perturbation has a negligible dependence upon the photon momentum magnitude. 

In a realistic universe, the unperturbed distribution function $f_{0}$ is expected to be of the Bose-Einstein form. Motivated by this, we introduce a new field $\Theta(\bar{P}_{i},x^{i},t)$
of first order in smallness as follows:

\begin{eqnarray}
f &=& \frac{1}{e^{\frac{p}{T(t)(1+\Theta)}}-1}
\end{eqnarray}
where $T$ is the background field temperature and we recall that $p$ is the magnitude of the \emph{physical} photon 3-momentum. From the background geodesic equation we recover the solution $p=p_{0}/a$ where $p_{0}$ is a constant. 
A simple calculation shows that the background Boltzmann equation $df/dt=0$ is satisfied if $T=T_{0}/a$, a familiar result.

Our definition of the fields $\Theta$ and $\delta f$ implies that:

\begin{eqnarray}
\delta f = -p\frac{\partial f_{0}}{\partial p}\Theta
\end{eqnarray}
We now look to find the collisionless Boltzmann equation to first order in perturbations. We note that as each of $\frac{\partial f}{\partial \alpha}$, 
$\frac{d\alpha}{dt}$, $\frac{\partial f}{\partial (\hat{p}^{a})}$, and $\frac{d (\hat{p}^{a})}{dt}$ are quantities of order 1 or greater in perturbations,
the combined terms appearing in (\ref{dfdt}) may be neglected to first order. We have then that:

\begin{eqnarray}
\nonumber\frac{d\delta f}{dt} &=& -p\partial_{t}\left(\frac{\partial f_{0}}{\partial p}\Theta\right)-p\frac{dx^{i}}{dt}\frac{\partial f_{0}}{\partial p}{\partial_{i}\Theta}\\
   && +\left(\frac{\partial}{\partial p}\left(-p\frac{\partial f_{0}}{\partial p}\Theta\right)\right)\frac{dp_{(0)}}{dt}+\frac{\partial f_{0}}{\partial p}\frac{dp_{(1)}}{dt}
\end{eqnarray}
where $\frac{dp_{(0)}}{dt}$ and $\frac{dp_{(1)}}{dt}$ denote background and perturbed time derivatives of $p$, respectively.
Furthermore, it may be checked that 

\begin{eqnarray}
-p\partial_{t}\frac{\partial f_{0}}{\partial p}=\frac{\partial}{\partial p}\left(p\frac{\partial f_{0}}{\partial p}\right)\frac{dp_{(0)}}{dt}
\end{eqnarray}
The ensuing cancellation of terms implies that:

\begin{eqnarray}
\frac{d\delta f}{dt} &=& -p\frac{\partial f_{0}}{\partial p}\partial_{t}\Theta-p\frac{dx^{i}}{dt}\frac{\partial f_{0}}{\partial p}{\partial_{i}\Theta}+\frac{\partial f_{0}}{\partial p}\frac{dp_{(1)}}{dt}
\end{eqnarray}
Additionally, $dx^{i}/dt=p\hat{P}^{i}/\bar{P}^{t}=\hat{P}^{i}$ to zeroth order. Then, reading off $\frac{dp_{(1)}}{dt}$ from (\ref{dpdt}) we have:

\begin{eqnarray}
\nonumber 0 &=& \partial_{t}\Theta+\alpha\partial_{z}\Theta+\sqrt{1-\alpha^{2}}\hat{p}^{a}\partial_{a}\Theta \\
\nonumber && +\sum\frac{1}{2}(\alpha^{3}iw F+2\alpha E'+\alpha iwV+2(1-\alpha^{2}) U' \nonumber\\
&&
   +2\alpha^{2}F'+\alpha(1-\alpha^{2}) iwU){\mathcal E} \nonumber\\
         &&  -\sum\frac{\sqrt{1-\alpha^{2}}}{2} ((\alpha^{2}-1)U -2\alpha^{2}iw B
          \nonumber  \\
       &&     -2C'-4\alpha B'-V-\alpha^{2}F)\hat{p}^{a}\nabla_{a}{\mathcal E} \nonumber\\
       &&  -\sum(1-\alpha^{2})(-B\alpha- X'-\frac{1}{2}iw X\alpha)\hat{p}^{a}\hat{p}^{b}\nabla_{a}\nabla_{b}{\mathcal E}\nonumber\\
       && +\sum\frac{\sqrt{1-\alpha^{2}}}{2}(1-\alpha^{2})X\hat{p}^{a}\hat{p}^{b}\hat{p}^{c}\nabla_{a}\nabla_{b}\nabla_{c}{\mathcal E}\nonumber\\
       && -\sum \sqrt{1-\alpha^{2}}(-R'-2\alpha S'-\alpha^{2}iwS)\hat{p}^{a}\bar{\nabla}_{a}{\mathcal E}\nonumber\\
       &&-\sum (1-\alpha^{2})(-\alpha  iw Q-2 Q'-\alpha S) \hat{p}^{a}\hat{p}^{b}\nabla_{a}\bar{\nabla}_{b}{\mathcal E}\nonumber\\
       && +\sum \sqrt{1-\alpha^{2}}(1-\alpha^{2})Q \hat{p}^{a}\hat{p}^{b}\hat{p}^{c}\nabla_{a}\nabla_{b}\bar{\nabla}_{c}{\mathcal E} \label{bolt2}
\end{eqnarray}
Much as it is often more convenient to decompose a position dependent field
into harmonic components, it may be additionally convenient to decompose a field's dependence upon the unit vector $\hat{P}^{i}$ 
into `moments' so that the partial differential equation (\ref{bolt2}) may be cast as a system of ordinary differential equations in conformal time for
a set of moments for each harmonic mode. 
For the field $\Theta$, the appropriate decomposition of this dependence shall be expected to depend on the symmetries 
of the problem and the nature of the source terms in the equation. 

We first note that the source terms from the metric perturbations are a combination of sums of polar and axial harmonic modes. It shall
be useful then to split $\Theta(x^{i},\bar{P}^{i},t)$ into contributions $\Theta^{(P)}(x^{i},\bar{P}^{i},t)$ and  $\Theta^{(A)}(x^{i},\bar{P}^{i},t)$
such that $\Theta(x^{i},\bar{P}^{i},t)=\Theta^{(P)}(x^{i},\bar{P}^{i},t)+\Theta^{(A)}(x^{i},\bar{P}^{i},t)$. As the labels suggest, these are
polar and axial contributions to the temperature perturbation. We will now consider these contributions in detail.

\subsection{The polar temperature perturbation}

We propose that for a given curvature $K$, the field $\Theta^{(P)}(x^{i},\bar{P}^{i},t)$ may be decomposed as follows:

\begin{eqnarray}
\Theta^{(P)}(x^{i},\hat{P}^{i},t) &=& \sum_{A,0\leq M \leq A} \sum_{k,m,w} c^{(P)}_{MA}(w,k) \Theta^{(P)}_{MA}(w,k,t){\mathcal B}^{(P)}_{MA} \label{pe} 
\end{eqnarray}
where

\begin{eqnarray}
{\mathcal B}^{(P)}_{MA}(z,w,k,m,\alpha,\hat{p}^{a}) &=&  \frac{e^{iwz}P^{A}_{M}(\alpha)C^{ab..M}(\hat{p}^{c},\gamma^{de})\nabla_{ab..M}{\mathcal E}^{k}_{m}}{(-k)^{M}} \label{basisf}\\
\nabla_{ab..M} &\equiv & \nabla_{a}\nabla_{b}....\nabla_{M}
\end{eqnarray}
$P^{A}_{M}(\alpha)$ is an associated Legendre function, $A$ is a non-negative integer and we introduce the $M$th rank tensor $C^{ab..M}$, which we shall
call a polar Chebyshev tensor for reasons made clear below. This choice of basis function ${\mathcal B}^{(P)}_{MA}$ is strongly motivated by the treatment of 
cosmological perturbations on curved FRW backgrounds, where one decomposes the angular part of the photon distribution (i.e. $\alpha$ \emph{and} $\bar{p}^{a}$) in
terms of Legendre tensors (see \cite{White:1995qm},\cite{Gebbie:1998fe},\cite{Ellis1983487}).

Note that the value $k$ shall refer the positive square root of $kk^{*}$. The properties of this tensor will be discussed below. We shall seek to ultimately relate
the moments $\Theta^{(P)}_{MA}(w,k,t)$ to observables, and so the functions $c^{(P)}_{MA}(w,k)$ shall be chosen so as to simplify the resulting expressions.

\subsubsection{Polar Chebyshev Tensors}

We define polar Chebyshev tensors via the following scalar, vector, and recursion relation to generate tensors of higher rank:
\begin{eqnarray}
C &=& 1 \\
C^{a} &=& \hat{p}^{a} \\
\label{recur1}
C^{ab..N+1}&=& 2 \hat{p}^{(a}C^{b..N)}-\gamma^{(ab}C^{..N-1)}
\end{eqnarray}
where brackets denote symmetrization and the notation $C^{ab..N}$ denotes a tensor with $N$ indices.
The tensors satisfy the following properties:

\begin{eqnarray}
C^{.a.b.N} &=& C^{.b.a.N}\\
C^{.a.b.N}\gamma_{ab} &=& 0 \\
C^{..b..N}\hat{p}_{b} &=& C^{..N-1}\\
C^{abc..N}e_{a}e_{b}e_{c}..e_{N} &=& C_{N}(e^{a}\hat{p}_{a}) = \cos(N\cos^{-1}(e^{a}\hat{p}_{a}))
\end{eqnarray}
where $e^{a}$ is a unit vector with respect to the comoving $\gamma_{ab}$ and  $C_{N}(e^{a}\hat{p}_{a})$
is a \emph{Chebyshev polynomial} of the first kind. For notational convenience we define the following function:

\begin{eqnarray}
{\mathcal V}^{(P)}_{M} &=&  (-k)^{-M}(1-\alpha^{2})^{\frac{M}{2}}C^{ab..M}\nabla_{ab..M}{\mathcal E}^{k}_{m}
\end{eqnarray}
If $\Theta^{(P)}(x^{i},\hat{P}^{i},t) $
allows the decomposition (\ref{pe}), a relevant quantity to evaluate will be the directional derivative $\sqrt{1-\alpha^{2}}\hat{p}^{a}\nabla_{a}$ acting on
${\mathcal B}^{(P)}_{AM}$ in (\ref{bolt2}).
The part of this derivative independent of $\alpha$ may be developed as follows:

\begin{eqnarray}
\hat{p}^{a}C^{bc..N}\nabla_{a}\nabla_{bc..N}{\mathcal E}^{k}_{m}&=& \frac{1}{N+1}\hat{p}^{a}C^{bc..N}\nabla_{a}\nabla_{bc..N}{\mathcal E}^{k}_{m} +\frac{N}{N+1} \hat{p}^{a}C^{bc..N}\nabla_{bac..N+1}{\mathcal E}^{k}_{m} \nonumber\\
&&+\frac{N}{N+1} \sum\hat{p}^{a}C^{b.f..N}R_{abf}^{\phantom{abf}d}\nabla_{..d..N-1}
{\mathcal E}^{k}_{m} \\ 
  &=& \frac{1}{N+1}\hat{p}^{a}C^{bc..N}\nabla_{abc..N+1}{\mathcal E}^{k}_{m}  +\frac{1}{N+1}\hat{p}^{a}C^{bc..N}\nabla_{bac..N+1}{\mathcal E}^{k}_{m}\nonumber \\
  && +\frac{N-1}{N+1} \hat{p}^{a}C^{bc..N+1}\nabla_{bca..N+1}{\mathcal E}^{k}_{m} \nonumber \\
  &&+ \frac{N}{N+1} K\sum_{j} C^{..\beta..N-1}\nabla_{..\beta..}{\mathcal E}^{k}_{m}
\end{eqnarray}

where $R_{abf}^{\phantom{abf}d}$ is the Riemann curvature tensor on the comoving 2-surface and we use the curvature convention of \cite{Wald:1984rg}.
After the appropriate number of exchanges of indices, we recover: 

\begin{eqnarray}
\nonumber(1-\alpha^{2})^{\frac{N+1}{2}}\hat{p}^{a}C^{bc..N}\nabla_{abc..N+1}{\mathcal E}^{k}_{m} &=&(1-\alpha^{2})^{\frac{N+1}{2}} \hat{p}^{(a}C^{bc..N)}\nabla_{abc..N+1}{\mathcal E}^{k}_{m}\\
 && +\frac{(1-\alpha^{2})}{3}N(N-1)K(-k)^{N-1}{\mathcal V}_{N-1}\nonumber\\
\end{eqnarray}
The first term on the right hand side is closely related to a term from the recursion relation (\ref{recur1}). We then seek to express the other term from (\ref{recur1}), $\gamma^{(ab}C^{c...N-1)}\nabla_{abc..N+1}$, in terms of our mode functions. This is indeed possible, a lengthy calculation yielding the following result:

\begin{eqnarray}
(1-\alpha^{2})^{\frac{N-1}{2}}\gamma^{(ab}C^{c..N-1)}\nabla_{abc..N+1}{\mathcal E}^{l}_{m} &=& \left(-k^{2}+\frac{N(N^{2}-1)}{3(N+1)}K\right)(-k)^{N-1}{\mathcal V}_{N-1}\nonumber \\
\end{eqnarray}
Therefore, collecting results:

\begin{eqnarray}
\nonumber\hat{p}^{a}\nabla_{a}{\mathcal V}^{(P)}_{N} &=&(-k)^{-N}  \hat{p}^{(a}C^{bc..N)}\nabla_{a,bc..N}{\mathcal E}^{k}_{m} +\frac{\sqrt{1-\alpha^{2}}}{3}N(N-1)K(-k)^{-1}{\mathcal V}_{N-1} \\
&=&\frac{(-k)^{-N}\sqrt{1-\alpha^{2}}}{2}\nabla_{ab..N+1}{\mathcal E}^{k}_{m}\left(\frac{1}{1-\alpha^{2}}C^{ab..N+1}+\gamma^{(ab}C^{c..N-1)}\right) \nonumber\\
\nonumber&&+ \frac{\sqrt{1-\alpha^{2}}}{3}N(N-1)K(-k)^{-1}{\mathcal V}_{N-1} \\
&=& \frac{1}{\sqrt{1-\alpha^{2}}}\frac{k}{2}{\mathcal V}^{(P)}_{N+1}+\frac{\sqrt{1-\alpha^{2}}}{2}\left(-k^{2}+\frac{N(N^{2}-1)}{3(N+1)}K\right)(-k)^{-1}{\mathcal V}^{(P)}_{N-1}\nonumber \\
\nonumber&& +\frac{\sqrt{1-\alpha^{2}}}{3}N(N-1)K(-k)^{-1}{\mathcal V}^{(P)}_{N-1}  \\
&=& \frac{k}{2}\left(\frac{1}{\sqrt{1-\alpha^{2}}}{\mathcal V}^{(P)}_{N+1}+\kappa_{N}^{2}\sqrt{1-\alpha^{2}}{\mathcal V}^{(P)}_{N-1}\right)
\end{eqnarray}
where

\begin{eqnarray}
\label{kappa}
\kappa_{N}^{2} &\equiv & 1-N(N-1)\frac{K}{k^{2}} 
\end{eqnarray}
This implies that 

\begin{eqnarray}
\nonumber\sqrt{1-\alpha^{2}}\hat{p}^{a}\nabla_{a}{\mathcal B}^{(P)}_{NA} &=& \frac{k}{2}\left({\mathcal V}^{(P)}_{N+1}+\kappa_{N}^{2}(1-\alpha^{2}){\mathcal V}^{(P)}_{N-1}\right)(-1)^{N}\frac{d^{N}P_{A}(\alpha)}{d\alpha^{N}}e^{iwz}\\
\nonumber &=&\frac{k}{2}\kappa_{N}^{2}\left(\alpha(A-(N-1)){\mathcal B}^{(P)}_{N-1,A}-(A+N-1){\mathcal B}^{(P)}_{N-1,A-1}\right)\\
  && +\frac{k}{2}\frac{1}{2A+1}\left({\mathcal B}^{(P)}_{N+1,A-1}-{\mathcal B}^{(P)}_{N+1,A+1}\right) 
\end{eqnarray}
Furthermore from the recurrence relations for associated Legendre functions we have that:

\begin{eqnarray}
(2A+1)\alpha {\mathcal B}^{(P)}_{N,A} &=& (A-N+1){\mathcal B}^{(P)}_{N,A+1}+(A+N){\mathcal B}^{(P)}_{N,A-1}
\end{eqnarray}
and therefore

\begin{eqnarray}
\nonumber\sqrt{1-\alpha^{2}} \hat{p}^{a}\nabla_{a}{\mathcal B}^{(P)}_{NA}&=& \frac{k}{2}\left({\mathcal V}^{(P)}_{N+1}+\kappa_{N}^{2}(1-\alpha^{2}){\mathcal V}^{(P)}_{N-1}\right)(-1)^{N}\frac{d^{N}P_{A}(\alpha)}{d\alpha^{N}}e^{iwz}\\
\nonumber &=& \frac{k}{2}\frac{1}{(2A+1)}\left({\mathcal B}^{(P)}_{N+1,A-1}-{\mathcal B}^{(P)}_{N+1,A+1}\right) \\
\nonumber &&+\frac{k}{2}\kappa_{N}^{2}\frac{(A-N+1)(A-N+2)}{(2A+1)}{\mathcal B}^{(P)}_{N-1,A+1}\\
 && -\frac{k}{2}\kappa_{N}^{2}\frac{(A+N)(A+N-1)}{(2A+1)}{\mathcal B}^{(P)}_{N-1,A-1}
\end{eqnarray}
The directional derivative in the Boltzmann equation also involves contributions from variations of $B_{MA}$ along the $z$ direction. This is position dependence is more simple due to the 
plane wave decomposition along this direction. Again utilizing recurrence relations for associated Legendre functions, we have that:

\begin{eqnarray}
\alpha\partial_{z}{\mathcal B}^{(P)}_{N,A} &=& i\frac{w}{(2A+1)}\left((A-N+1){\mathcal B}^{(P)}_{N,A+1}+(A+N){\mathcal B}^{(P)}_{N,A-1}\right)
\end{eqnarray}
Collecting terms, we have that the three dimensional directional derivative of the basis function ${\mathcal B}^{(P)}_{NA}$ is given by:

\begin{eqnarray}
\label{dirdevalf}
\hat{P}^{i}\nabla_{i}{\mathcal B}^{(P)}_{NA}  \equiv \sum_{B=A-1}^{B=A+1}\sum_{M=N-1}^{M=N+1}{\mathcal M}^{(N-1-M,A-1-B)}(N,A){\mathcal B}^{(P)}_{MB}
\end{eqnarray}
where

\[
{\mathcal M} =
\left( {\begin{array}{ccc}
\frac{-k\kappa^{2}_{N}}{2}\frac{(A+N)(A+N-1)}{(2A+1)}& 0 &\frac{k\kappa^{2}_{N}}{2}\frac{(A-N+1)(A-N+2)}{(2A+1)}\\
\frac{iw(A+N)}{(2A+1)}  & 0  & \frac{iw(A-N+1)}{(2A+1)} \\
\frac{k}{2}\frac{1}{2A+1}& 0 & -\frac{k}{2}\frac{1}{2A+1}
 \end{array} } \right)
\]
We designate rows and columns so that the $00$ index position (upper left in the matrix) corresponds to $M=N-1$ and
$B=A-1$. Advancing down a column increases $M$ by 1, advancing to the right across a row advances $B$ by 1. Therefore we have that 

\begin{eqnarray}
\nonumber
\hat{P}^{i}\nabla_{i}\Theta^{(P)}(x^{i},\hat{P}^{i},t) &=& \sum_{w,k,m} \sum_{A,M,B}\sum^{A}_{N=0} c^{(P)}_{NA}\Theta^{(P)}_{NA}(w,l,t)\\
&&\cdot{\mathcal M}^{(N-1-M,A-1-B)}(NA){\mathcal B}^{(P)}_{MB}
\label{dirdiv}
\end{eqnarray}
This is a crucial result as it enables the partial differentials to be replaced in (\ref{bolt2}) by sums of other moments of $\Theta^{(P)}(x^{i},\hat{P}^{i},t)$.
Additionally, we note that all of the polar metric source terms in (\ref{bolt2}) may be expressed in terms of the ${\mathcal B}^{(P)}_{NA}$ functions. Consequently, 
the polar Boltzmann equation (\ref{bolt2}) may be expressed in terms of a sum of ${\mathcal B}^{(P)}_{NA}$ functions. 
It is desirable now to find an operator that, when applied
to the mode and moment expansion of $\Theta^{(P)}(x^{i},\bar{P}^{i},t) $ yields a term involving only a single moment $\Theta^{(P)}_{NA}(w,l,t)$. This will
enable us to express the partial time derivative of $\Theta^{(P)}(x^{i},\bar{P}^{i},t)$ the time derivative of the moment  $\Theta^{(P)}_{NA}(w,l,t)$ along with terms of metric perturbations and other moments of the photon distribution. Collectively these ordinary differential equations in time will be equivalent to (\ref{bolt2}). 

\subsubsection{A polar projection operator}

Consider a field $V(x^{i},\hat{P}^{i},t)$ may be decomposed into \emph{polar} moments $V_{NA}(k,,w,t)$ as follows:

\begin{eqnarray}
V(\hat{P}^{i},x^{i},t) &=& \sum_{k,m,w} \sum_{A,N} V_{NA}(k,w,t){\mathcal B}^{(P)}_{NA}
\label{modo}
\end{eqnarray}
and recall that $\hat{P}^{i}=(\alpha,\sqrt{1-\alpha^{2}}\hat{p}^{a})$.

We seek to find an operator  that can operate on $V(\hat{P}^{i},x^{i},t)$ to isolate a single moment $V_{AN}(k,w,t)$.

Let us consider the operator ${\mathcal O}^{(P)}(M',M,B,k',m',w') $ defined as:

\begin{eqnarray}
\nonumber
{\mathcal O}^{(P)}\equiv \sum_{m'}\int e^{i(M'-M)\psi}d\psi \int d^{2}S\int dz e^{-iw'z} \int d\Omega \frac{C^{ab..M}\nabla_{ab..M}({\mathcal E}^{k'}_{m'})^{*}}{(-k')^{M}} P^{B}_{M}(\alpha)\\
\label{projp}
\end{eqnarray}
where $d^{2}S$ is an area element in the co-moving 2-surface and $d\Omega=-d\alpha d\chi$ is a solid angle element in the space of photon momentum i.e. $\chi$ is the angle between $\hat{p}^{a}$ and
an arbitrary unit vector at the origin of the momentum space surface orthogonal to $(\partial_{z})^{i}$. The meaning of the newly introduced parameter $\psi$ will be discussed later.
Consider the following operation:

\begin{eqnarray}
{\mathcal O}^{(P)}V(\hat{P}^{i},x^{i},t) &=& \sum_{A,N,k,m,m',w} V_{NA}(k,w,t){\mathcal O}^{(P)}(N,M,B,k',w',m) {\mathcal B}^{(P)}_{NA} \nonumber \\
\end{eqnarray}

We will evaluate this integral in stages. Examining (\ref{basisf}) and (\ref{modo}), we see that (\ref{projp}) will contain the following integral:

\begin{eqnarray}
\label{int1}
\int d\chi C^{ab..M}C^{cd..N}
\end{eqnarray}
We introduce a tensor $W^{ab...m}_{\phantom{ab...m}cd...m}$, defined as follows:

\begin{eqnarray}
\label{wdef}
C^{ab..m} =W^{ab...m}_{\phantom{ab...m}cd...m} \hat{p}^{c}\hat{p}^{d}...\hat{p}^{m}
\end{eqnarray}
For instance:

\begin{eqnarray}
W^{ab}_{\phantom{ab}ef} =2\gamma^{a}_{\phantom{a}(e}\gamma^{b}_{\phantom{a}f)}-\gamma^{ab}\gamma_{ef}
\end{eqnarray}
Therefore (\ref{int1}) may be written as follows:

\begin{eqnarray}
\int d\chi C^{ab..M}C^{ef..N}= W^{ab..M}_{\phantom{ab..M}cd..M}W^{ef..N}_{\phantom{ef..N}gh..N}\int  d\chi \hat{p}^{a}\hat{p}^{g}..\hat{p}^{M}\hat{p}^{N}
\label{intcc}
\end{eqnarray}
The integral over $\hat{p}^{a}\hat{p}^{b}..$ in (\ref{intcc}) may be evaluated by expressing each $\hat{p}^{a}$ in terms of the introduced angle $\chi$ an a 
local orthonormal basis i.e.
$\hat{p}^{a}=(\cos\chi,\sin\chi)$. The integral may then be evaluated using standard methods, yielding:

\begin{eqnarray}
\int_{0}^{2\pi} \hat{p}^{a}\hat{p}^{b}...\hat{p}^{M}\hat{p}^{N} d\chi = \frac{2\pi(M+N)!\gamma^{(ab..(M+N))}}{2^{(M+N)}(((M+N)/2)!)^{2}}
\end{eqnarray}
where we have defined the tensor $\gamma^{(abc..m)}\equiv \gamma^{(ab}\gamma^{cd}...\gamma^{.m)}$, it is taken that
the result is only nonvanishing when $(M+N)$ is an even number.

By the definition of the tensor $W^{ab...m}_{\phantom{ab...m}cd...m}$, it must be traceless with respect
to contraction with $\gamma^{ef}$ across any two lower indices. Therefore, (\ref{int1}) is only nonvanishing
if $\gamma^{(ab..(M+N))}$ is contracted with one index of each component $\gamma^{ef}$ acting on one of the two
$W$ tensors, and the other acting on the other. This implies that $M=N$. By a counting argument, there're $2^{Q}Q!(Q)!$ permutations of
$\gamma^{ab..2Q}$ that fulfil this condition. Therefore:

\begin{eqnarray}
\nonumber\int d\chi C^{ab..M}C^{wf..M}&=& W^{ab..M}_{\phantom{ab..M}cd..M}W^{ef..M}_{\phantom{ef..N}gh..M} \frac{(2M)!}{2^{(2M)-1}((M)!)^{2}}\pi\gamma^{(cg..2M)} \\
\nonumber  &=& \pi W^{ab..M,cd..M}W^{ef..M}_{\phantom{ef..N}cd..M} \frac{(2M)!}{2^{(2M)-1}((M)!)^{2}}\frac{2^{M}(M)!(M)!}{(2M)!}\\
  &=& \frac{2\pi}{2^{M}} W^{ab..M,cd..M}W^{ef..M}_{\phantom{ef..N}cd..M} 
\end{eqnarray}
Given the condition that $N=M$, we may now simply evaluate the integral from standard results:

\begin{eqnarray}
-\int d\alpha P^{A}_{M}(\alpha)P^{B}_{M}(\alpha) &=& -2\frac{(A+M)!}{(2A+1)(A-M)!}\delta_{AB}
\end{eqnarray}
Similarly the following integral holds:

\begin{eqnarray}
\int_{-\infty}^{+\infty} dz e^{-iw'z}e^{iwz}= \delta(w-w')
\end{eqnarray}
We have then reduced the expression to:

\begin{eqnarray}
-\delta(w-w')2\frac{(A+M)!}{(2A+1)(A-M)!}\delta_{AB}\delta_{MN}\frac{2\pi}{2^{M}}
\end{eqnarray}

The final integral to evaluate is:

\begin{eqnarray}
\label{over2}
\int  d^{2}S W^{ab..M,cd..M}W^{ef..M}_{\phantom{ef..N}cd..M} \nabla_{ab..M}{\mathcal E}^{*k'}_{m'} \nabla_{ef..M}{\mathcal E}^{k}_{m}
\end{eqnarray}
This calculation is simplified by the introduction of some new quantities:

\subsubsection{The $\textbf{A}^{\pm}$ vectors}

Consider the vectors $(\textbf{e}_{\theta})^{a}=(\partial_{\theta})^{a}$ and
$(\textbf{e}_{\phi})^{a}=(1/\sqrt{\tilde{\gamma}_{\phi\phi}})(\partial_{\phi})^{a}$ , where the coordinates $\theta$ and $\phi$ are 
angular coordinates on the unit sphere for the case $K>0$, cylindrical coordinates for the case $K=0$, and hyperbolic
angular coordinates in the case $K>0$.
This set of vectors are orthonormal with respect to the metric $\tilde{\gamma}^{ab}$ except at the origin.
We define the following complex vectors:

\begin{eqnarray}
\label{ap}
(\textbf{A}^{+})^{a}  \equiv \frac{|K|^{1/2}}{\sqrt{2}}\left((\textbf{e}_{\theta})^{a}-i(\textbf{e}_{\phi})^{a}\right) \\
\label{am}
(\textbf{A}^{-})^{a}  \equiv \frac{|K|^{1/2}}{\sqrt{2}}\left((\textbf{e}_{\theta})^{a}+i(\textbf{e}_{\phi})^{a}\right)
\end{eqnarray}
Note that in the case $|K|=0$, the relations (\ref{ap}) and (\ref{am}) reduce to $(\textbf{A}^{\pm})^{a} =
(\textbf{e}_{r})^{a}\mp i(\textbf{e}_{\phi})^{a}$ where $r$ and $\phi$ are cylindrical coordinates on a flat
2-surface.

The following identities then hold for arbitrary values of $K$:

\begin{eqnarray}
(\textbf{e}_{\theta})^{a} &=& \frac{|K|^{-1/2}}{\sqrt{2}}\left((\textbf{A}^{+})^{a}+(\textbf{A}^{-})^{a} \right) \\
(\textbf{e}_{\phi})^{a} &=& \frac{i|K|^{-1/2}}{\sqrt{2}}\left((\textbf{A}^{+})^{a}-(\textbf{A}^{-})^{a} \right) \\
\label{prop1}
(\textbf{A}^{+})^{a}(\textbf{A}^{+})_{a} &=& (\textbf{A}^{-})^{a}(\textbf{A}^{-})=0 \\
\label{prop2}
(\textbf{A}^{-})^{a}(\textbf{A}^{+})_{a} &=& (\textbf{A}^{+})^{a}(\textbf{A}^{-})=1 \\
\gamma_{ab} &=& (\textbf{A}^{+})_{a} (\textbf{A}^{-})_{b} +(\textbf{A}^{-})_{a} (\textbf{A}^{+})_{b} \\
i\epsilon_{ab} &=&  (\textbf{A}^{+})_{a} (\textbf{A}^{-})_{b} -(\textbf{A}^{-})_{a} (\textbf{A}^{+})_{b} 
\end{eqnarray} 
Furthermore, for compactness of notation we define tensors

\begin{eqnarray}
\textbf{A}^{(\pm)ab..M}\equiv(\textbf{A}^{\pm})^{a}(\textbf{A}^{\pm})^{b}..(\textbf{A}^{\pm})^{M}
\end{eqnarray}

We may use $(\textbf{A}^{\pm})^{a}$ as basis vectors with which to construct tensors. For instance, a component $X_{++}$ of a tensor
$X_{ab}$ is given by $(\textbf{A}^{-})^{a}(\textbf{A}^{-})^{a}X_{ab}$. As a result of the properties (\ref{prop1}) and (\ref{prop2}), a
tensor $X_{ab..}$ expressed in terms of $(\textbf{A}^{\pm})^{a}$ has only components which are traceless with respect to any possible contraction of indices with $\gamma^{ab}$: the components $X_{\pm\pm..}$.

Consider a tensor operator $\textbf{\o}$ which acts on another tensor to produce a tensor proportional to part of the tensor which is traceless for all possible contractions
and symmetric for all exchanges of indices. $W^{ab..m}_{\phantom{ab}cd..m}$  is an example of such a tensor but need not be the only possibility- we shall see that a different tensor may be constructed using the metric and the volume form on the two surface. The operation of $\textbf{\o}$ on a tensor
$X_{ab..M}$ will yield the following result:

\begin{eqnarray}
\nonumber
(\textbf{\o})^{ab..M}_{\phantom{ab..M}cd..M}(\textbf{X})_{ab..M} &=& \beta_{+}\cdot (X_{++..M}) (\textbf{A}^{(+)})_{cd..M}\\
&&+\beta_{-} \cdot(X_{--..M})(\textbf{A}^{(-)})_{cd..M}
\end{eqnarray}
where $\beta_{+}$ and $\beta_{-}$ are constants independent of $\textbf{X}$. We now find $\beta_{+}$ and $\beta_{-}$ for the case $(\textbf{\o})^{ab..M}_{\phantom{ab..M}cd..M}=W^{ab..M}_{\phantom{ab..M}cd..M}$. By its definition we have that:

\begin{eqnarray}
W^{ab..M}_{\phantom{ab..M}cd..M}\hat{p}_{a}\hat{p}_{b}..\hat{p}_{M} &=& C_{cd..M}
\end{eqnarray}
For notational convenience we introduce the notation $\hat{p}^{ab..M}\equiv \hat{p}^{a}\hat{p}^{b}..\hat{p}^{M}$

By explicit calculation, we have 

\begin{eqnarray}
C_{a} &=& \hat{p}_{+}(\textbf{A}^{+})_{a}+\hat{p}_{-}(\textbf{A}^{-})_{a}
\end{eqnarray}
Then repeatedly applying the recurrence relation (\ref{recur1}), we then have that for $m\geq 1$:

\begin{eqnarray}
C_{cd..M} =2^{M-1}\hat{p}_{++..M(+)}(\textbf{A}^{+})_{cd..M}+2^{M-1}\hat{p}_{--..M(-)}(\textbf{A}^{-})_{cd..M}
\end{eqnarray}
This implies that $\beta_{(P)+}= 2^{M-1}$, $\beta_{(P)-}=2^{M-1}$ where $M\geq 1$, and hence:

\begin{eqnarray}
\nonumber
W^{ab..M}_{\phantom{ab..M}cd..M}\nabla_{ab..M}{\mathcal E}^{k}_{m}&=& 2^{M-1}\cdot \nabla_{++..M}{\mathcal E}^{k}_{m} (\textbf{A}^{(+)})_{cd..M}\\
&&+2^{M-1}\cdot\nabla_{--..M}{\mathcal E}^{k}_{m}(\textbf{A}^{(-)})_{cd..M}
\end{eqnarray}
The integral (\ref{over2}), now accompanied by the integration over $\psi$ reduces to:

\begin{eqnarray}
\nonumber
&&\int e^{i(M'-M)\psi}d\psi\int d^{2}S  2^{2M-1}(\nabla^{--..M}{\mathcal E}^{*k'}_{m'}\nabla_{++..M}{\mathcal E}^{k}_{m}\\
&&
+\nabla^{++..M}{\mathcal E}^{*k'}_{m'}\nabla_{--..M}{\mathcal E}^{k}_{m})\nonumber\\
&=&\int e^{i(M'-M)\psi}d\psi\int d^{2}S  2^{2M-1}((-1)^{m'+m}\nabla^{++..M}{\mathcal E}^{*k'}_{-m'}\nabla_{--..M}{\mathcal E}^{k}_{-m}\nonumber\\
&&+\nabla^{++..M}{\mathcal E}^{*k'}_{m'}\nabla_{--..M}{\mathcal E}^{k}_{m}) 
\end{eqnarray}
where we have used the fact that ${\mathcal E}^{*k}_{p}=(-1)^{p}{\mathcal E}^{k}_{-p}$ for each of our cases.

The total expression is then proportional to:

\begin{eqnarray}
&&-2\pi\sum_{k,w,m,m'}2^{M} \sum_{N}\frac{(A+M)!\delta_{AB}\delta_{MN}\delta(w-w')V_{MA}}{(2A+1)(A-M)!(-k)^{2M}}\int d\psi\int d^{2}S\nonumber\\
&& \cdot ((-1)^{m'+m}\nabla^{++..M}{\mathcal E}^{*k'}_{-m'}\nabla_{--..M}{\mathcal E}^{k}_{-m}
+\nabla^{++..M}{\mathcal E}^{*k'}_{m'}\nabla_{--..M}{\mathcal E}^{k}_{m}) \nonumber\\
&=&-\pi\sum_{k,w,m,m'}2^{M} \sum_{N}\frac{(A+M)!\delta_{AB}\delta_{MN}\delta(w-w')V_{MA}}{(2A+1)(A-M)!(-k)^{2M}}\int d\psi\int d^{2}S \nonumber \\
&&\cdot((-1)^{-(m'+m)}+1)(\nabla^{++..M}{\mathcal E}^{*k'}_{m'}\nabla_{--..M}{\mathcal E}^{k}_{m}) 
\
\end{eqnarray}
where we have used the fact that for a value of $|p|$, the total expression is a sum over $-|p|$ and $+|p|$, enabling the above rearrangement terms in the summation above.

We first consider the closed $K>0$ case.
By the results of \ref{appa}, we have that 

\begin{eqnarray}
 \nabla^{++..M}{\mathcal E}^{*k'}_{m'}\nabla_{--..M}{\mathcal E}^{k}_{m}&=& \left(\frac{|K|}{2}\right)^{M}\left(\frac{2L+1}{4\pi}\right)T^{(L')*}_{-m',M}T^{L}_{-m,M}i^{-m'+m} \nonumber \\
 && \cdot \left(\prod^{M}_{q=0}\sqrt{(L+q)(L-q+1)}\right) \left(\prod^{M}_{q=0}\sqrt{(L'+q)(L'-q+1)}\right) \nonumber
 \end{eqnarray}
The matrix elements $T^{(L')*}_{-p',M'}T^{L}_{-p,M}$ are orthogonal with respect to integration over the group \cite{Vilenkin1} i.e.

\begin{eqnarray}
\int T^{(L')*}_{-m',M'}T^{L}_{-m,M} \sin\theta d\psi d\phi d\theta \propto \frac{1}{2L+1}\delta_{ll'}\delta_{MM'}\delta_{mm'}
\end{eqnarray}
Therefore, collecting results we see that the operator ${\mathcal O}^{(P)}$ applied to the function $V(x^{i},\hat{P}^{i},t)$ in the closed case yields
a term proportional to:

\begin{eqnarray}
\nonumber
{\mathcal O}^{(P)}V(x^{i},\hat{P}^{i},t) &\propto & \left(\frac{|K|}{k^{2}}\right)^{M}\frac{(A+M)!}{(2A+1)(A-M)!}V_{MA}(w,L,t)
  \left(\prod^{M}_{q=0}(L+q)(L-q+1)\right)\\
  &&\cdot \delta_{AB}\delta_{MN}\delta(w-w')\delta_{ll'}\delta_{mm'} \\
   &=& \frac{(A+M)!}{(2A+1)(A-M)!}V_{MA}(w,L,t)
  \left(\prod^{M}_{q=0}\kappa_{q}^{2}\right)\nonumber\\
  &&\cdot \delta_{AB}\delta_{MN}\delta(w-w')\delta_{ll'}\delta_{mm'}
 \end{eqnarray}
 where the constant of proportionality may depend on labels other than A and M and we have used the definition of $\kappa_{q}$ (\ref{kappa}). The terms which depend
 on $A$ and $M$ are important as we shall seek to operate on (\ref{bolt2}) with ${\mathcal O}^{(P)}$ which, via
 the directional derivative, links $\Theta_{MA}$ to basis functions with different labels (for instance $M+1,A$).
 Terms which depend upon the value of harmonic labels $k,m,w$ however will be common to all terms
 and may be ignored if nonzero.

 For the $K<0$ case, the above argument may be repeated, relating derivatives of the harmonic functions to 
 successive derivatives projected along $(\textbf{A}^{\pm})^{a}$. The operator ${\mathcal O}^{(P)}$ applied to a function $V$ in
 the open case yields a term proportional to:
 
 \begin{eqnarray}
\nonumber
{\mathcal O}^{(P)}V &\propto& \left(\frac{|K|}{k^{2}}\right)^{M}\frac{(A+M)!}{(2A+1)(A-M)!}V_{MA}(w,\rho,t)
  \left(\prod^{M}_{q=0}(i\rho-\frac{1}{2}+q)(-i\rho-\frac{1}{2}+q))\right)\\
  &&\cdot\delta_{AB}\delta_{MN}\delta(w-w')\delta(\rho-\rho')\delta_{mm'} \\
 &=& \frac{(A+M)!}{(2A+1)(A-M)!}V_{MA}(w,\rho,t)
  \left(\prod^{M}_{q=0}\kappa_{q}^{2}\right)\nonumber\\
  &&\cdot\delta_{AB}\delta_{MN}\delta(w-w')\delta(\rho-\rho')\delta_{mm'} 
 \end{eqnarray}
 For the $K=0$ case, the above argument may be repeated, relating derivatives of the harmonic functions to 
 successive derivatives projected along $(\textbf{A}^{\pm})^{a}$. The operator ${\mathcal O}^{(P)}$ applied to a function $V$ in
 the open case yields a term proportional to:
 
  \begin{eqnarray}
\nonumber
{\mathcal O}^{(P)}V &\propto& \frac{(A+M)!}{(2A+1)(A-M)!}V_{MA}(w,\rho,t)\cdot\delta_{AB}\delta_{MN}\delta(w-w')\delta(\rho'-\rho)\delta_{mm'} 
 \end{eqnarray}
 We now consider the effect of applying the polar projection operator, so as to pick out moments labelled $N'$ and $A'$, onto
the directional derivative (\ref{dirdiv}) i.e.

\begin{eqnarray}
{\mathcal O}^{(P)}\hat{P}^{i}\nabla_{i}V(x^{i},\hat{P}^{i},t)
\end{eqnarray}

 A straightforward calculation shows that 
\begin{eqnarray}
{\mathcal O}^{(P)}\hat{P}^{i}\nabla_{i}V(x^{i},\hat{P}^{i},t)&\propto&\prod^{N'}_{q=0}\kappa_{q}^{2}\sum_{\alpha=0}^{2}\sum_{\beta=0}^{2}
\Theta^{(P)}_{N'-1+\alpha,A'-1+\beta}c^{(P)}_{N'-1+\alpha,A'-1+\beta} \nonumber\\
&&\cdot {\mathcal M}^{(2-\alpha,2-\beta)}_{N'-1+\alpha,A'-1+\beta}\frac{(A'+N')!}{(2A'+1)(A'-N')!}
\end{eqnarray}
where again functions that are common to both the results of ${\mathcal O}^{(P)}V(x^{i},\hat{P}^{i},t)$ and 
 ${\mathcal O}^{(P)}\hat{P}^{i}\nabla_{i}V(x^{i},\hat{P}^{i},t)$ are absorbed into the proportionality.

 \subsection{The axial temperature perturbation}

We propose that for a given curvature $K$, the axial temperature perturbation field $\Theta^{(A)}(x^{i},\bar{P}^{i},t)$ may be decomposed as follows:

\begin{eqnarray}
\Theta^{(A)}(x^{i},\hat{P}^{i},t) &=& \sum_{A,0\leq M \leq A} \sum_{k,m,w} c^{(A)}_{MA}(w,k) \Theta^{(A)}_{MA}(w,k,t){\mathcal B}^{(A)}_{MA} \label{ae} 
\end{eqnarray}
where

\begin{eqnarray}
{\mathcal B}^{(A)}_{MA}(z,w,k,m,\alpha,\hat{p}^{a}) &=&  \frac{e^{iwz}P^{A}_{M}(\alpha){\mathcal G}^{ab..M}(\hat{p}^{c},\gamma^{de})\nabla_{ab..M}{\mathcal E}^{k}_{m}}{(-k)^{M}} \label{basisf}
\end{eqnarray}
The difference between the axial and polar case is encoded in the appearance of the tensor ${\mathcal G}^{ab..M}$ instead of $C^{ab..M}$.  We shall
call the former tensor an  \emph{axial} Chebyshev tensor. The properties of this tensor will be discussed below. As in the polar case, we shall seek to ultimately relate
the moments $\Theta^{(A)}_{MA}(w,k,t)$ to observables, choosing the functional form of the quantity $c^{(A)}_{MA}(w,k)$ so as to simplify the appearance and

\subsubsection{Axial Chebyshev Tensors}

We define the set of axial Chebyshev tensors via the following scalar, vector, and recurrence relation
\begin{eqnarray}
{\mathcal G} &=& 0 \\
{\mathcal G}^{a} &=& \epsilon_{b}^{\phantom{b}a}\hat{p}^{b} \\
\label{recur}
{\mathcal G}^{ab..N+1}&=& 2 \hat{p}^{(a}{\mathcal G}^{b..N)}-\gamma^{(ab}{\mathcal G}^{..N-1)}
\end{eqnarray}
The tensors satisfy the following properties

\begin{eqnarray}
{\mathcal G}^{.a.b.N} &=& {\mathcal G}^{.b.a.N}\\
{\mathcal G}^{.a.b.N}\gamma_{ab} &=& 0 \\
{\mathcal G}^{..b..N}\hat{p}_{b} &=& {\mathcal G}^{..N-1}\\
{\mathcal G}^{ab..N}e_{a}e_{b}..e_{N} &=& \sin(N\cos^{-1}(\hat{p}^{c}e_{c}))
\end{eqnarray}
where again $e_{a}$ satisfies $e^{a}e^{b}\gamma_{ab}=1$.

An immediate implication of the identical recurrence relations obeyed by polar and axial Chebyshev tensors is that the axial equivalents of the functions ${\mathcal V}_{NA}$ and ${\mathcal B}^{(P)}_{NA}$ 
are affected by the directional derivative in the Boltzmann equation in an identical manner. Therefore we have that:

\begin{eqnarray}
\nonumber
\hat{P}^{i}\nabla_{i}\Theta_{(A)}(x^{i},\bar{P}^{i},t) =\sum_{k,m,w}\sum_{A}\sum^{A}_{N=0} c^{(A)}_{NA}\Theta^{(A)}_{NA}(w,k,t){\mathcal M}^{MB}(N,A){\mathcal B}^{(A)}_{MB}
\end{eqnarray}
Again we note that the axial metric source terms in the Boltzmann equation (\ref{bolt2}) are simply expressed in the basis functions of the above expansion. As in the polar case, we
now seek to find an operator that can isolate individual moments of a function which may be expressed in terms of a sum of axial moments.

\subsubsection{Axial projection operator}
 
Motivated by the similar formalism between polar and axial cases, consider an axial projection operator ${\mathcal O}^{(A)}$,
obtained by taking the operator ${\mathcal O}^{(P)}$ and replacing the polar Chebyshev tensor $C^{ab..M}$ with an axial
Chebyshev tensor ${\mathcal G}^{ab..M}$ i.e.

\begin{eqnarray}
\nonumber
{\mathcal O}^{(A)}\equiv \left(\int e^{i(M'-M)\psi}d\psi \int d^{2}S\int dz e^{-iw'z} \int d\Omega \frac{{\mathcal G}^{ab..M'}\nabla_{ab..M'}{\mathcal E}^{*k'}_{p'}}{(-k')^{M'}} P^{B}_{M'}(\alpha)\right)\\
\label{proj}
\end{eqnarray}
 where again ${\mathcal O}^{(A)}$ is simply related to ${\mathcal O}$ by swapping polar with axial Chebyshev tensors.
 
 The analysis of the effect of ${\mathcal O}^{(A)}$ proceeds as in the ${\mathcal O}^{(P)}$ case. The significant difference is that one may no longer use the tensor $W^{ab..M}_{\phantom{ab..}cd..M}$ to related ${\mathcal G}^{ab..M}$ 
 to $\hat{p}^{a}\hat{p}^{b}..\hat{p}^{M}$. We instead introduce a new tensor $Z^{ab..M}_{\phantom{ab..}cd..M}$, which is a second variety of the operator $\textbf{\o}$, defined via the relation:
 
 \begin{eqnarray}
 \label{zdef}
{\mathcal G}^{ab..M} &=&  Z^{ab..M}_{\phantom{ab..M}cd..M}\hat{p}^{cd..M} 
 \end{eqnarray}
 By explicit calculation, we have 

\begin{eqnarray}
{\mathcal G}_{a} &=& \frac{1}{i}\left(\hat{p}_{+}(\textbf{A}^{+})_{a}-\hat{p}_{-}(\textbf{A}^{-})_{a}\right)
\end{eqnarray}
Then repeatedly applying the recurrence relation (\ref{recur}), we then have that for $M\geq 1$:

\begin{eqnarray}
{\mathcal G}_{cd..M} &=&\frac{1}{i}(2^{M-1}\hat{p}_{++..M(+)}(\textbf{A}^{+})_{cd..M}\\
&&-2^{M-1}\hat{p}_{--..M(-)}(\textbf{A}^{-})_{cd..M})
\end{eqnarray}
This implies that ${\beta}_{(A)+}=2^{M-1}/i$ and ${\beta}_{(A)+}=-2^{M-1}/i$, and therefore:

 \begin{eqnarray}
\nonumber
Z^{ab..M}_{\phantom{ab..M}cd..M}\nabla_{ab..M}{\mathcal E}^{k}_{p}&=& \frac{1}{i}(2^{M-1}\cdot \nabla_{++..M}{\mathcal E}^{k}_{p} (\textbf{A}^{(+)})_{cd..M}\\
   && -2^{M-1}\cdot\nabla_{--..M}{\mathcal E}^{k}_{p}(\textbf{A}^{(-)})_{cd..M})
\end{eqnarray}
However, it is readily seen upon evaluation that the integrals produce the same result i.e. 
a result proportional to 
 
\begin{eqnarray}
&=& \frac{(A+M)!}{(2A+1)(A-M)!}V^{(A)}_{MA}(w,l,t)
  \left(\prod^{M}_{q=0}\kappa_{q}^{2}\right)\nonumber\\
  &&\cdot \delta_{AB}\delta_{MN}\delta(w-w')\delta_{ll'}\delta_{mm'}
 \end{eqnarray}
 for $K>0$
 
 \begin{eqnarray}
&&\frac{(A+M)!}{(2A+1)(A-M)!}V^{(A)}_{MA}(w,k,t)
  \left(\prod^{M}_{q=0}\kappa_{q}^{2}\right)\nonumber\\
  &&\cdot\delta_{AB}\delta_{MN}\delta(w-w')\delta(\rho-\rho')\delta_{mm'} 
 \end{eqnarray}
 for $K<0$
 
  \begin{eqnarray}
\nonumber
&& \frac{(A+M)!}{(2A+1)(A-M)!}V^{(A)}_{MA}(w,\rho,t)\\
&&\cdot\delta_{AB}\delta_{MN}\delta(w-w')\delta(\rho'-\rho)\delta_{mm'} 
 \end{eqnarray}
 for $K=0$.
 
 Similarly we have that
 
 \begin{eqnarray}
{\mathcal O}^{(A)}\hat{P}^{i}\nabla_{i}\Theta(x^{i},\hat{P}^{i},t)&\propto&\prod^{N'}_{q=0}\kappa_{q}^{2}\sum_{\alpha=0}^{2}\sum_{\beta=0}^{2}
\Theta^{(A)}_{N'-1+\alpha,A'-1+\beta}c^{(A)}_{N'-1+\alpha,A'-1+\beta} \nonumber\\
&&\cdot {\mathcal M}^{(2-\alpha,2-\beta)}_{N'-1+\alpha,A'-1+\beta}\frac{(A'+N')!}{(2A'+1)(A'-N')!}
\end{eqnarray}
Finally consider a function $V^{(P,A)}(x^{i},\hat{P}^{i},t)$ which is a sum of axial and polar components i.e.

\begin{eqnarray}
V^{(P,A)}(x^{i},\hat{P}^{i},t)  = \sum V^{(P)}_{MA}{\mathcal B}^{(P)}_{MA} + \sum V^{(A)}_{MA} {\mathcal B}^{(A)}_{MA}
\end{eqnarray}
where the sum is over harmonic components and moments.

It is readily checked that applying ${\mathcal O}^{(P)}V^{(A)}_{MA} {\mathcal B}^{(A)}_{MA}=0$ as it involves the following integral and sum:

\begin{eqnarray}
\label{over2}
&&\int \sum_{m',m} d\psi d^{2}S W^{ab..M,cd..M}Z^{ef..M}_{\phantom{ef..N}cd..M} \nabla_{ab..M}{\mathcal E}^{*k'}_{m'} \nabla_{ef..M}{\mathcal E}^{k}_{m}\nonumber \\
&\propto &\int\sum_{m',m}d\psi d^{2}S \left(\nabla_{++..M}{\mathcal E}^{*k}_{m'} (\textbf{A}^{(+)})_{cd..M}+\nabla_{--..M}{\mathcal E}^{*k}_{m'}(\textbf{A}^{(-)})_{cd..M}\right)\cdot\nonumber \\
&&  \left(\nabla_{++..M}{\mathcal E}^{k}_{m} (\textbf{A}^{(+)})^{cd..M}-\nabla_{--..M}{\mathcal E}^{k}_{m}(\textbf{A}^{(-)})^{cd..M}\right) \\
\nonumber&=&\int \sum_{m',m}d\psi d^{2}S\left(-\nabla_{++..M}{\mathcal E}^{*k}_{m'}\nabla_{--..M}{\mathcal E}^{k}_{m}+(-1)^{m+m'}\nabla_{--..M}{\mathcal E}^{k}_{-m'}\nabla_{++..M}{\mathcal E}^{*k}_{-m}\right) \\
\nonumber&=&\int \sum_{m',m}d\psi d^{2}S\left(-\nabla_{++..M}{\mathcal E}^{*k}_{m'}\nabla_{--..M}{\mathcal E}^{k}_{m}+(-1)^{-(m+m')}\nabla_{--..M}{\mathcal E}^{k}_{m'}\nabla_{++..M}{\mathcal E}^{*k}_{m}\right) \\
\nonumber&\propto&  \sum_{m',m}\delta_{mm'}\left(-\nabla_{++..M}{\mathcal E}^{*k}_{m'}\nabla_{--..M}{\mathcal E}^{k}_{m'}+\nabla_{++..M}{\mathcal E}^{*k}_{-m'}\nabla_{--..M}{\mathcal E}^{k}_{-m'}\right) \\
&=& 0
\end{eqnarray}
A similar argument implies that ${\mathcal O}^{(A)}V^{(P)}_{MA} {\mathcal B}^{(P)}_{MA}=0$. Therefore the operators ${\mathcal O}^{(P)}$ and ${\mathcal O}^{(A)}$ are suitable for picking
out moments from functions composed of a combination of both.

The Botzmann equation (\ref{bolt2}) with no collision terms may then be cast as a system of coupled evolution equations for the moments
$\Theta^{(P)}_{NA}(k,w,t)$ and $\Theta^{(A)}_{NA}(k,w,t)$ as follows:

\begin{enumerate}
\item Use the result (\ref{dirdevalf}) to express all metric source terms (and collision terms if appropriate) in
equation (\ref{bolt2}) in terms of the functions ${\mathcal B}^{(P)}_{NA}$ and ${\mathcal B}^{(A)}_{NA}$.
\item Expand $\Theta(x^{i},t)$ in terms of harmonic mode moments $\Theta^{(A)}_{NA}(k,w,t)$ and
$\Theta^{(P)}_{NA}(k,w,t)$ using the expansions (\ref{pe}) and (\ref{ae}).\\
\item Then use the operators ${\mathcal O}^{(P)}$ and ${\mathcal O}^{(A)}$ to extract a particular term
$d\Theta^{(P)}_{N'A'}(k,w,t)/dt$ or $d\Theta^{(A )}_{N'A'}(k,w,t)/dt$ in (\ref{bolt2}).  This 
will be accompanied by terms in adjacent moments as dictated by the directional derivative of $\Theta$
and (\ref{dirdevalf}) as well as terms due to the metric.
\end{enumerate}

The presence of collisional terms ${\mathcal C}\neq 0$  may be simply accounted for in the manner that the metric source terms are are if they are expressable
in terms of the functions ${\mathcal B}^{(P)}_{MA}$ and ${\mathcal B}^{(A)}_{MA}$ and do not violate the earlier assumption that the time component of the geodesic
equation holds.

 \section{Comparison of moment expansions in the FRW limit}
 \label{moment}
 
 Having set up the Boltzmann equation, and found an appropriate projection operator to isolate polar and axial $M$ and $A$ components
 of the moment expansion, we now seek to motivate a particular choice for the functions $c^{(P)}_{MA}$. The polar case is of particular interest
 as it is likely that the source of the anisotropic, shearless background will only have polar perturbations (for instance the scalar field in Section \ref{sf}).
 
 Towards these ends, we consider the typical choice of moment expansion for scalar
 contributions to the temperature perturbation in the flat FRW case:
 
\begin{eqnarray}
\label{frwd}
\Theta(\hat{P}_{i},q,t) &=& \sum_{A}(-i)^{A}(2A+1)\Theta_{A}(q,t){\mathcal P}_{A}(\mu)
\end{eqnarray}
where ${\mathcal P}_{A}$ is a Legendre polynomial, $\mu$ is the cosine of the angle between the unit three dimensional co-moving wave-vector $\hat{\textbf{q}}$ of a plane
wave perturbation (with magnitude $q$) and the photon 3-momentum. 
However, in the spirit of the 2+1 decomposition of space, we may decompose the wavevector $q\hat{\textbf{q}}$ as follows:

\begin{eqnarray}
q\hat{\textbf{q}}=q\beta \partial_{z}+q\sqrt{1-\beta^{2}}\hat{k}^{a}\partial_{a}
\end{eqnarray}
where the 2-wave vector is unit with respect to the co-moving background 2-surface metric: $\gamma_{ab}\hat{k}^{a}\hat{k}^{b}=1$. This implies that

\begin{equation}
\mu =\hat{P}_{i}\hat{q}^{i}=\alpha\beta+\sqrt{1-\beta^{2}}\sqrt{1-\alpha^{2}}\cos(\eta)
\end{equation}
where $\eta$ is the angle between $\hat{k}^{a}$ and $\hat{p}^{a}$.

Noting the group multiplication property implied by (mat1), the addition theorem for matrix elements (add1)
implies that:

\begin{eqnarray}
\nonumber
{\mathcal P}_{A}(\mu) &=& \sum_{M=-l}^{M=l}(-1)^{M}e^{-iM(\eta+\pi)}\frac{(A-|M|)!}{(A+|M|)!}
   P^{A}_{|M|}(\alpha)P^{A}_{|M|}(\beta)\\
     &=& {\mathcal P}_{A}(\alpha){\mathcal P}_{A}(\beta)  \nonumber \\
     && +2 \sum_{M=1}^{A}\cos(M\eta)\frac{(A-M)!}{(A+M)!}
   P^{A}_{M}(\alpha)P^{A}_{M}(\beta)\nonumber \\
    &=& {\mathcal P}_{A}(\alpha){\mathcal P}_{A}(\beta)  \nonumber \\
     && + 2\sum_{M=1}^{A}C_{M}(\eta)\frac{(A-M)!}{(A+M)!}
   P_{M}^{A}(\alpha)P_{M}^{A}(\beta)
   \label{exp1}
\end{eqnarray}
A temperature perturbation $\Theta(\bar{P}_{i},x^{i},t)$ is expressed in terms of basis mode-moment functions $P_{A}(\mu)e^{i\textbf{q}\cdot\textbf{x}}$.
Using equation (\ref{exp1}) we have that:

\begin{eqnarray}
{\mathcal P}_{A}(\mu)e^{i\textbf{q}\cdot\textbf{x}} &=&{\mathcal P}_{A}(\alpha){\mathcal P}_{A}(\beta)e^{i\textbf{q}\cdot\textbf{x}}   \nonumber \\
     && + 2\sum_{M=1}^{A}\frac{(A-M)!}{(A+M)!}P_{M}^{A}(\alpha)P_{M}^{A}(\beta)e^{iwz}(-k)^{-M}C^{abc..M}\nabla_{abc..M}e^{ik_{a}x^{a}}
   \nonumber\\
   &=& \sum_{n}i^{n}e^{in\phi_{k}}{\mathcal P}_{A}(\alpha){\mathcal P}_{A}(\beta) e^{iwz}{\mathcal E}^{k}_{m} \nonumber \\
     &&\nonumber + 2\sum_{n}i^{n}e^{in\phi_{k}}\sum_{M=1}^{A}\frac{(A-M)!}{(A+M)!}P_{M}^{A}(\alpha)P_{M}^{A}(\beta)e^{iwz}(-k)^{-M}C^{abc..M}\nabla_{abc..M}{\mathcal E}^{k}_{m}
\end{eqnarray}
where we have used the Jacobi-Anger expansion:

\begin{eqnarray}
e^{ikr\cos\phi} &=& \sum_{m} i^{m}J_{m}(kr) e^{im\phi}
\end{eqnarray}

We see then that the mode-moment functions in the flat FRW case may indeed be related to the proposed mode functions ${\mathcal B}^{(P)}_{MA}$ as follows:

\begin{eqnarray}
{\mathcal P}_{A}(\mu)e^{i\textbf{q}\cdot\textbf{x}} &=& \sum_{n}i^{n}e^{in\phi_{k}}{\mathcal P}_{A}(\beta) {\mathcal B}^{(P)}_{A0} \nonumber \\
     && + 2\sum_{n}\sum_{M=1}^{A}\frac{(A-M)!}{(A+M)!}i^{n}e^{in\phi_{k}}P_{M}^{A}(\beta){\mathcal B}^{(P)}_{MA} \label{decpl}
\end{eqnarray}
The temperature perturbation $\Theta(P^{i},x^{i},t)$ around a spacetime with FRW symmetries admits the following decomposition:

\begin{eqnarray}
\Theta(P^{i},x^{i},t) &=& \int d^{3}q \sum_{A}(-i)^{A}(2A+1){\mathcal P}_{A}(\hat{P}^{i}\hat{q}_{i})\Theta_{A}(\textbf{q},t) e^{i\textbf{q}\cdot\textbf{x}}
\end{eqnarray}
Therefore using (\ref{decpl}) we may relate the moments $\Theta_{A}(\textbf{q},t)$ to the moments $\Theta_{NA}$, yielding:

\begin{eqnarray}
\label{oatoa}
c^{(P)}_{0A}\Theta_{0A} &=& (-i)^{A}(2A+1)\Theta_{A}i^{m}e^{im\phi_{k}}{\mathcal P}_{A}(\beta)
\end{eqnarray}
and 

\begin{eqnarray}
\label{natoa}
c^{(P)}_{NA}\Theta_{NA} &=& 2\frac{(A-N)!}{(A+N)!}(-i)^{A}(2A+1)\Theta_{A}(\textbf{q},t)i^{m} e^{im\phi_{k}}P^{A}_{N}(\beta) 
\end{eqnarray}
for $N\geq 1$.

We make the following choice for the functions $c^{(P)}_{NA}\equiv c^{(P)f}_{NA}$ in the flat case:
 

 
 

\begin{eqnarray}
 c^{(P)f}_{NA}= \frac{(-i)^{A}(2A+1)(A-N)!}{(A+N)!} 
 \end{eqnarray}
 Note that this choice in the flat case implies via (\ref{oatoa}) and (\ref{natoa}) that
 
 \begin{eqnarray}
 \Theta_{0A} &=& \Theta_{A}(\textbf{q},t) i^{m} e^{im\phi_{k}}{\mathcal P}_{A}(\beta) \label{reltheta1} \\
 \Theta_{NA} &=& 2\Theta_{A}(\textbf{q},t) i^{m} e^{im\phi_{k}}P^{A}_{N}(\beta) \label{reltheta2}
 \end{eqnarray}

\section{Free-streaming equation}
\label{freest}

The linearity of the perturbed equations implies that we can isolate the impact of the anisotropically curved background itself 
on the evolution of inhomgeneities and isotropies of the photon distribution in the absence of perturbational source 
such as those due to the perturbed metric in the Boltzmann equation.
In the absence of collisional terms and metric source terms, each moment of $\Theta(\hat{P}^{i},x^{i},t)$ 
obeys the following equation.

\begin{eqnarray}
\nonumber0 &=& c_{NA}\Theta'_{NA} + c_{N+1,A-1}\Theta_{N+1,A-1}{\mathcal M}^{02}(N+1,A-1)\\
&&+ c_{N+1,A+1}\Theta_{N+1,A+1}{\mathcal M}^{00}(N+1,A+1) + c_{N,A-1}\Theta_{N,A-1}{\mathcal M}^{12}(N,A-1)\nonumber \\
&&+ c_{N,A+1}\Theta_{N,A+1}{\mathcal M}^{10}(N,A+1) + c_{N-1,A-1}\Theta_{N-1,A-1}{\mathcal M}^{22}(N-1,A-1)\nonumber \\
&&+ c_{N-1,A+1}\Theta_{N-1,A+1}{\mathcal M}^{20}(N-1,A+1) \label{freebolt}
\end{eqnarray}
The effect of the anisotropic curvature of space is via the appearance of the functions $\kappa_{\alpha}^{2}$ in the matrix entries ${\mathcal M}^{00}(\alpha,\beta)$ and
${\mathcal M}^{02}(\alpha,\beta)$.

On the other hand, in the flat case $K=0$, given an initial monopole perturbation $\Theta_{00}(k,w,t_{0})$, the solution at a later time $t$ is known:

\begin{eqnarray}
\label{21sol}
\Theta_{NA}(k,w,m,t)  &=& 2j_{A}(q(t-t_{0})) i^{m} e^{im\phi_{k}}P^{A}_{N}(\beta)
\end{eqnarray}
where $j_{A}$ is a \emph{spherical} Bessel function. This implies that (\ref{21sol}) is the solution of (\ref{freebolt}) when $K=0$ given our above choice for the functional form of $c^{f}_{NA}$. 

Henceforth we adopt the following choice for $c_{NA}$:

\begin{eqnarray}
c_{NA} &=& \frac{(-i)^{A}(2A+1)(A-N)!}{(A+N)!\kappa_{N}\kappa_{N-1}\kappa_{N-2}...} \\
              &\equiv & \frac{c^{f}_{NA}}{\prod_{q=0}^{N}\kappa_{N}} \label{cchoice}
\end{eqnarray}
where $c^{f}_{NA}$ simply means the choice for $c_{NA}$ in the spatially flat case. It shall be seen that this choice simplifies the appearance of the photon temperature-temperature
power spectrum (see equation (\ref{colpola})).  Given this choice, we may write (\ref{freebolt}) as :

\begin{eqnarray}
\nonumber(2A+1)\frac{(A-N)!}{(A+N)!}\Theta'_{NA} &=&  - \kappa_{N+1}i\frac{k}{2}\frac{(A-N-2)!}{(A+N-2)!}\Theta_{N+1,A-1}\\
\nonumber&&-\kappa_{N+1}i\frac{k}{2}\frac{(A-N+2)!}{(A+N+2)!}\Theta_{N+1,A+1} \\
\nonumber&& +w\frac{(A+N)!}{(A+N-1)!}\Theta_{N,A-1} \\
\nonumber&& -w\frac{(A+1-N)!}{(A-N)!}\Theta_{N,A+1} \\
\nonumber&& +\kappa_{N}i\frac{k}{2}\frac{(A-N)!}{(A+N-2)!}\Theta_{N-1,A-1} \\
&&+\kappa_{N}i\frac{k}{2}\frac{(A-N+2)!}{(A+N)!}\Theta_{N-1,A+1}
\label{fs2}
\end{eqnarray}
where $\Theta_{NA}$ is nonzero only for $A\geq 0$, $0\leq N \leq A$.

Care must be taken in the curved cases. When $K<0$, the eigenvalue $k^{2}$ of the 2-d Laplace-Beltrami operator need only take values greater than $1/4$ to describe square integrable functions. This is not 
so in the flat case. Additionally, when $K>0$, the function ${\mathcal E}^{k}_{m}$ is simply a constant when $k=0$ and so, in defining the basis mode-moment functions ${\mathcal B}_{NA}$ via
gradients of this function, only $N=0$ is permitted. In standard cosmology it is typically assumed that at recombination the photon distribution is described by a monopole and dipole $\Theta_{0}$ and $\Theta_{1}$, each of which depend on the 3 dimensional $q^{i}$ only via its magnitude $q$. Inevitably then, contributions are considered from wavenumbers which lie entirely along the $(\partial_{z})^{\mu}$ direction and so
have no component in the orthogonal direction i.e. $k=0$. We have seen above though that such a setup is complicated in the curved setup - if, for instance, $\Theta_{00}$ depends on the combination $q=\sqrt{w^{2}+k^{2}}$ then inevitably scales large compared to the curvature are probed as a constant value of $q$ becomes more and more dominated by $w$; this does not appeal to an intuitive notion 
of the curvature being irrelevant at early times. 

By way of example of the effects of curvature, we initially restrict ourselves to the open case. We again restrict ourselves to an initial monopole perturbation, but now consider it to be sharply peaked at some value $k_{*}^{2} \gg |K|/4$ with no z-dependence. Furthermore we solve for a perturbation $\Pi_{NA}=\Theta_{NA}\prod_{q=0}^{N}\kappa_{N}/(\kappa^{2}_{N}\kappa^{2}_{N-2}\kappa^{2}_{N-4}...)$, and assume that $N\gg1 $ so that even though we are essentially restricted to contributions from
wavenumbers far greater than those associated with the curvature scale, we may still have non-negligible deviations of $\kappa_{N}$ from unity.

\begin{eqnarray*}
(2A+1)\frac{(A-N)!}{(A+N)!}\Pi'_{NA} &=&  - i\frac{k_{N}}{2}\frac{(A-N-2)!}{(A+N-2)!}\Pi_{N+1,A-1}\\
&&-i\frac{k_{N}}{2}\frac{(A-N+2)!}{(A+N+2)!}\Pi_{N+1,A+1} \\
&& +i\frac{k_{N}}{2}\frac{(A-N)!}{(A+N-2)!}\Pi_{N-1,A-1} \\
&&+i\frac{k_{N}}{2}\frac{(A-N+2)!}{(A+N)!}\Pi_{N-1,A+1}
\label{fs3}
\end{eqnarray*}
where we have introduced the variable $k_{N}\equiv k \frac{\kappa^{2}_{N}\kappa^{2}_{N-2}\kappa^{2}_{N-4}...}{\kappa^{2}_{N-1}\kappa^{2}_{N-3}\kappa^{2}_{N-3}...}$.
This equation for the evolution of the `moments' $\Pi_{NA}$ is, for each value of $N$, equivalent to the system in a flat universe and so admits the following solution:

\begin{eqnarray}
\label{21solgess}
\Pi_{NA}(k,w,m,t)  &=& 2j_{A}(k_{N}(t-t_{0})) i^{m} e^{im\phi_{k_{N}}}P^{A}_{N}(0)
\end{eqnarray}
and so

\begin{eqnarray}
\label{21break}
\Theta_{NA} &=& \frac{(\kappa^{2}_{N}\kappa^{2}_{N-2}\kappa^{2}_{N-4}...)}{\prod_{q=0}^{N}\kappa_{N}}2j_{A}(k_{N}(t-t_{0})) i^{m} e^{im\phi_{k_{N}}}P^{A}_{N}(0)
\end{eqnarray}
Now instead consider a case where instead the initial monopole perturbation has no dependence upon the coordinates on the co-moving 2-surface, now peaked
at a value $w=k_{*}$. By the above arguments  we then have
$N=0$ and the Boltzmann equation takes the following form:

\begin{eqnarray*}
(2A+1)\Theta'_{0A} &=& wA\Theta_{0A-1}- w(A+1)\Theta_{0A+1} 
\label{fs2}
\end{eqnarray*}
There is again an analogue to the flat case, with solution

\begin{eqnarray}
\label{21break2}
\Theta_{0A} &=& j_{A}(w(t-t_{0}))i^{m} P^{A}_{0}(1)
\end{eqnarray}
Given that the initial monopole perturbation is sharply peaked at a value $k_{*}$, we expect the $\Theta$ moments in each case to be peaked at the lowest value of $A$
for $j_{A}$ is maximized by its argument. For a function $j_{A}(x)$ this happens when $A\sim x$. We note then that this peak happens at 
$A\sim k_{*}(t-t_{0})$ in the latter case but $A\sim k_{*} \frac{\kappa^{2}_{A}\kappa^{2}_{A-2}\kappa^{2}_{A-4}...}{\kappa^{2}_{A-1}\kappa^{2}_{A-3}\kappa^{2}_{A-3}...}(t-t_{0})$ in the former case.

Though far from exhaustive, these limited results seem to indicate that the notion of the position of an acoustic peak in A-space in the case 
of nonvanishing curvature can vary depending upon the value of $N$ in question and the form of initial perturbations. This could have important
observational implications.

\section{The Boltzmann Equation for massive particles}
\label{boltmassive}

We now consider the Boltzmann equation for fields which can be described statistically in terms of a distribution function $f^{T}$ for
massive particles. It's to be expected that this description is appropriate for baryons, cold dark matter, or massive neutrinos. The collisional
Boltzmann equation is then

\begin{eqnarray}
\frac{df^{T}}{dt} ={\mathcal C}^{T}
\end{eqnarray}

The time derivative $df^{T}/dt$ is then decomposed as follows: 

\begin{eqnarray}
\frac{df^{T}}{dt} &=& \frac{\partial f^{T}}{\partial t}+ \frac{\partial f^{T}}{\partial x^{i}}\frac{dx^{i}}{dt}+\frac{\partial f^{T}}{\partial p}\frac{dp}{dt}  +\frac{\partial f^{T}}{\partial \alpha}\frac{d\alpha}{dt}+\frac{\partial f^{T}}{\partial (\hat{p}^{a})}\frac{d (\hat{p}^{a})}{dt} \label{dfmdt}
\end{eqnarray}
As in the massless case, we expect dependence of $f^{T}$ upon $x^{i}$,$\hat{p}^{a}$ and $\alpha$ as well as $d\alpha/dt$ and $d\hat{p}^{a}/dt$ to 
be first order or great in smallness. Therefore up to first order in perturbations, the collisional Boltzmann equation for massive particles is

\begin{eqnarray}
\label{boltm}
 \frac{\partial f^{T}}{\partial t}+ \frac{\partial f^{T}}{\partial x^{i}}\frac{dx^{i}}{dt}+\frac{\partial f^{T}}{\partial p}\frac{dp}{dt} ={\mathcal C}^{T}
 \end{eqnarray}

We may define the field four-momentum in terms of the proper time $\tau$ as follows $P^{\mu}=dx^{\mu}/d\tau$.
We explicitly decompose the momentum as follows:

\begin{eqnarray}
P^{\mu}\partial_{\mu}&=& \frac{1}{a}(\bar{P}^{t}\partial_{t}+p\hat{P}^{i}\partial_{i}) \\
\bar{P}^{t} &\equiv& aP^{t}\\
\hat{P}^{i}\partial_{i} &\equiv & \alpha \partial_{z}+p\sqrt{1-\alpha^{2}}\hat{p}^{a}\partial_{a} 
\end{eqnarray}
Therefore $dx^{j}/dt=\bar{P}^{j}/\bar{P}^{t}$. The timelike constraint $P^{\mu}P_{\mu}=-m^{2}$, up to first order, is as follows:

\begin{eqnarray}
-m^{2} &=& -(1+V)(\bar{P}^{t})^{2}+2p(\alpha  E+(C_{a}+R_{a})\hat{p}^{a}\sqrt{1-\alpha^{2}})\bar{P}^{t} \nonumber\\
    && +p^{2}+U(1-\alpha^{2})p^{2}+F\alpha^{2}p^{2} +2(S_{a}+B_{a})\hat{p}^{a}\alpha\sqrt{1-\alpha^{2}}p^{2}\nonumber \\
    && +(1-\alpha^{2})p^{2}(X_{ab}+2Q_{ab})\hat{p}^{a}\hat{p}^{b}
     \end{eqnarray}
Where the metric perturbations here are the actual metric perturbations and not their harmonic modes.

Up to linear order in perturbations we have:

\begin{eqnarray}
\label{ptmm}
\nonumber\bar{P}^{t} &=& \sqrt{p^{2}+m^{2}}+p(\alpha E+(C_{a}+R_{a})\hat{p}^{a}\sqrt{1-\alpha^{2}})\\
  && +\frac{p^{2}}{\sqrt{p^{2}+m^{2}}}(\frac{U}{2}(1-\alpha^{2})
 +\frac{F}{2}\alpha^{2} 
                  +(S_{a}+B_{a})\hat{p}^{a}\alpha\sqrt{1-\alpha^{2}}\nonumber\\
                  &&+\frac{1}{2}(1-\alpha^{2})(X_{ab}+2Q_{ab})\hat{p}^{a}\hat{p}^{b})-\sqrt{p^{2}+m^{2}}\frac{V}{2}
\end{eqnarray}
As in the massless particle case, we restrict ourselves to cases where $\bar{P}^{t}$ is not influenced by the 
collisional terms that exist.
From the time component of the geodesic equation we have then that:

\begin{eqnarray}
\frac{dP^{t}}{d\tau} =\frac{\bar{P}^{t}}{a}\frac{d}{dt}\left(\frac{\bar{P}^{t}}{a}\right)= -\Gamma^{t}_{\mu\nu}P^{\mu}P^{\nu}
\end{eqnarray}
Yielding: 

\begin{eqnarray}
\nonumber\frac{dp}{dt} &=& -p{\mathcal H}-\sum\frac{1}{2\epsilon}(\alpha^{3}p^{2}\partial_{z}F+2\epsilon^{2}\alpha E'+\epsilon^{2}\alpha\partial_{z}V+2(1-\alpha^{2})p\epsilon U' \\
&&
 \nonumber  +2p\epsilon\alpha^{2}F'+\alpha(1-\alpha^{2})p^{2} \partial_{z}U +2 (\epsilon^{2}-p^{2}){\mathcal H}\alpha E){\mathcal E} \\
     \nonumber    &&  +\sum\frac{\sqrt{1-\alpha^{2}}}{2\epsilon} (2{\mathcal H}(p^{2}-\epsilon^{2})C+p^{2}(\alpha^{2}-1)U -2\alpha^{2}p^{2}\partial_{z}B
            \\
   \nonumber    &&     -2\epsilon^{2}C'-4p\epsilon\alpha B'-\epsilon^{2}V-p^{2}\alpha^{2}F)\hat{p}^{a}\nabla_{a}{\mathcal E} \\
     \nonumber  &&  +\sum\frac{(1-\alpha^{2})}{\epsilon}(-B p^{2}\alpha-p\epsilon X'-\frac{p^{2}}{2}\partial_{z}X\alpha)\hat{p}^{a}\hat{p}^{b}\nabla_{a}\nabla_{b}{\mathcal E}\\
     \nonumber  && -\sum\frac{\sqrt{1-\alpha^{2}}}{2\epsilon}(1-\alpha^{2})p^{2}X\hat{p}^{a}\hat{p}^{b}\hat{p}^{c}\nabla_{a}\nabla_{b}\nabla_{c}{\mathcal E}\\
     \nonumber  && +\sum \frac{\sqrt{1-\alpha^{2}}}{\epsilon}(-\epsilon^{2}R'-2p\epsilon\alpha S'-\alpha^{2}p^{2}\partial_{z}S+{\mathcal H}(p^{2}-\epsilon^{2})R)\hat{p}^{a}\bar{\nabla}_{a}{\mathcal E}\\
    \nonumber   &&+\sum \frac{(1-\alpha^{2})}{\epsilon}(-\alpha p^{2} \partial_{z}Q-2p\epsilon Q'-\alpha p^{2}S) \hat{p}^{a}\hat{p}^{b}\nabla_{a}\bar{\nabla}_{b}{\mathcal E}\\
   && -\sum \frac{\sqrt{1-\alpha^{2}}}{\epsilon}(1-\alpha^{2})p^{2}Q \hat{p}^{a}\hat{p}^{b}\hat{p}^{c}\nabla_{a}\nabla_{b}\bar{\nabla}_{c}{\mathcal E} \label{mdpdt}
\end{eqnarray}
where $\epsilon^{2}\equiv p^{2}+m^{2}$.

For a given species of particles, we can decompose $f^{T}(x^{i},P^{i},t)$ as follows:

\begin{eqnarray}
f^{T}(x^{i},P^{i},t)= f^{T}_{0(M)}(p,t) +\delta f^{T}(x^{i},\alpha,\hat{p}^{a},t)
\end{eqnarray}
where $\delta f^{T}$ is of first order in smallness.
We propose the following harmonic and mode decomposition of the perturbation $\delta f^{T}$:

\begin{eqnarray}
\nonumber
\delta f^{T}(x^{i},\alpha,\hat{p}^{a},t) &=& \sum_{k,m,w}\sum_{A,0\leq M \leq A}d^{(P)}_{MA}(w,k,t) {\mathcal F}^{(P)}_{MA}(w,k,t) \\
&& \cdot e^{iwz}P^{A}_{M}(\alpha)\frac{C^{ab..M}\nabla_{ab..M}{\mathcal E}^{k}_{m}}{(-k)^{M}}  \nonumber \\
&&+\sum_{k,m,w}\sum_{A,0\leq M \leq A}d^{(A)}_{MA}(w,k,t) {\mathcal F}^{(A)}_{MA}(w,k,t)\nonumber \\
&& \cdot e^{iwz}P^{A}_{M}(\alpha)\frac{{\mathcal G}^{ab..M}\nabla_{ab..M}{\mathcal E}^{k}_{m}}{(-k)^{M}} \label{qe}  
\end{eqnarray}
where ${\mathcal F}^{(P)}_{MA}(w,k,t)$ and $ {\mathcal F}^{(A)}_{MA}(w,k,t)$ are polar and axial moments of the massive particle distribution function. As such, the results from the massless
case carry over, allowing for the additional terms that appear in (\ref{mdpdt}) due to the presence of a a nonzero mass.
It is readily checked that the metric source terms appearing in (\ref{boltm}) via (\ref{mdpdt})  can be expressed in terms of the functions ${\mathcal B}_{NA}^{(A)}$ and
${\mathcal B}_{NA}^{(P)}$. If this is also the case for the collisional terms (having assumed already that they allow the use of the time component of the massive particle geodesic equation) then
a set of time  evolution equations for ${\mathcal F}^{(P)}_{NA}$ and ${\mathcal F}^{(A)}_{NA}$ may be deduced precisely in the manner of the massless case i.e. by application of the appropriate polar or axial projection operator.

The components of the energy-momentum tensor for the massive and massive fields may then be related to the distribution function and momentum 4-vectors $P^{\mu}$
by standard methods, see for instance \cite{Ma:1995ey}.
In principle, given a cosmological background, the collection of perturbed Boltzmann equations along with the perturbed Einstein equations are sufficient to calculate predictions for observables on scales where linear perturbation theory holds.

\section{Polar correlations in the photon distribution}
\label{polcorr}

A very useful observable is the angular correlation function $\left<\Theta^{(P)}(\textbf{n}_{(1)}^{i})\Theta^{(P)}(\textbf{n}_{(2)}^{j})\right>$ for perturbations to the photon temperature:

\begin{eqnarray*}
\left<\Theta^{(P)}(\textbf{n}_{(1)}^{i})\Theta^{(P)}(\textbf{n}_{(2)}^{j})\right> &=& \sum_{lm}\sum_{l'm'}   \left<a_{l'm'}a_{lm}\right> Y^{*l}_{m}(\textbf{n}_{(1)}^{i})Y^{*l}_{m}(\textbf{n}_{(2)}^{j})
\end{eqnarray*}
Where $Y^{l}_{m}$ are spherical harmonics and $\textbf{n}_{(1)}^{j}$ and $\textbf{n}_{(2)}^{j}$ are angles in the sky. Henceforth we will use the notation $\Omega=(\theta,\phi)$ to parameterize the angular vector
$n^{i}$. We now express the `spherical harmonic space' correlation function $ \left<a_{l'm'}a_{lm}\right>$ :

\begin{eqnarray}
\nonumber\left<a_{l'm'}a_{lm}\right> &= & \int d\Omega \int d\Omega' \sum_{N',A',N,A}\sum_{w',w,k',k}Y^{*l}_{m}(\Omega)Y^{l'}_{m'}(\Omega')\\
\nonumber   && {\mathcal B}^{(P)}_{NA}(x^{i},k,w,\Omega)({\mathcal B}^{(P)}_{NA})^{*}(x^{i},k',w',\Omega') F^{(P)}_{N'A'NA}(k,k',w,w',t)\\
        &=& \frac{1}{2\pi}\int dw \int d\Omega \int d\Omega' \sum_{N',A',N,A}\sum_{w,k}Y^{*m}_{l}(\Omega)Y^{m'}_{l'}(\Omega')\cdot \nonumber\\
  \nonumber        && {\mathcal B}^{(P)}_{NA}(z=0,k,w,\Omega)({\mathcal B}^{(P)}_{NA})^{*}(z=0,k,w,\Omega') \cdot \\
          && F^{(P)}_{N'A'NA}(k,w,t) \label{correl}
\end{eqnarray}
where

\begin{eqnarray*}
F^{(P)}_{N'A'NA}(k,w,t) &\equiv & c^{(P)}_{NA}c^{*(P)}_{N'A'}\frac{\Theta^{(P)}_{NA}\Theta^{*(P)}_{N'A'}}{|\delta(k,w)|^{2}}{\mathcal P}(k,w)
\end{eqnarray*}
and we have defined the dark matter power spectrum $\left<\delta(w',k',m')\delta(w,k,m)\right>\equiv {\mathcal P}(k,w)\delta(w-w')\delta_{k'k}\delta_{mm'}$, 
where $\delta(k,w,t)$ is the dark matter overdensity and $\delta_{kk'}$ is  either an appropriate delta function or Kronecker delta symbol depending on whether $k$ is discrete or continuous. The step to express the expectation of $\Theta^{(P)}_{NA}\Theta^{*(P)}_{N'A'}$ in terms of the dark matter power spectrum
is a standard one in perturbation theory around FRW backgrounds and is also appropriate here; its utility is to isolate the contribution to due to the initial amplitude and phase
of primordial perturbation but that has no dependence upon $\hat{P}^{i}$. One expects this lack of momentum dependence to be the case for the dark matter
power spectrum. Assuming adiabaticity of primordial perturbations, ratio $\Theta/\delta$ is not expected to depend upon the initial amplitude of perturbations.

After lengthy calculation (detailed in  \ref{calpolar}), one may carry out the integrations over $\Omega$ and $\Omega'$ in (\ref{correl}) as well as the
summations over $N'$,$A'$,$N$, and $A$ yielding the following expressions for $\left<a_{l'm'}a_{lm}\right>$: 

$K>0$: 

\begin{eqnarray}
\nonumber\left<a_{l'm'}a_{lm}\right>&=& \delta_{mm'} \frac{K}{2}\int dw\sum_{L}  (2L+1)\frac{\Theta^{(P)}_{ml}\Theta^{*(P)}_{ml'}}{|\delta(k,w)|^{2}}{\mathcal P}(k,w)e(l,l',m)\\
\label{colclosed}
\end{eqnarray}

$K<0$: 
\begin{eqnarray}
\left<a_{l'm'}a_{lm}\right>&=&\delta_{mm'}2\pi|K|\int dw \int d\lambda \lambda \tanh\left(\pi \lambda\right)\frac{\Theta^{(P)}_{ml}\Theta^{*(P)}_{ml'}}{|\delta(k,w)|^{2}}{\mathcal P}(k,w)e(l,l',m)\nonumber\\
\label{colopen}
\end{eqnarray}

$K=0$:
\begin{eqnarray}
\left<a_{l'm'}a_{lm}\right>&=&   \delta_{mm'}2\pi\int dw \int k dk\frac{\Theta^{(P)}_{ml}\Theta^{*(P)}_{ml'}}{|\delta(k,w)|^{2}}{\mathcal P}(k,w)e(l,l',m)\nonumber \\
\label{colflat}
\end{eqnarray}
where the time dependence of perturbational quantities is implicit and recall that $k^{2}=kk^{*}=KL(L+1)$ in the case $K>0$, $k^{2}=kk^{*}= |K|(\lambda^{2}+\frac{1}{4}$)) in the case $K<0$.

\section{Statistical anisotropy}
\label{secstatanis}

The CMB temperature anisotropy is said to be statistically isotropic if the following property holds:

\begin{eqnarray}
\left<\Theta(\textbf{n}_{(1)}^{i})\Theta(\textbf{n}_{(2)}^{j})\right> &=& f(\textbf{n}_{(1)}^{i}\textbf{n}_{(2)i})
\end{eqnarray}
i.e. that the two point angular correlation function depends only on the projection of the direction in the sky $\textbf{n}_{(1)}$ along 
the second direction $\textbf{n}_{(2)}$.  
Decomposing the angular functions, as before,  in terms of spherical harmonics we have that:

\begin{eqnarray}
\nonumber
\left<\Theta(\textbf{n}_{(1)}^{i})\Theta(\textbf{n}_{(2)}^{j})\right> &=& \sum_{m,l,m',l'}\left<a_{l'm'}a_{lm}\right>Y^{*l'}_{m'}(\textbf{n}_{(1)}^{i})Y^{l}_{m}(\textbf{n}_{(2)}^{i})  \\
\label{2pc}
\end{eqnarray}
If $\left<a_{l'm'}a_{lm}\right>= C_{l}\delta_{ll'}\delta_{mm'}$ and $C_{l}$ is independent of $m$, we have:

\begin{eqnarray}
\left<\Theta(\textbf{n}_{(1)}^{i})\Theta(\textbf{n}_{(2)}^{j})\right> &=& \sum_{m,l} C_{l} Y^{*l}_{m}(\textbf{n}_{(1)}^{i})Y^{l}_{m}(\textbf{n}_{(2)}^{i})\nonumber \\
     &=& \sum_{l} \frac{2l+1}{4\pi}C_{l}{\mathcal P}_{l}(\textbf{n}_{(1)}^{i}\textbf{n}_{(2)i}) \nonumber\\
     &=&  f(\textbf{n}_{(1)}^{i}\textbf{n}_{(2)i})
\end{eqnarray}
In general  $\left<a_{l'm'}a_{lm}\right>$ may not in fact not be diagonal in $l$ and $l'$ or may depend on $m$, indicating then that the condition for statistical isotropy is not 
fulfilled. A natural origin for this in the event of background anisotropic curvature would be due to the presence of an additional direction in the 
problem: $(\partial_{z})^{i}$. 
In the spirit of our 2+1 approach, we decompose the angles in the sky as follows $\textbf{n}_{(1)}=(\alpha_{(1)},\sqrt{1-\alpha^{2}_{(1)}}\hat{n}_{(1)}^{a})$,
$\textbf{n}_{(2)}=(\alpha_{(2)},\sqrt{1-\alpha^{2}_{(2)}}\hat{n}_{(2)}^{a})$. The condition of statistical isotropy may then be written as follows:

\begin{eqnarray}
\nonumber
\left<\Theta(\textbf{n}_{(1)}^{i})\Theta(\textbf{n}_{(2)}^{j})\right> &=& f\left(\alpha_{(1)}\alpha_{(2)}+\sqrt{1-\alpha^{2}_{(1)}}\sqrt{1-\alpha^{2}_{(2)}}\hat{n}_{(1)}^{a}\hat{n}_{(2)a}\right)
\end{eqnarray}
We expect there to be a limited statistical isotropy in that the dependence of the correlation function upon 
$\hat{n}_{(1)}^{a}$ and $\hat{n}_{(2)}^{a}$ will only be via the projection of $\hat{n}_{(1)}^{a}$ along $\hat{n}_{(2)}^{a}$, there being no other preferred
direction existing in the co-moving 2-surface. If we allow $\left<a_{l'm'}a_{lm}\right>$ to take the following form:

\begin{eqnarray}
\left<a_{l'm'}a_{lm}\right> &=& \delta_{mm'}C_{ll'}(m) 
\end{eqnarray}
then we have that

\begin{eqnarray}
\nonumber
\left<\Theta(\textbf{n}_{(1)}^{i})\Theta(\textbf{n}_{(2)}^{j})\right> &=& \sum_{m,l,m',l'} C_{ll'}(m)\delta_{mm'}Y^{*l'}_{m'}(\textbf{n}_{(1)}^{i})Y^{l}_{m}(\textbf{n}_{(2)}^{i})  \\
&\propto & \sum_{m,l,l'} C_{ll'}(m) e^{im(\phi_{2}-\phi_{1})}P^{l}_{m}(\alpha_{(2)})P^{l'}_{m}(\alpha_{(1)}) \label{mush}
\end{eqnarray}
where $\phi_{1}$ and $\phi_{2}$ are angles made between  $\hat{n}_{(1)}^{a}$ and $\hat{n}_{(2)}^{a}$ with respect to a fixed direction in the co-moving
2-surface. Clearly though (\ref{mush}) depends only upon the difference between these two angles and so only upon the projection of  $\hat{n}_{(1)}^{a}$ along $\hat{n}_{(2)}^{a}$.
We see then that the functional form of the correlation functions (\ref{colclosed}),(\ref{colopen}), and (\ref{colflat}) in being diagonal in $m$ and $m'$ reflect
the limited statistical isotropy of the problem. Clearly in general though, these functional forms need not be diagonal in $l$ and $l'$ nor independent of
$m$ in the event that diagonality may exist. These are observational signatures of a preferred spatial direction in cosmology.

One possibility is that this preferred spatial direction is encoded entirely in the primordial power spectrum i.e the subsequent evolution equations are 
not sensitive to a preferred direction. This may happen for instance if there is a period of primordial anisotropic cosmic inflation giving way to an
isotropic FRW expansion afterwards. This possibility has been examined in detail in \cite{Pitrou:2008gk} and \cite{Gumrukcuoglu:2007bx}.
We may recover their results for the angular correlation function by using the result (\ref{colflat}) along with equations
 (\ref{reltheta1}) and (\ref{reltheta2}) which relate the two-index moments $\Theta_{NA}$ to the more familiar $\Theta_{A}$ for scalar perturbations in the event
 that the background spacetime is spatially flat with FRW symmetry:

\begin{eqnarray}
\left<a_{l'm'}a_{lm}\right>&=&  \delta_{mm'}2\pi\int dw \int k dk\frac{\Theta_{l}\Theta^{*}_{l'}}{|\delta|^{2}}(q,t_{0}){\mathcal P}(k,w)P^{l'}_{m}(\beta)P^{l}_{m}(\beta)\nonumber \\
&& \cdot \sqrt{\frac{(2l+1)(2l'+1)(l-m)!(l'-m)!}{16\pi^{2}(l+m)!(l'+m)!}}
\end{eqnarray}
We now write the wave-vector space integral in terms of the magnitude of the 3-dimensional wavevector $\textbf{q}$: $q= |\sqrt{w^{2}+k^{2}}|$ and $\beta$,
the projection of of $q^{i}$ along the z direction, yielding:

\begin{eqnarray}
\left<a_{l'm'}a_{lm}\right>&=&  \delta_{mm'}2\pi\int dq q^{2} \int d\beta \frac{\Theta_{l}\Theta^{*}_{l'}}{|\delta|^{2}}(q,t_{0}){\mathcal P}(q,\beta)P^{l'}_{m}(\beta)P^{l}_{m}(\beta)\nonumber \\
&& \cdot \sqrt{\frac{(2l+1)(2l'+1)(l-m)!(l'-m)!}{16\pi^{2}(l+m)!(l'+m)!}}
\end{eqnarray}
This is identical to the result of \cite{Gumrukcuoglu:2007bx}. The ratio $\Theta_{l}\Theta^{*}_{l'}/|\delta|^{2}$ is taken to be independent of the angular
quantity $\beta$ and so statistical anisotropy may occur solely from the following angular integral:

\begin{eqnarray}
\label{o1}
\sqrt{\frac{(2l+1)(2l'+1)(l-m)!(l'-m)!}{16\pi^{2}(l+m)!(l'+m)!}}\int d\beta{\mathcal P}(q,\beta)P^{l}_{m}(\beta)P^{l'}_{m}(\beta)
\end{eqnarray}
Only if ${\mathcal P}(q,\beta)={\mathcal P}(q)$ is the equation (\ref{o1}) independent of $m$ and diagonal in $l$ and $l'$, therefore any angular dependence present will yield statistical anisotropy. 
The period anisotropic inflation considered in  \cite{Pitrou:2008gk} and \cite{Gumrukcuoglu:2007bx} corresponds to an example of a background where $a\neq b$ and $K=0$. It has been argued \cite{Gumrukcuoglu:2007bx} that one should expect the power spectrum  
${\mathcal P}(q,\beta)$ to respect the symmetry of the background spacetime. The absence of a dependence upon the unit wavevector $\hat{k}^{a}$ in the co-moving 2-surface  
is a reflection of this. Additionally though, there background has a reflection symmetry $(z,x^{a})\rightarrow(-z,-x^{a})$  which suggests that one should additionally insist that ${\mathcal P}(q,\beta)={\mathcal P}(q,-\beta)$. This restriction implies that (\ref{o1}) is only non-vanishing when the difference
between $l$ and $l'$ is an even number  \cite{Gumrukcuoglu:2007bx}. It is interesting to speculate whether this is expected to hold 
for the results (\ref{colclosed}) and (\ref{colopen}) which arise from anisotropically curved backgrounds with the same reflection symmetry.

\section{The Inclusion Of Shear}
\label{inclshear}

The preceding formalism has assumed that the cosmological background is shearless. There are advantages to these scenarios as they avoid
constraints typically associated with shear such as those due to the anisotropy of the CMB present in the background and constraints on the anisotropy of the expansion
rate. As discussed in Section \ref{sf}, shearless solutions with anisotropic curvature may exist only in the presence of a new matter field which allows the background solution to exist
via the particular form of its anisotropic stress.  In the absence of such a matter field, models with background anisotropic curvature concomitantly possess shear \cite{Graham:2010hh},
and therefore a more general analysis of the effects of background anisotropic curvature should allow for background shear. Recall that allowing for shear, the background
metric can be written by as:

\begin{eqnarray}
ds^{2} = -a^{2}\left(dt^{2} + \frac{b^{2}}{a^{2}}dz^{2}+\gamma_{cd}dx^{c}dx^{d}\right)
\end{eqnarray}
In the presence of shear (i.e. $b(t)\neq a(t)$), the co-moving metric is clearly no longer static. However, with this choice of co-moving metric we may as before decompose metric
perturbations in terms of eigenfunctions, eigenvectors, and eigentensors of the the Laplace-Beltrami operator associated with the static metric $\gamma_{cd}$. For instance, a function may be again
decomposed as follows:

\begin{eqnarray}
\label{scalard2}
y(t,z,x^{c}) = \sum_{l,m}\tilde{y}_{k}(t,z){\mathcal E}^{k}_{m}(x^{d})
\end{eqnarray}
It is then desirable, as before, to decompose the functions $\tilde{y}_{k}(t,z)$ into harmonic components and then obtain field equations for modes which depend
only on wavenumber labels and time. 

A choice for such harmonics are the eigenfunctions $\epsilon_{w}(z,t)$ of the Laplace-Beltrami operator on the co-moving 3-space projected along
the z-direction, which are thus defined by the following equation:

\begin{eqnarray}
\left(\frac{a^{2}}{b^{2}}\right)\partial^{2}_{z}\epsilon_{w}(z,t)= -w^{2}\epsilon_{w}(z,t)
\end{eqnarray}
where the $w^{2}$ are constants. Solutions to this equation are then simply plane waves but with time-dependent frequency $\omega\equiv (b/a)w$, i.e.

\begin{eqnarray}
\label{szum}
\tilde{y}_{k}(t,z)= \frac{1}{2\pi} \int d \omega(t) \tilde{y}_{k,\omega}(t) e^{i\omega(t)z}
\end{eqnarray}
Therefore one may decompose a perturbed tensor such as the Einstein equation in terms of generalized sums of the functions $\epsilon_{w}$ multiplied by the eigenfunctions, eigenvectors, and eigentensors of the the Laplace-Beltrami operator associated with the static metric $\gamma_{cd}$. Care must be taken with partial time derivatives of (\ref{szum}) as the integration measure and `wavenumber' now have a time dependence. 
Consequently, the components of the perturbed Einstein tensor $\delta G^{a}_{\phantom{a}b}$ will involve additional terms due to the presence of shear in the background and its effect on the definition of 
harmonic components.

If the shear is sufficiently small, it may be regarded as a small homogeneous perturbation to a shearless background spacetime. For instance, in the closed case this may be
realized by considering the polar perturbation $F= F(t)\delta(w)\delta_{L0}$, for which the collisionless photon Boltzmann equation becomes:

\begin{eqnarray}
(\Theta_{02}+\frac{2}{3}F)' &=& 0\\
(\Theta_{00}+\frac{1}{3}F)' &=& 0 
\end{eqnarray}
The contribution to $\Theta_{00}$, assumed to have no spatial dependence, can be absorbed into the definition of the background temperature, leaving a time-dependent contribution to
$\left<a_{lm}a_{l'm'}\right>= C_{l}(m,t)\delta_{ll'}\delta_{l2}\delta_{mm'}\delta_{m0}$, i.e. a quadrupole contribution to the CMB. Note however that the perturbations
to the CMB are not statistically isotropic as the function $C_{l}$ depends upon $m$.  

In treating the presence of shear as a background quantity, further complications emerge. Consider decomposing the momentum of a photon propagating in the background spacetime. We may decompose the particle momentum $P^{\mu}$ as follows:

\begin{eqnarray}
P^{\mu}\partial_{\mu} &=& P^{t}\partial_{t}+\frac{1}{a}\left(\left(\frac{a}{b}\right)p\alpha \partial_{z}+p\sqrt{1-\alpha^{2}}\hat{p}^{c}\partial_{c}\right)\\
    &=& \frac{1}{a}\left(\bar{P}^{t}\partial_{t}+\bar{P}^{i}\partial_{i}\right)
\end{eqnarray}
The null-norm constraint immediately enforces $\bar{P}^{t}=p$ if it is required that $\gamma_{cd}\hat{p}^{c}\hat{p}^{d}=1$. The geodesic equations then imply the following relations:

\begin{eqnarray}
\frac{dp}{dt} &=& -p\left(\alpha^{2}\frac{b'}{b}+(1-\alpha^{2})\frac{a'}{a}\right) \\
\frac{d\alpha}{dt} &=& \alpha(\alpha^{2}-1)\left(\frac{b'}{b}-\frac{a'}{a}\right) \\
\frac{d\hat{p}^{c}}{dt} &=& 0
\end{eqnarray}
Therefore one can no longer assume that $\frac{d\alpha}{dt} =0$ at the background level, and this must be accounted for in the Boltzmann equation along with the background angular dependence of $f$ implied
by the presence of shear. However, it seems that the expansion of temperature perturbations in terms of ${\mathcal B}^{(P)}_{NA}$ and ${\mathcal B}^{(A)}_{NA}$ will remain appropriate, and that the expressions 
for the angular correlation function (\ref{colclosed}),(\ref{colopen}), and (\ref{colflat}) will still apply- the difference from the shearless case appearing in a generically different dependence of the moments $\Theta_{NA}$ upon the labels $w,k$, and $t$.

\section{Conclusions}
\label{conc}

In this paper we have presented a formalism for the analysis of cosmological perturbations to an anisotropically curved but shearless background.
In particular, a harmonic decomposition of spacetime dependent fields has been introduced, as well as a moment decomposition of the dependence
these harmonics may in turn have on the direction of particle momentum. One then obtains via the massless and massive particle Boltzmann equations
a system of coupled ordinary differential equations in conformal time. These equations are coupled to one another by the presence of adjacent moments of an identical species, possible collisional terms, and to
the spacetime via metric source terms. In turn, the Einstein equations describing the evolution of metric perturbations are coupled to the matter fields
via their stress energy tensors which can be determined from the fields' collected harmonic moments. Given a set of initial data, these equations are sufficient to describe
the evolution of cosmological perturbations. 

Of particular interest is the temperature-temperature power spectrum. Explicit expressions for the angular correlation function for polar
perturbations have been determined in Section \ref{polcorr}, and the conditions under which they represent statistical anisotropy have been discussed.
In particular, these expressions have been written in a manner which seeks to separate statistical anisotropy which is present via initial conditions (primordial
statistical anisotropy) from that which ensues from the evolution of perturbations on an anisotropically curved background and in the presence of perturbations to which fields produce this curvature (which
may be termed emergent statistical anisotropy). Indeed, from our formalism we have recovered antecedent results regarding statistical anisotropy of temperature fluctuations 
induced by an anisotropic primordial power spectrum in an FRW universe (an example of primordial anisotropy, see \cite{Pitrou:2008gk},\cite{Gumrukcuoglu:2007bx})
as well as statistical anisotropy induced by a background with anisotropic curvature and anisotropic expansion rate (emergent anisotropy, see \cite{Graham:2010hh}). Furthermore,
the present formalism in principle can be used to see the free-streaming evolution of the imprint of certain anisotropic physical processes in the early universe (e.g. see \cite{BlancoPillado:2010uw}).

Although in principle the formalism may be used to determine the evolution of cosmological perturbations, there are yet several steps to take in order to be able to accurately assess
what kind of observations to expect in a universe with background anisotropic curvature supported by a field such as that discussed in Section \ref{sf}. 
Firstly, it is important to provide a careful treatment of primordial power spectra in the presence of anisotropic curvature, where intuitively it would seem that the 
effects of curvature should be unimportant for the co-moving scales probed today. Secondly, the results in this paper have been presented in a general gauge and the existence of particular gauge choices
which may offer considerable simplification has not been determined. Thirdly, polarization in the photon field has not yet been allowed for in calculations, and possible collisional terms (for instance Thompson scattering between electrons and photons) have not yet been explicitly included in the Boltzmann equations.
Additionally, it would be desirable to extend the formalism to the case where shear is present in the background. The modifications to the formalism that this should
involve have been discussed in Section \ref{inclshear}.

Acknowledgements: We would like to thank Miguel Quartin, Tomi Koivisto, David Mota, Matthew Mewes, Tim Clifton, Chris Clarkson, Pedro Ferreira, Joe Zuntz, and Mark Wyman for useful discussions.

\section{Bibliography}
\bibliographystyle{unsrt}

\appendix

\section{Special functions and addition theorems}
\label{appa}

In the decomposition of spacetime dependent perturbational functions, vectors, and tensors we have made use of the eigenfunctions ${\mathcal E}^{k}_{m}$ of the Laplace-Beltrami operator $\nabla^{2}$ on the co-moving two surface. We now discuss properties of these functions in each curvature case in more detail. For simplicity we discuss eigenfunctions of the Laplace-Beltrami operator $\tilde{\nabla}^{2}$ which is associated with the co-moving surface with metric $\tilde{\gamma}_{cd}$ (see Section \ref{intro} for discussion of this quantity).

\subsection{Positive curvature case}

In the positively curved case eigenfunctions ${\mathcal E}^{l}_{m}(\tilde{x}^{a})$ of the Laplace-Beltrami operator $\tilde{\nabla}^{2}$ are the familiar spherical harmonics $Y^{l}_{m}$, 
which satisfy:

\begin{eqnarray}
\tilde{\nabla}^{2}Y^{l}_{m}(\theta,\phi) &=& -l(l+1)Y^{l}_{m}(\theta,\phi) \\
-i\frac{\partial Y^{l}_{m}(\theta,\phi)}{\partial\phi} &=& m Y^{l}_{m}(\theta,\phi)
\end{eqnarray}
where $0\leq l \leq \infty$ ($l\in \mathbf{Z}$) and $-l \leq m \leq l$.

\subsubsection{Connection to representations of the group SU(2)}

Recall that the group $SU(2)$  has a realization as the group of all 2$\times$2 matrices $M^{A}_{\phantom{A}B}$ with determinant equal to one and satisfying the following condition:

\begin{eqnarray}
\delta_{BB'}  =  \delta_{AA'}M^{A}_{\phantom{A}B}M^{(*)A'}_{\phantom{*A'}B'}
\end{eqnarray}
where the star denotes complex conjugation and un-primed and primed indices are indices in a two-dimensional vector space ${\mathcal W}$ and two-dimensional conjugate vector space ${\mathcal W}^{*}$ respectively (e.g. for instance see Chapter 12 of \cite{Wald:1984rg}).
It may be shown \cite{Miller1} that a generic 
element $g$ of the group may be parameterized by three numbers $\phi,\theta,\psi$ such that $M(g)^{A}_{\phantom{A}B}$ takes the form:

\[
\left( {\begin{array}{cc}
 \cos\left(\frac{\theta}{2}\right)e^{i(\phi+\psi)/2} & i\sin\left(\frac{\theta}{2}\right)e^{i(\phi-\psi)/2} \\
i\sin\left(\frac{\theta}{2}\right)e^{i(\psi-\phi)/2}& \cos\left(\frac{\theta}{2}\right)e^{-i(\phi+\psi)/2}
 \end{array} } \right)
\]
where $0\leq\phi<2\pi$, $0\leq\theta\leq 2\pi$, $-2\pi\leq\psi<2\pi$ and the identity element $e$ of the group
is the matrix  $M^{A}_{\phantom{A}B}(\theta,\phi,\psi=0)=\delta^{A}_{\phantom{A}B}$. With this realization, the group multiplication operation
which takes two group elements $g_{1}$ and $g_{2}$ to produce another element $g_{1}g_{2}$ is simply the process
of matrix multiplication: $M^{A}_{\phantom{A}B}(g_{1}g_{2})=M^{A}_{\phantom{A}C}(g_{1})M^{C}_{\phantom{C}B}(g_{2})$ and so, for instance $\psi_{12}$ may 
be found in terms of $\psi_{1}$ and $\psi_{2}$ etc. We now adopt the following definitions of \cite{Miller1}:

\begin{itemize}
\item Let V be a Hilbert space. A representation of a group G with representation space V is a homomorphism $\mathbf{T}:g\rightarrow
\mathbf{T}(g)$ of G into the space of bounded linear operators on V. From this it follows that:

\begin{eqnarray}
\mathbf{T}(g_{1})\mathbf{T}(g_{2}) &=& \mathbf{T}(g_{1}g_{2}) \label{gm}\\
\mathbf{T}(g)^{-1} &=& \mathbf{T}(g^{-1}) \\
\mathbf{T}(e) &=& \textbf{I} \;\;\;\;\;\;\;\; g_{1},g_{2},g,g^{-1} \in G
\end{eqnarray}
where $\textbf{I}$ is the identity operator.
\item The representation $\mathbf{T}$ is reducible if there is a proper subspace W of V which is invariant under $\mathbf{T}$.
Otherwise $\textbf{T}$ is irreducible.
\item Let $\{v_{n}\}$ be an orthogonal basis of V with inner product $\left<v_{m},v_{n}\right>=\delta_{mn}$. The \emph{matrix element}
$T_{mn}(g)$ may then be defined as follows: $T_{mn}(g)\equiv \left<v_{m},(\mathbf{T}(g)v)_{n}\right>$. If $T^{*}_{nm}(g)=T_{mn}(g^{-1})$ then
the representation is said to be unitary.
\end{itemize}

It may be shown \cite{Vilenkin1} that for the group SU(2)), there're a set of irreducible representations $\textbf{T}^{l}$ labelled by the number $l$ which is an integer ranging 
taking values $0,1,2...\infty$, with basis vectors $e^{im\alpha}$ ($-l\leq m\leq l$, $0\leq \alpha < 2\pi$) and accompanying inner product $\left<e^{im\alpha},e^{in\alpha}\right>\equiv \frac{1}{2\pi}\int (e^{im\alpha})^{*}e^{in\alpha}d\alpha$. In this case the matrix elements $T^{(l)}_{mn}$
take the following form:

\begin{eqnarray}
T^{(l)}_{mn}(g)= i^{m-n}e^{-i(m\phi+n\psi)}{\mathcal P}^{(l)}_{mn}(\cos\theta) \label{tlmn}
\end{eqnarray}
where

\begin{eqnarray}
\nonumber {\mathcal P}^{(l)}_{mn}(\eta) &\equiv & \left[\frac{(l-n)!(l+m)!}{(l+n)!(l-m)!}\right]^{\frac{1}{2}}\frac{(1-\eta)^{\frac{m-n}{2}}(1+\eta)^{\frac{m+n}{2}}}{2^{m}(m-n)!}\\
 && \times _{2}F_{1}(l+m+1,-l+m;m-n+1;\frac{1-\eta}{2}) \label{plmn}
\end{eqnarray}
where $_{2}F_{1}$ is Gauss's hypergeometric function. The above expression applies when $m-n\geq 0$. If $m-n< 0$ then 
one must replace $m$ and $n$ by $-m$ and $-n$ respectively. This function has the following symmetry properties:

\begin{eqnarray}
{\mathcal P}^{(l)}_{mn}(\eta) &=& (-1)^{m+n}{\mathcal P}^{(l)}_{nm}(\eta) \\
{\mathcal P}^{(l)}_{mn}(\eta) &=& (-1)^{m-n}{\mathcal P}^{(l)}_{-m,-n}(\eta) \\
{\mathcal P}^{(l)}_{mn}(\eta) &=& {\mathcal P}^{(l)}_{-n,-m}(\eta)
\end{eqnarray}
It follows immediately that

\begin{eqnarray}
{\mathcal P}^{(l)}_{00}(\cos\theta) &=& P_{(l)}(\cos\theta)
\end{eqnarray}
where $P_{l}(\cos\theta)$ is the familiar Legendre polynomial. Furthermore we adopt the following definition
of the associated Legendre function $P^{m}_{(l)}$:

\begin{eqnarray}
\nonumber P^{m}_{(l)}(\eta) &=& (-1)^{m}\frac{\Gamma(l+m+1)}{m!\Gamma(l-m+1)}\left[\frac{1-\eta}{1+\eta}\right]^{m/2} \\
      && \times _{2}F_{1}(1+l,-l;m+1;\frac{1-\eta}{2})
\end{eqnarray}
for $m\geq 0$ and

\begin{eqnarray}
P^{m}_{(l)}(\eta) &=& (-1)^{m}\frac{\Gamma(l+m+1)}{\Gamma(l-m+1)}P^{-m}_{(l)}(\eta)
\end{eqnarray}
if $m<0$. Therefore it follows that

\begin{eqnarray}
P^{n}_{(l)}(\eta)=\left[\frac{(n+l)!}{(l-n)!}\right]^{1/2}{\mathcal P}^{(l)}_{-n,0}
\end{eqnarray}
where $n>0$, and hence:

\begin{eqnarray}
\label{spherdef}
T^{(l)}_{-n0}(g) &=& i^{-n}\sqrt{\frac{4\pi}{2l+1}}Y_{n,l}(g)
\end{eqnarray}
where $Y^{m}_{l}(g)$ is the spherical harmonic function, defined as follows:

\begin{eqnarray}
Y^{m}_{l}(g)= \sqrt{\frac{(2l+1)(l-m)!}{4\pi(l+m)!}}P^{m}_{l}(\cos\theta)e^{im\phi}
\end{eqnarray}
The group multiplication property (\ref{gm}) produces the following result for the matrix elements:

\begin{eqnarray}
T^{(l)}_{mn}(g_{1}g_{2})=\sum^{l}_{p=-l}T^{(l)}(g_{1})_{mp}\delta^{pq}T^{(l)}_{qn}(g_{2})
\end{eqnarray}
As the representation is unitary, we have $T_{ab}^{(l)}(g_{1})=T_{ba}^{*(l)}(g_{1}^{-1})$ and so we have:

\begin{eqnarray}
\label{addclosed}
T^{(l)}_{mn}((\theta,\phi,\psi)=0)=\sum^{l}_{q=-l}T^{*(l)}_{qm}(g_{1})\delta^{qp}T^{(l)}_{pn}(g_{1})
\end{eqnarray}
Applying the definition (\ref{spherdef}) we then have that:

\begin{eqnarray}
\label{addclosed}
\left(\frac{4\pi}{2l+1}\right)^{\frac{1}{2}} Y^{l}_{0}((\theta,\psi)=0)&=& \sum^{l}_{-l}T^{*(l)}_{0q}(g_{1})\delta^{qp}T^{(l)}_{0n}(g_{1}) \\
   &=& \sum^{l}_{m=-l} \left(\frac{4\pi}{2l+1}\right)Y^{*l}_{m}(g)Y^{l}_{m}(g)
\end{eqnarray}
This result is known as Uns\"{o}ld's theorem and will be useful in our calculations. An additional summation that appears is the following:

\begin{eqnarray}
\label{spinsum}
\sum_{M=-L}^{M=L}\tilde{A}^{(-)ab..n}\tilde{\nabla}_{ab..n}Y^{L}_{M}\tilde{A}^{(+)ab..n}\tilde{\nabla}_{ab..n}Y^{*L}_{M}
\end{eqnarray}
We now show that when $\tilde{A}^{(-)a}$ is defined in using $\textbf{e}_{\theta}$ and $\textbf{e}_{\phi}$ as basis
vectors, the summation (\ref{spinsum}) can be shown to be an example of the addition theorem.
The matrices ${\mathcal P}^{(l)}_{mn}$ satisfy the following recurrence relations:

\begin{eqnarray*}
\sqrt{1-\eta^{2}}\frac{d}{d\eta}{\mathcal P}^{(l)}_{mn}+\frac{(n\eta-m)}{\sqrt{1-\eta^{2}}}{\mathcal P}^{(l)}_{mn} &=& -\sqrt{(l-n)(l+n+1)}{\mathcal P}^{(l)}_{m,n+1}\\
\sqrt{1-\eta^{2}}\frac{d}{d\eta}{\mathcal P}^{(l)}_{mn}-\frac{(n\eta-m)}{\sqrt{1-\eta^{2}}}{\mathcal P}^{(l)}_{mn} &=& \sqrt{(l+n)(l-n+1)}{\mathcal P}^{(l)}_{m,n-1}
\end{eqnarray*}
We write these equations in terms of matrix elements:

\begin{eqnarray*}
\left(\sqrt{1-\eta^{2}}\frac{\partial}{\partial \eta}+n\frac{\eta}{\sqrt{1-\eta^{2}}}-i\frac{\partial}{\partial\phi}\right)T^{(l)}_{mn}=-\sqrt{(l-n)(l+n+1)}ie^{i\Psi}T^{(l)}_{m,n+1}\\
\left(\sqrt{1-\eta^{2}}\frac{\partial}{\partial \eta}-n\frac{\eta}{\sqrt{1-\eta^{2}}}+i\frac{\partial}{\partial\phi}\right)T^{(l)}_{mn}= -\sqrt{(l+n)(l-n+1)}ie^{-i\Psi}T^{(l)}_{m,n-1}
\end{eqnarray*}
We now consider the tensor $X_{bcd..m} \equiv \tilde{\nabla}_{bcd..m}f$ where $f$ is an appropriate function. Now consider taking the gradient of this tensor and projecting all indices along the vector field $\tilde{A}^{(-)a}=|K|^{-1/2}A^{(-)a}=((\textbf{e}_{\theta})^{a}+i(\textbf{e}_{\phi})^{a})$. Therefore we have the identity:

\begin{eqnarray}
\nonumber
\tilde{A}^{(-)a}\tilde{A}^{(-)bc..m}\tilde{\nabla}_{a}X_{bc..m} &=& \tilde{A}^{(-)a}\tilde{\nabla}_{a}(\tilde{A}^{(-)bc..m}X_{bc..m})\\
&&-X_{bc..m}\tilde{A}^{(-)a}\tilde{\nabla}_{a}\tilde{A}^{(-)bc..m}
\end{eqnarray}
The gradient of $\tilde{A}^{(-)a}$ is:

\begin{eqnarray}
\tilde{\nabla}_{b}\tilde{A}^{(-)a}= \frac{\cos\theta}{\sqrt{2}}(\partial_{\phi})_{b}(-i(\textbf{e}_{\theta})^{a}+(\textbf{e}_{\phi})^{a})
\end{eqnarray}
It follows then that  $\tilde{\nabla}_{a}\tilde{A}^{(-)bc..m}$ is:

\begin{eqnarray}
\tilde{\nabla}_{a}\tilde{A}^{(-)bc..m} &=& \sum_{j}\tilde{A}^{....(m-1)}\tilde{\nabla}_{a}\tilde{A}^{(-)j}\\
             &=& (\textbf{e}_{\phi})_{a}\frac{\cos\theta}{\sqrt{2}\sin\theta}\sum_{j}\tilde{A}^{(-)....(m-1)}(-i(\textbf{e}_{\theta})^{j}+(\textbf{e}_{\phi})^{j})\\
             &=& -i(\textbf{e}_{\phi})_{a}\frac{\cos\theta}{\sin\theta}\sum_{j}\tilde{A}^{(-)....(m-1)}\tilde{A}^{(-)j}\\
             &=& -im(\textbf{e}_{\phi})_{a}\frac{\cos\theta}{\sin\theta}\tilde{A}^{(-)bc..m}
\end{eqnarray}
where the last line follows from the symmetry of $\tilde{A}^{bc..m}$ with respect to the interchange of all indices. Therefore:

\begin{eqnarray*}
\nonumber
\tilde{A}^{(-)a}\tilde{A}^{(-)bc..m}\tilde{\nabla}_{a}X_{bc..m} &=& \tilde{A}^{(-)a}\tilde{\nabla}_{a}(\tilde{A}^{(-)bc..m}X_{bc..m}) \\
&& \nonumber-\frac{m\cos\theta}{\sqrt{2}\sin\theta}\tilde{A}^{(-)bc..m}X_{bc..m} \\
  &=& \left(\tilde{A}^{(-)a}\tilde{\nabla}_{a}-\frac{m\cos\theta}{\sqrt{2}\sin\theta}\right)\tilde{A}^{(-)bc..m}X_{bc..m}
\end{eqnarray*}
similarly it may be shown that

\begin{eqnarray*}
\nonumber
\tilde{A}^{(+)a}\tilde{A}^{(+)bc..m}\tilde{\nabla}_{a}X_{bc..m} &=& \tilde{A}^{(+)a}\tilde{\nabla}_{a}(\tilde{A}^{(+)bc..m}X_{bc..m})\\
\nonumber &&-\frac{m\cos\theta}{\sqrt{2}\sin\theta}\tilde{A}^{(+)bc..m}X_{bc..m} \\
  &=& \left(\tilde{A}^{(+)a}\tilde{\nabla}_{a}-\frac{m\cos\theta}{\sqrt{2}\sin\theta}\right)\tilde{A}^{(+)bc..m}X_{bc..m}
\end{eqnarray*}
Additionally, on some function $y$ we have

\begin{eqnarray}
\label{p1}
\partial_{+}y &=& \frac{1}{\sqrt{2}}(\partial_{\theta}+\frac{i}{\sin\theta}\partial_{\phi})y = \tilde{A}^{(-)a}\tilde{\nabla}_{a}y \\
\label{p2}
\partial_{-}y &=& \frac{1}{\sqrt{2}}(\partial_{\theta}-\frac{i}{\sin\theta}\partial_{\phi})y = \tilde{A}^{(+)a}\tilde{\nabla}_{a}y
\end{eqnarray}
Therefore, projection of the gradient of $X_{bcd..p}$ along the vector $\tilde{A}^{(+)a}$ for all indices is equal to the differential operator ${\mathcal D}_{+}(p)\equiv (1/\sqrt{2})(\partial_{\theta}+(i/\sin\theta)\partial_{\phi}-p(\cos\theta)/(\sin\theta))$ applied to the projection of $X_{bcd..p}$ along the vector $\tilde{A}^{(+)a}$ for all indices.

 However, recalling that $\eta=\cos\theta$, we similarly see that the recursion relations imply that the matrix element index $n$ is raised by operating on $T^{(l)}_{mn}$ with a differential operator proportional to ${\mathcal D}_{+}(p)$:

\begin{eqnarray}
{\mathcal D}_{+}(n)T^{(l)}_{mn} &=& \frac{1}{\sqrt{2}}\sqrt{(l-n)(l+n+1)}ie^{i\Psi}T^{(l)}_{m,n+1}
\end{eqnarray}
Now recall that $Y_{n,l}(g)=i^{n}\sqrt{(2l+1)/4\pi} T^{(l)}_{-n0}(g)$. This implies that 

\begin{eqnarray}
\tilde{A}^{(-)ac..p}\tilde{\nabla}_{abc..p}Y_{nl} &=& \prod^{p}_{q=0}{\mathcal D}_{+}(q)Y^{l}_{n}\\
     &=& \prod^{p}_{q=0}{\mathcal D}_{+}(q)i^{n}\sqrt{\frac{2l+1}{4\pi}}T^{l}_{-n0}\\
     &=& \left(\frac{1}{\sqrt{2}}\right)^{p}i^{n+p}e^{ip\Psi}\sqrt{\frac{2l+1}{4\pi}}T^{(l)}_{-n,p}\cdot \nonumber \\
     && \prod^{p}_{q=0}\sqrt{(l+q)(l-q+1)} \label{a33}
\end{eqnarray}
Note the close relation of these projected derivatives to the spin-weighted spherical harmonics  \cite{Newman:1966ub},\cite{Kostelecky:2009zp}.
Then rearranging (\ref{a33}) we have that:

\begin{eqnarray}
T^{(l)}_{-n,p} &=&  \frac{\left(\frac{1}{\sqrt{2}}\right)^{-p}i^{-n-p}e^{-ip\Psi}\sqrt{\frac{4\pi}{2l+1}}}{\prod^{|p|}_{q=0}\sqrt{(l+q)(l-q+1)}}\tilde{A}^{(-)ac..|p|}\tilde{\nabla}_{abc..|p|}Y^{l}_{n}
\end{eqnarray}
The addition theorem for matrix elements immediately implies then that:

\begin{eqnarray*}
\nonumber
\frac{2l+1}{4\pi}T^{l}_{p,p}(0)&=& \frac{2l+1}{4\pi}\cdot 1\\
 &=& \frac{\left(\frac{1}{\sqrt{2}}\right)^{-2p}}{\prod_{q=0}^{p}(l+q)(l+1-q)}\sum_{m=-l}^{m=l}\tilde{A}^{(-)ab..p}\tilde{\nabla}_{ab..p}Y^{l}_{m}\tilde{A}^{(+)ab..p}\tilde{\nabla}_{ab..p}Y^{*l}_{m}
\end{eqnarray*}
\subsection{Negative curvature case}

As in the positively curved case, we characterize perturbations in terms of eigenfunctions of the Laplace-Beltrami operator $\tilde{\nabla}^{2}$ now defined on the upper sheet of the two dimensional hyperboloid of two-sheets possessing metric $\tilde{\gamma}_{ab}$. We shall call these functions $Z^{k}_{m}$,  defined as solutions to the following equations:

\begin{eqnarray}
\nabla^{2}Z^{\tilde{k}}_{m}(\theta,\phi) &=& -\tilde{k}^{2}Z^{\tilde{k}}_{m}(\theta,\phi)= -\left(\frac{1}{4}+\lambda^{2}\right)Z^{\tilde{k}}_{m}(\theta,\phi) \\
-i\frac{\partial Z^{\tilde{k}}_{m}(\theta,\phi)}{\partial\phi} &=& m Z^{\tilde{k}}_{m}(\theta,\phi)
\end{eqnarray}
Writing $\tilde{k}^{2}=\tilde{k}\cdot \tilde{k}^{*}$,  solutions appear in three branches. The first covers values $\tilde{k}^{2}>1/4$, for 
which $\tilde{k}$ takes the value $\tilde{k}=-1/2+i\lambda$ where $0<\lambda<\infty$. The second covers values
$\tilde{k}=-1/2+i\lambda$ where $0i<\lambda<-(1/2)i$.  For these two cases the label $m$ satisfies $m\in\mathbf{Z}$. The third
branch covers the case $\tilde{k}=0$, allows only $m=0$ and is given by $Z^{0}_{0}=\textrm{constant}$.

Note that a square-integrable function $f(\theta)$ on the interval $[0,\infty)$ can be expanded as follows:

\begin{eqnarray}
f(\theta)= \int_{0}^{\infty} c(\lambda) {\mathcal Q}^{-\frac{1}{2}+i\lambda}_{00}(\theta)\lambda\tanh(\pi\lambda)d\lambda
\end{eqnarray}
where 

\begin{eqnarray}
c(\lambda)= \int_{0}^{\infty}f(\theta){\mathcal Q}^{-\frac{1}{2}-i\lambda}_{00}(\theta)\sinh\theta d\theta
\end{eqnarray}
i.e. to describe square integrable functions on the hyperboloid we need only consider the first branch $0\leq\lambda\leq \infty$, up
to a constant component to the function.

We now outline how eigenfunctions of the Laplace-Beltrami operator on the upper sheet of the hyperboloid of two sheets are related to matrix elements of a unitary, irreducible
representation of the group SU(1,1). This group has a realization as the group of all 2$\times$2 matrices $M^{A}_{\phantom{A}B}$ with determinant equal to one and satisfying the following condition:

\begin{eqnarray}
\eta_{BB'}  =  \eta_{AA'}M^{A}_{\phantom{A}B}M^{(*)A'}_{\phantom{(*)A'}B'}
\end{eqnarray}
where $\eta_{BB'}= \mathrm{diag}(-1,1)$. An element $g$ of the group may be parameterized as follows:

\[
\left( {\begin{array}{cc}
 \cosh\left(\frac{\xi}{2}\right)e^{i(\phi+\psi)/2} & \sinh\left(\frac{\xi}{2}\right)e^{i(\phi-\psi)/2} \\
\sinh\left(\frac{\xi}{2}\right)e^{i(\psi-\phi)/2}& \cosh\left(\frac{\xi}{2}\right)e^{-i(\phi+\psi)/2}
 \end{array} } \right)
\]
where $0\leq \phi< 2\pi$, $-2\pi \leq \psi < 2\pi$, $0\leq \xi <\infty$. By inspection the identity element is associated with $\phi=\psi=t=0$.

 It may be shown \cite{Vilenkin1} that there exist irreducible representations $T_{\chi=(\tilde{k},\epsilon)}$ where
where $\tilde{k}= i\lambda-1/2$ $(0\leq \lambda \leq \infty) $ and $\epsilon$ may take the value 0 or 1/2. The basis for each of these representations may be taken to be eigenvectors $e^{im\alpha}$ 
($-\infty< m < \infty$, $m\in \mathbf{Z}$, $0\leq \alpha < 2\pi$) and accompanying inner product $\left<e^{i\xi\alpha},e^{i\mu\alpha}\right>\equiv \frac{1}{2\pi}\int e^{-i\xi\alpha}e^{i\nu\alpha}d\alpha$. As before,
matrix elements $T^{\chi}_{ab}$ may be defined as follows:

\begin{eqnarray}
T^{(\chi)}(g)e^{-i\xi\alpha}=\sum_{b=-\infty}^{+\infty}T_{\xi b}^{(\chi)}(g)e^{-ib\alpha}
\end{eqnarray}
they are found to be \cite{Vilenkin1};

\begin{eqnarray}
T_{ab}^{(\chi)}(g) = e^{-i(a+\epsilon)\phi-i(b+\epsilon)\psi}{\mathcal Q}^{\tilde{k}}_{a+\epsilon,b+\epsilon}(\cosh(\xi))
\end{eqnarray}

For $a\geq b$ we have:

\begin{eqnarray}
\nonumber
{\mathcal Q}^{\tilde{k}}_{ab}(\cosh \xi) &=& \frac{\Gamma(\tilde{k}-b+1)(\cosh(\frac{\xi}{2}))^{2\tilde{k}}(\tanh(\frac{\xi}{2}))^{a-b}}{(a-b)!\Gamma(\tilde{k}-a+1)}\\
&& \times _{2}F_{1}(-\tilde{k}-b,-\tilde{k}+a;a-b+1,\tanh^{2}(\frac{\xi}{2})) \label{qdef}
\end{eqnarray}
For $a<b$ it is necessary to replace $a$ and $b$ by $-a$ and $-b$ respectively. The ${\mathcal Q}^{\tilde{k}}_{ab}$ satisfy the following symmetry
relations:

\begin{eqnarray}
{\mathcal Q}^{\tilde{k}}_{mn}(\cosh \xi) &=& {\mathcal Q}^{\tilde{k}}_{-m,-n}(\cosh \xi) \\
{\mathcal Q}^{\tilde{k}}_{nm} &=& \frac{\Gamma(\tilde{k}+m+1)\Gamma(\tilde{k}-m+1)}{\Gamma(\tilde{k}+n+1)\Gamma(\tilde{k}-n+1)}{\mathcal Q}^{\tilde{k}}_{mn}(\cosh\xi)
\end{eqnarray}

The addition theorem for the matrix elements is then:

\begin{eqnarray}
T_{ab}^{(\chi)}(g_{1}g_{2})= \sum_{c=-\infty}^{+\infty}T_{ac}^{(\chi)}(g_{1})\delta^{cd}T_{db}^{(\chi)}(g_{2})
\end{eqnarray}
Again, as the representation is unitary, we have $T_{ab}^{(\chi)}(g_{1})=T_{ba}^{*(\chi)}(g_{1}^{-1})$ and so we have:

\begin{eqnarray}
\label{hypadd}
T_{ab}^{(\chi)}(\phi=\psi=\xi=0)= \sum_{c=-\infty}^{+\infty}T_{ac}^{(\chi)}(g_{1})\delta^{cd}T_{bd}^{*(\chi)}(g_{1})
\end{eqnarray}
From hereon in we shall look exclusively at the case $\epsilon=0$. It may be checked that the function $T_{m0}^{i\lambda-\frac{1}{2}}$ is an eigenfunction of the Laplace-Beltrami operator on the upper sheet of the hyperboloid 
of two sheets:

\begin{eqnarray}
\tilde{\nabla}^{2}T_{m0}^{i\lambda-\frac{1}{2}}(\cosh(\theta))=-\left(\frac{1}{4}+\lambda^{2}\right)T_{m0}^{i\lambda-\frac{1}{2}}(\cosh(\theta))
\end{eqnarray}
where

\begin{eqnarray}
\tilde{\nabla}^{2} = \frac{\partial^{2}}{\partial \theta^{2}}+\frac{\cosh\theta}{\sinh\theta} \frac{\partial}{\partial \theta}+\frac{1}{\sinh^{2} \theta}\frac{\partial^{2}}{\partial\phi^{2}}
\end{eqnarray}
The matrix elements $T_{-m0}^{i\lambda-\frac{1}{2}}\equiv e^{im\phi}Q^{\lambda}_{m}(\cosh \theta) \equiv Z^{\tilde{k}}_{m}$ may then form a basis for a harmonic expansion. The 
addition theorem for matrix elements then implies that:

\begin{eqnarray}
Z^{\tilde{k}}_{0}((\theta,\phi)=0) = 1=\sum_{-\infty}^{\infty}Z^{*\tilde{k}}_{m}(\cosh(\theta))Z^{\tilde{k}}_{m}(\cosh(\theta))
\end{eqnarray}
This is an analogue of Uns\"{o}ld's theorem theorem for the case of functions on the hyperboloid of two sheets. Another quantity that appears in the calculations is the following summation:

\begin{eqnarray}
\sum_{m=-\infty}^{m=\infty}\tilde{A}^{(-)ab..p}\tilde{\nabla}_{ab..p}Z^{\tilde{k}}_{m}\tilde{A}^{(+)ab..p}\tilde{\nabla}_{ab..p}Z^{*\tilde{k}}_{m}
\end{eqnarray}
Analogously to the positively curved case, we will seek to calculate this quantity by relating projections of $\tilde{\nabla}_{ab..p}Z^{\tilde{k}}_{m}$ along
the vector $\tilde{A}^{(-)a}$.

In the hyperboloid case, expressions for $\tilde{A}^{(\pm)a}$ in terms of unit basis vectors $\textbf{e}_{\theta}$ and $\textbf{e}_{\phi}$ are identical, however
now $(\textbf{e}_{\phi})^{a}=(1/\sinh\theta)(\partial_{\phi})^{a}$. Explicit calculation shows that:

\begin{eqnarray}
\tilde{\nabla}_{b}\tilde{A}^{(-)a}= \frac{\cosh\theta}{\sqrt{2}\sinh\theta}(\textbf{e}_{\phi})_{b}(-i(\textbf{e}_{\theta})^{a}+(\textbf{e}_{\phi})^{a})
\end{eqnarray}
Following the calculation of the positively curved case, we find that for the tensor $X_{bc..m}\equiv \tilde{\nabla}_{bc..m}f$

\begin{eqnarray*}
\nonumber
\tilde{A}^{(-)a}\tilde{A}^{(-)bc..m}\tilde{\nabla}_{a}X_{bc..m} &=& \tilde{A}^{(-)a}\tilde{\nabla}_{a}(\tilde{A}^{(-)bc..m}X_{bc..m})\nonumber \\
\nonumber && -\frac{m\cosh\theta}{\sqrt{2}\sinh\theta}\tilde{A}^{(-)bc..m}X_{bc..m} \\
  &=& \left(\tilde{A}^{(-)a}\tilde{\nabla}_{a}-\frac{m\cosh\theta}{\sqrt{2}\sinh\theta}\right)\tilde{A}^{(-)bc..m}X_{bc..m}
\end{eqnarray*}
similarly we have that

\begin{eqnarray*}
\nonumber
\tilde{A}^{(+)a}\tilde{A}^{(+)bc..m}\tilde{\nabla}_{a}X_{bc..m} &=& \tilde{A}^{(+)a}\tilde{\nabla}_{a}(\tilde{A}^{(+)bc..m}X_{bc..m})\\
\nonumber && -\frac{m\cosh\theta}{\sqrt{2}\sinh\theta}\tilde{A}^{(+)bc..m}X_{bc..m} \\
  &=& \left(\tilde{A}^{(+)a}\tilde{\nabla}_{a}-\frac{m\cosh\theta}{\sqrt{2}\sinh\theta}\right)\tilde{A}^{(+)bc..m}X_{bc..m}
\end{eqnarray*}

Additionally, on some function $y$ we have

\begin{eqnarray}
\label{p3}
\partial_{+}y &=& \frac{1}{\sqrt{2}}(\partial_{\theta}+\frac{i}{\sinh\theta}\partial_{\phi})y = \tilde{A}^{(-)a}\tilde{\nabla}_{a}y \\
\label{p4}
\partial_{-}y &=& \frac{1}{\sqrt{2}}(\partial_{\theta}-\frac{i}{\sinh\theta}\partial_{\phi})y = \tilde{A}^{(+)a}\tilde{\nabla}_{a}y
\end{eqnarray}
Therefore we have the recursion relations:

\begin{eqnarray}
\tilde{A}^{(-)ab...m+1}\tilde{\nabla}_{ab...m+1} &=& {\mathcal D}_{+}(m)\tilde{A}^{(-)ab...m}\tilde{\nabla}_{ab...m} \\
\tilde{A}^{(+)ab...m+1}\tilde{\nabla}_{ab...m+1} &=& {\mathcal D}_{-}(m)\tilde{A}^{(+)ab...m}\tilde{\nabla}_{ab...m} 
\end{eqnarray}
where

\begin{eqnarray}
{\mathcal D}_{+}(m) &\equiv &  \frac{1}{\sqrt{2}}\left(\partial_{\theta}+\frac{i}{\sinh\theta}\partial_{\phi}-m\frac{\cosh\theta}{\sinh\theta}\right) \label{dpl}\\
{\mathcal D}_{-}(m) &\equiv &  \frac{1}{\sqrt{2}}\left(\partial_{\theta}-\frac{i}{\sinh\theta}\partial_{\phi}-m\frac{\cosh\theta}{\sinh\theta}\right) \label{dmi}
\end{eqnarray}
It follows then that

\begin{eqnarray}
\tilde{A}^{(-)ab..p}\tilde{\nabla}_{ab..p}Z^{\tilde{k}}_{m} &=& \prod_{q=0}^{p}{\mathcal D}_{+}(p)Z^{\tilde{k}}_{m}
\end{eqnarray}

From \cite{Vilenkin1} we can read off the following recursion relations:

\begin{eqnarray}
\left(\frac{d}{d\theta}+\left(\frac{a-b\cosh\theta}{\sinh\theta}\right)\right){\mathcal Q}^{\tilde{k}}_{ab} &=& (\tilde{k}-b){\mathcal Q}^{\tilde{k}}_{a,b+1} \\
\left(\frac{d}{d\theta}+\left(\frac{b\cosh\theta-a}{\sinh\theta}\right)\right){\mathcal Q}^{k}_{ab} &=& (\tilde{k}+b){\mathcal Q}^{\tilde{k}}_{a,b-1} 
\end{eqnarray}
or in terms of the matrix elements:

\begin{eqnarray}
\left(\frac{\partial}{\partial\theta}+\left(\frac{a-b\cosh\theta}{\sinh\theta}\right)\right)T^{(i\lambda-1/2)}_{ab} &=& (\tilde{k}-b)e^{i\Psi}T^{(i\lambda-1/2)}_{a,b+1} \\
\left(\frac{\partial}{\partial\theta}+\left(\frac{b\cosh\theta-a}{\sinh\theta}\right)\right)T^{(i\lambda-1/2)}_{ab} &=& (\tilde{k}+b)e^{-i\Psi}T^{(i\lambda-1/2)}_{a,b-1} 
\end{eqnarray}
therefore, recalling the definition of $T^{(i\lambda-1/2)}_{ab}$ in terms of constituent functions we have that

\begin{eqnarray}
\left(\frac{\partial}{\partial\theta}+\left(\frac{a-b\cosh\theta}{\sinh\theta}\right)\right)T^{\tilde{k}}_{ab} &=& \sqrt{2}{\mathcal D}_{+}T^{(i\lambda-1/2)}_{ab} \\
&=& \sqrt{2}(\tilde{k}-b)e^{i\Psi}T^{\tilde{k}}_{a,b+1} 
\end{eqnarray}
and therefore

\begin{eqnarray}
T^{\tilde{k}}_{ap} &=&\left(\frac{1}{\sqrt{2}}\right)^{-p} \prod_{q=0}^{p}\frac{1}{\tilde{k}+q}{\mathcal D}_{+}(q)T^{\tilde{k}}_{a0}  \\
                                    &=&\left(\frac{1}{\sqrt{2}}\right)^{p} \prod_{q=0}^{|p|}\frac{e^{-i\Psi}}{\tilde{k}+q}{\mathcal D}_{+}(q)Z^{\tilde{k}}_{-a}\\
                                    &=& \left(\frac{1}{\sqrt{2}}\right)^{p} e^{-ip\Psi}\left(\prod_{q=0}^{|p|}\frac{1}{\tilde{k}+q}\right)\tilde{A}^{(-)bc..|p|}\tilde{\nabla}_{bc..|p|}Z^{\tilde{k}}_{-a}
\end{eqnarray}
Then applying this equivalence to the addition relation for matrix elements, we have that:

\begin{eqnarray}
\nonumber
 T^{(i\lambda-1/2)}_{pp} &=&\left(\frac{1}{2}\right)^{-p} \left|\prod^{p}_{q=0}\frac{1}{\tilde{k}+q}\right|^{2}\sum_{e=-\infty}^{+\infty}\delta^{ef}
 \tilde{A}^{(-)bc..p}\tilde{\nabla}_{bc..p}Z^{\tilde{k}}_{-f}\tilde{A}^{(+)bc..p}\tilde{\nabla}_{bc..p}Z^{*\tilde{k}}_{-e}\\
     &=& 1
\end{eqnarray}

\subsection{Flat Case}

Finally we consider the flat case. The appropriate group here is the group inhomogeneous unitary group $IU(1)$, elements of which can be represented by the following matrix:

\[
\left( {\begin{array}{cc}
e^{i\alpha} & z \\
0&  1
 \end{array} } \right)
\]
where $0\leq \alpha <2\pi$, $z\in \mathbb{C}$.
The basis for each of these representations may be taken to be eigenvectors $e^{im\alpha}$ 
($-\infty< m < \infty$,$m\in \mathbf{Z}$, $0\leq \alpha < 2\pi$) and accompanying inner product $\left<e^{i\xi\alpha},e^{i\mu\alpha}\right>\equiv \frac{1}{2\pi}\int e^{-i\xi\alpha}e^{i\nu\alpha}d\alpha$. As before,
matrix elements $T^{\chi}_{ab}$ may be defined as follows:

\begin{eqnarray}
T^{(\chi)}(g)e^{-i\xi\theta}=\sum_{b=-\infty}^{+\infty}T_{\xi b}^{(\chi)}(g)e^{-ib\theta}
\end{eqnarray}
they are found to be \cite{Vilenkin1}:

\begin{eqnarray}
T_{mn}^{(ik)}(g) =i^{n-m} e^{-in\phi-i(n-m)\psi}J_{n-m}(kr)
\end{eqnarray}
where $J_{n-m}(kr)$ are Bessel functions of the first kind. 
We define the following functions:

\begin{eqnarray}
{\mathcal J}^{k}_{n}\equiv T^{ik}_{0,-n}= i^{-n}e^{in\phi}J_{-n}(kr) \label{calj}
\end{eqnarray}
These functions are indeed eigenfunctions of the Laplace-Beltrami operator on the Euclidean 2-plane, and satisfy

\begin{eqnarray}
\tilde{\nabla}^{2}{\mathcal J}^{(k)}_{n}(k r) &=& -k^{2} {\mathcal J}^{(k)}_{n}(k r) \\
-i\frac{\partial{\mathcal J}^{(k)}_{n}(k r)}{\partial\phi}&=& n{\mathcal J}^{(k)}_{n}(k r)
\end{eqnarray}
The Bessel functions satisfy the following recursion relations:

\begin{eqnarray}
\left(\frac{d}{dz}+\frac{\nu}{z}\right)J_{\nu}(z) &=& J_{\nu-1}(z) \\
\left(\frac{d}{dz}-\frac{\nu}{z}\right)J_{\nu}(z) &=& -J_{\nu+1}(z) 
\end{eqnarray}
or in terms of matrix elements

\begin{eqnarray}
\left(\frac{d}{d\theta}+\frac{(n-m)}{\theta}\right)T^{(ik)}_{mn} &=&ik e^{i\psi}T^{(ik)}_{m+1,n}\\
\left(\frac{d}{d\theta}-\frac{(n-m)}{\theta}\right)T^{(ik)}_{mn}&=&i k e^{-i\psi}T^{(ik)}_{m-1,n}
\end{eqnarray}
Again it is useful to define vectors $\tilde{A}^{(\pm)a} = (1/\sqrt{2})((\textbf{e}_{\theta})^{a}\mp i(\textbf{e}_{\phi})^{a})$ where $\textbf{e}_{\theta}=\partial_{\theta}$ and $\textbf{e}_{\phi}=(1/\theta)\partial_{\phi}$  i.e.
they are unit vectors with respect to the metric $ds^{2}=d\theta^{2}+\theta^{2}d\phi^{2}$. Explicitly we have that 

\begin{eqnarray}
\tilde{\nabla}_{b}\tilde{A}^{(-)a}= \frac{1}{\sqrt{2}\theta}(\textbf{e}_{\phi})_{b}(-i(\textbf{e}_{\theta})^{a}+(\textbf{e}_{\phi})^{a})
\end{eqnarray}
Following the calculation of the positively curved case, we find that for the tensor $X_{bc..m}\equiv \tilde{\nabla}_{bc..m}f$

\begin{eqnarray*}
\nonumber
\tilde{A}^{(-)a}\tilde{A}^{(-)bc..m}\tilde{\nabla}_{a}X_{bc..m} &=& \tilde{A}^{(-)a}\tilde{\nabla}_{a}(\tilde{A}^{(-)bc..m}X_{bc..m})\\
&& -\frac{m}{\sqrt{2}\theta}\tilde{A}^{(-)bc..m}X_{bc..m} \\
  &=& \left(\tilde{A}^{(-)a}\tilde{\nabla}_{a}-\frac{m}{\sqrt{2}\theta}\right)\tilde{A}^{(-)bc..m}X_{bc..m}
\end{eqnarray*}
similarly we have that

\begin{eqnarray*}
\nonumber
\tilde{A}^{(+)a}\tilde{A}^{(+)bc..m}\tilde{\nabla}_{a}X_{bc..m} &=& \tilde{A}^{(+)a}\tilde{\nabla}_{a}(\tilde{A}^{(+)bc..m}X_{bc..m})\\
&& -\frac{m}{\sqrt{2}\theta}\tilde{A}^{(+)bc..m}X_{bc..m} \\
  &=& \left(\tilde{A}^{(+)a}\tilde{\nabla}_{a}-\frac{m}{\sqrt{2}\theta}\right)\tilde{A}^{(+)bc..m}X_{bc..m}
\end{eqnarray*}
Additionally, on some function $y$ we have

\begin{eqnarray}
\label{p3}
\partial_{+}y &=& \frac{1}{\sqrt{2}}(\partial_{\theta}+\frac{i}{\theta}\partial_{\phi})y = \tilde{A}^{(-)a}\tilde{\nabla}_{a}y \\
\label{p4}
\partial_{-}y &=& \frac{1}{\sqrt{2}}(\partial_{\theta}-\frac{i}{\theta}\partial_{\phi})y = \tilde{A}^{(+)a}\tilde{\nabla}_{a}y
\end{eqnarray}
Therefore we have the recursion relations:

\begin{eqnarray}
\tilde{A}^{(-)ab...m+1}\tilde{\nabla}_{ab...m+1} &=& {\mathcal D}_{+}(m)\tilde{A}^{(-)ab...m}\tilde{\nabla}_{ab...m} \\
\tilde{A}^{(+)ab...m+1}\tilde{\nabla}_{ab...m+1} &=& {\mathcal D}_{-}(m)\tilde{A}^{(+)ab...m}\tilde{\nabla}_{ab...m} 
\end{eqnarray}
where

\begin{eqnarray}
{\mathcal D}_{+}(m) &\equiv &  \frac{1}{\sqrt{2}}\left(\partial_{\theta}+\frac{i}{\theta}\partial_{\phi}-\frac{m}{\theta}\right) \\
{\mathcal D}_{-}(m) &\equiv &  \frac{1}{\sqrt{2}}\left(\partial_{\theta}-\frac{i}{\theta}\partial_{\phi}-\frac{m}{\theta}\right)
\end{eqnarray}
It follows then that

\begin{eqnarray}
\tilde{A}^{(-)ab..p}\tilde{\nabla}_{ab..p}{\mathcal E}^{k}_{m} &=& \prod_{q=0}^{p}{\mathcal D}_{+}(p){\mathcal E}^{k}_{m}
\end{eqnarray}
Therefore we have that 

\begin{eqnarray}
T^{(ik)}_{ap} &=& \left(\frac{1}{ik}\right)^{p}\left(\frac{1}{\sqrt{2}}\right)^{-p}i^{p-a}e^{-ip\psi}\tilde{A}^{bc..|p|}\tilde{\nabla}_{bc..|p|}{\mathcal J}^{(k)}_{-a}
\end{eqnarray}
Then applying this equivalence to the addition relation for matrix elements, we have that:

\begin{eqnarray}
\nonumber
 T^{(ik)}_{pp} &=&\left(\frac{1}{2}\right)^{-p}\left(\frac{1}{k^{2}}\right)^{p} \sum_{e=-\infty}^{+\infty}\delta^{ef}
 \tilde{A}^{(-)bc..p}\tilde{\nabla}_{bc..p}{\mathcal J}^{(k)}_{-f}\tilde{A}^{(+)bc..p}\tilde{\nabla}_{bc..p}{\mathcal J}^{*(k)}_{-e}\\
     &=& 1
\end{eqnarray}
In the case $p=0$ one recovers the familiar addition theorem for Bessel functions of the first kind.

\section{The calculation of polar correlations}
\label{calpolar}

\begin{eqnarray}
\nonumber\left<a_{l'm'}a_{lm}\right> &=& \int d\Omega \int d\Omega' \sum_{N',A',N,A}\sum_{w',w,k',k,p,p'}Y^{*l}_{m}(\Omega)Y^{l'}_{m'}(\Omega')\\
\nonumber   && {\mathcal B}^{(P)}_{NA}(x^{i},k,w,\Omega)({\mathcal B}^{(P)}_{NA})^{*}(x^{i},k',w',\Omega') F^{(P)}_{N'A'NA}(k,k',w,w',t)\\
        &=&\frac{1}{2\pi} \int dw \int d\Omega \int d\Omega' \sum_{N',A',N,A}\sum_{k,p}Y^{*m}_{l}(\Omega)Y^{m'}_{l'}(\Omega') \nonumber\\
  \nonumber        && {\mathcal B}^{(P)}_{NA}(z=0,k,w,\Omega)({\mathcal B}^{(P)}_{NA})^{*}(z=0,k,w,\Omega') \cdot \\
          && F^{(P)}_{N'A'NA}(k,w,t) \label{correla}\\
F^{(P)}_{N'A'NA}(k,w,t) &\equiv & c^{(P)}_{NA}c^{*(P)}_{N'A'}\frac{\Theta^{(P)}_{NA}\Theta^{*(P)}_{N'A'}}{|\delta(k,w)|^{2}}{\mathcal P}(k,w)
\end{eqnarray}
where recall that here  $k$ and $p$ are labels on the harmonic functions ${\mathcal E}^{k}_{p}$ and $\sum_{k,w}$ and
$\sum_{k}$ denote the appropriate sums/integrals over labels $k$ and $w$. If we work in the preferred frame, then a particularly convenient choice is $d\Omega=-d\alpha d\chi$ where $\chi$ is an angular coordinate in the tangent plane at s
some point in the co-moving 2 surface.

We now consider the integral over $\chi$. The integrand will contain terms $e^{-im\chi}$ from the spherical harmonic $Y^{*l}_{m}$ as well as terms
$C^{ab..M}=W^{ab..M}_{\phantom{ab..M}cd..M}\hat{p}^{cd..M}$ contained within the mode functions ${\mathcal B}_{NA}$. It will be helpful to express 
$e^{im\chi}$ in terms of tensors formed from  $\hat{p}^{c}$. We choose a local orthonormal basis $(\textbf{e}_{1},\textbf{e}_{2})$ in the co-moving
2-surface, such that

\begin{eqnarray}
\hat{p}^{c} &=&  \cos\chi (\textbf{e}_{1})^{c}+\sin\chi (\textbf{e}_{2})^{c} \\
  &=& \frac{1}{2}(e^{i\chi}+e^{-i\chi})(\textbf{e}_{1})^{c}+\frac{1}{2i}(e^{i\chi}-e^{-i\chi})(\textbf{e}_{2})^{c}
\end{eqnarray}
therefore we have that

\begin{eqnarray}
e^{i\chi} &=& \hat{p}^{c}\left((\textbf{e}_{1})_{c}+i(\textbf{e}_{2})_{c}\right) \\
e^{-i\chi} &=& \hat{p}^{c}\left((\textbf{e}_{1})_{c}-i(\textbf{e}_{2})_{c}\right)
\end{eqnarray}
We claim that without loss of generality we can choose the coordinate $\chi$ such that 

\begin{eqnarray}
\left((\textbf{e}_{1})_{c}+i(\textbf{e}_{2})_{c}\right)=\sqrt{2}(\textbf{A}^{-})_{c} \\
\left((\textbf{e}_{1})_{c}-i(\textbf{e}_{2})_{c}\right)=\sqrt{2}(\textbf{A}^{+})_{c} 
\end{eqnarray}
and so for a quantity $e^{-im\chi}$ we have:

\begin{eqnarray}
e^{i|m|\chi} &=& 2^{\frac{|m|}{2}}(\textbf{A}^{-})_{ab..|m|}\hat{p}^{ab..|m|}\\
e^{-i|m|\chi} &=& 2^{\frac{|m|}{2}}(\textbf{A}^{+})_{ab..|m|}\hat{p}^{ab..|m|}
\end{eqnarray}
The integral over $\chi$ is as follows:

\begin{eqnarray}
\int d\chi e^{-im\chi}C^{ab..|N|}
\end{eqnarray}
For $m>0$ we have:

\begin{eqnarray}
\nonumber
\int d\chi e^{-im\chi}C^{ab..|N|}=2^{\frac{|m|}{2}}(\textbf{A}^{+})_{ef..|m|}W^{ab..|N|}_{\phantom{ab..|N|}cd..|N|}\int d\chi \hat{p}^{efcd..|m|+|N|} 
\end{eqnarray}
whilst for $m<0$ we have:

\begin{eqnarray}
\nonumber
\int d\chi e^{-im\chi}C^{ab..|N|}=2^{\frac{|m|}{2}}(\textbf{A}^{-})_{ef..|m|}W^{ab..|N|}_{\phantom{ab..|N|}cd..|N|}\int d\chi \hat{p}^{efcd..|m|+|N|} 
\end{eqnarray}
The above integral may be evaluated using standard methods, yielding:

\begin{eqnarray}
\nonumber
\int_{0}^{2\pi} \hat{p}^{ab..|m|+|N|}d\chi = \frac{(|m|+|N|)!}{2^{(|m|+|N|)}(((|m|+|M|)/2)!)^{2}}2\pi\gamma^{(ab..(|m|+|N|))}
\end{eqnarray}
which is nonvanishing only for $(|m|+|N|)$ being an even number. In general then we will be left with terms like:

\begin{eqnarray}
\label{gam2}
&&\gamma^{(cdef..(|m|+|N|))}W^{ab..|N|}_{\phantom{ab..|N|}cd..|N|}(\textbf{A}^{\pm})_{ef..|m|}\nabla_{ab..|N|}{\mathcal E}^{k}_{p} \\
&=& \gamma^{(cdef..(|m|+|N|))}(\textbf{A}^{\pm})_{ef..|m|}(2^{|N|-1}\nabla_{++..|N|}{\mathcal E}^{k}_{p}(\textbf{A}^{+})_{cd..|N|} \nonumber \\
&& +2^{|N|-1}\nabla_{--..|N|}{\mathcal E}^{k}_{p}(\textbf{A}^{-})_{cd..|N|})
\end{eqnarray}
Again we note that by the definition of the tensor $W^{ab...|N|}_{\phantom{ab...|N|}cd...|N|}$, it must be traceless with respect
to contraction with $\gamma^{ef}$ across any two lower indices. Therefore, (\ref{gam2}) is only nonvanishing
if $\gamma^{(ab..(|m|+|N|))}$ is contracted with one index of each component $\gamma^{ef}$ acting on one index of the $W$ tensor and one index of the $\textbf{A}^{\pm}$ tensor. This implies that $|m|=|N|$. By a counting argument, there're $2^{Q}Q!(Q)!$ permutations of
$\gamma^{ab..2Q}$ that fulfil this condition. Therefore:

\begin{eqnarray}
\label{gam3}
&&\gamma^{(cdef..(|m|+|N|))}W^{ab..|N|}_{\phantom{ab..|N|}cd..|N|}(\textbf{A}^{\pm})_{ef..|m|}\nabla_{ab..|N|}{\mathcal E}^{k}_{p}\nonumber \\
&=& 2^{|m|}\frac{|m|!|m|!}{(2m)!}2^{|m|-1}\nabla_{\mp\mp..|m|}{\mathcal E}^{k}_{p}
\end{eqnarray}
In total then we have that:

\begin{eqnarray}
\nabla_{ab..|N|}{\mathcal E}^{k}_{p} \int d\chi e^{-im\chi}C^{ab..|N|} = 2^{\frac{|m|}{2}}\pi\nabla_{\mp\mp..|m|}{\mathcal E}^{k}_{p}\delta_{|m||N|}
\end{eqnarray}
where $-$ and $+$ correspond to the cases $m>0$ and $m<0$ respectively. The case $m=0$ is trivially
evaluated, yielding:

\begin{eqnarray}
{\mathcal E}^{k}_{p}\int C^{ab..|N|} d\chi = 2\pi {\mathcal E}^{k}_{p}\delta_{0|N|}
\end{eqnarray}
We may immediately then evaluate the similar integral (again assuming $m\neq 0$):

\begin{eqnarray}
\nabla_{ab..|N'|}{\mathcal E}^{*k}_{p} \int d\chi e^{im'\chi}C^{ab..|N'|} = 2^{\frac{|m'|}{2}}\pi\nabla_{\pm\pm..|m'|}{\mathcal E}^{*k}_{p}\delta_{|m'||N'|}
\end{eqnarray}
We next consider the sum over $p$:

\begin{eqnarray}
\label{sum3}
\sum_{p} 2^{\frac{|m'|}{2}}\pi\nabla_{\pm\pm..|m'|}{\mathcal E}^{*k}_{p}\delta_{|m'||N'|}2^{\frac{|m|}{2}}\pi\nabla_{\mp\mp..|m|}{\mathcal E}^{k}_{p}\delta_{|m||N|}
\end{eqnarray}
We first make the physical assumption that $m=m'$ (and hence $N=N'$). Following the arguments in Section \ref{secstatanis}, this is a statement about the 
statistical isotropy of correlations in the co-moving 2-surface. This assumption simplifies 
(\ref{sum3}) to:

\begin{eqnarray}
\label{sum4}
\sum_{p} 2^{|m|}\pi^{2}\nabla_{\pm\pm..|m|}{\mathcal E}^{*k}_{p}\nabla_{\mp\mp..|m|}{\mathcal E}^{k}_{p}
\end{eqnarray}
We now may make use of the results of \ref{appa}, relating the derivative terms in (\ref{sum3}) to matrix elements for group representations, where it was found that:

\begin{eqnarray*}
\nonumber
\frac{2k+1}{4\pi}T^{l}_{|m||m|}(0)&=& \frac{2k+1}{4\pi}\cdot 1\\
 &=& \frac{2^{|m|}|K|^{-|m|}}{\prod_{q=0}^{|m|}(k+q)(k-q+1)}\sum_{p=-k}^{p=k}\nabla_{++..|m|}Y^{k}_{p}\nabla_{--..|m|}Y^{*k}_{p}
\end{eqnarray*}
It follows then that in the case $K>0$:

\begin{eqnarray}
\nonumber
\sum_{p=-k}^{l} 2^{|m|}\pi^{2}\nabla_{\pm\pm..|m|}{\mathcal E}^{*k}_{p}\nabla_{\mp\mp..|m|}{\mathcal E}^{k}_{p}=\frac{(2k+1)\pi|K|^{|m|}}{4}\prod_{q=0}^{|m|}(l+q)(l-q+1)
\end{eqnarray}
Again we may directly use the results of \ref{appa} in the manner above, to find:

\begin{eqnarray}
\nonumber
\sum_{p=-\infty}^{+\infty} 2^{|m|}\pi^{2}\nabla_{\pm\pm..|m|}{\mathcal E}^{*k}_{p}\nabla_{\mp\mp..|m|}{\mathcal E}^{k}_{p}=\pi^{2}|K|^{|m|}\prod_{q=0}^{|m|}(k+q)(k^{*}+q)
\end{eqnarray}
Additionally we must integrate over the $\alpha$ and $\alpha'$. In the preferred frame there're two functions with dependence upon $\alpha$:  the associated
Legendre function $P^{l}_{m}(\alpha)$ appearing in the spherical harmonic $Y^{*l}_{m}(\alpha,\phi)$ and the associated Legendre function $P^{A}_{N=|m|}(\alpha)$ 
appearing in ${\mathcal B}^{(P)}_{NA}$. An identical combination appears for the dependence upon $\alpha'$. The form of (\ref{correla}) dictates that we must integrate over the product of
these functions, which is a standard integral:

\begin{eqnarray}
\int P^{l}_{m}P^{A}_{m} d\alpha&=& \frac{2(l+m)!}{(2l+1)(l-m)!}\delta_{Al}\\
\int P^{l'}_{m}P^{A'}_{m} d\alpha' &=&  \frac{2(l'+m)!}{(2l'+1)(l'-m)!}\delta_{A'l'}
\end{eqnarray}
for $m>0$ and

\begin{eqnarray}
\int P^{l}_{m}P^{A}_{|m|} d\alpha&=&(-1)^{|m|}\frac{(l+|m|)!}{(l-|m|)!} \int P^{l}_{m}P^{A}_{|m|} d\alpha  \\
  &=& \frac{2}{(2l+1)} (-1)^{|m|}\left(\frac{(l+|m|)!}{(l-|m|)!}\right)^{2}\delta_{Al} \\
\int P^{l'}_{m}P^{A'}_{|m|} d\alpha' &=& (-1)^{|m|}\frac{(l'+|m|)!}{(l'-|m|)!} \int P^{l'}_{m}P^{A'}_{|m|} d\alpha \\
  &=& \frac{2}{(2l'+1)} (-1)^{|m|}\left(\frac{(l'+|m|)!}{(l'-|m|)!}\right)^{2}\delta_{A'l'} 
\end{eqnarray}
for $m<0$. Finally, assembling the previous results, we have for the closed case that :

\begin{eqnarray}
\nonumber<a_{l'm'}a_{lm}>&=&\frac{1}{2\pi}K\int dw \sum_{L}\delta_{mm'} \frac{2(l'+m)!}{(2l'+1)(l'-m)!}\frac{2(l+m)!}{(2l+1)(l-m)!}\\
 \nonumber     &&\cdot \frac{(2L+1)\pi}{4}\prod_{q=0}^{|m|}\kappa_{q}^{2}F^{(P)}_{ll'}(k,w,t)e(l,l',m) \\
 \nonumber        &=&\delta_{mm'} \frac{K}{2}\int dw\sum_{L}  (2L+1)\frac{\Theta^{(P)}_{ml}\Theta^{*(P)}_{ml'}}{|\delta(k,w)|^{2}}{\mathcal P}(k,w)e(l,l',m) \\
 \label{colpola}
\end{eqnarray}
where we have used the functional form (\ref{cchoice}) for the function $c^{(P)}_{ml}$. Recall that $k^{2}=KL(L+1)$ in the closed case. The function $e(l,l',m)$ reflects the normalization of the spherical harmonics and allows for the particular case $m=0$. It is defined as follows:

\begin{eqnarray}
e(l,l',m>0) &=& \sqrt{\frac{(2l+1)(2l'+1)(l-m)!(l'-m)!}{16\pi^{2}(l+m)!(l'+m)!}}\label{e1} \\
e(l,l',m=0) &=& 4\sqrt{\frac{(2l+1)(2l'+1)}{16\pi^{2}}}\label{e2}\\
e(l,l',m<0) &=&\sqrt{\frac{(2l+1)(2l'+1)}{16\pi^{2}}\left(\frac{(l-m)!(l'-m)!}{(l+m)!(l'+m)!}\right)^{3}} \label{e3}
\end{eqnarray}
In the open case we have that (where $k^{2}=|K|(\lambda^{2}+\frac{1}{4}$)):

\begin{eqnarray}
<a_{l'm'}a_{lm}>&=&\frac{1}{2\pi}|K|\int dw \int d\lambda \lambda \tanh\left(\pi \lambda\right) \delta_{mm'} \frac{2(l'+m)!}{(2l'+1)(l'-m)!}\frac{2(l+m)!}{(2l+1)(l-m)!}\nonumber\\
 \nonumber     &&\cdot \pi^{2}\prod_{q=0}^{|m|}\kappa_{q}^{2}F^{(P)}_{ll'}(k,w,t)e(l,l',m)\\
 \nonumber     &=& \delta_{mm'}2\pi |K|\int dw \int d\lambda \lambda \tanh\left(\pi \lambda\right)\frac{\Theta^{(P)}_{ml}\Theta^{*(P)}_{ml'}}{|\delta(k,w)|^{2}}{\mathcal P}(k,w)e(l,l',m)\\
\end{eqnarray}
Finally in the flat case we have that:

\begin{eqnarray}
<a_{l'm'}a_{lm}>&=&   \delta_{mm'}2\pi\int dw \int k dk\frac{\Theta^{(P)}_{ml}\Theta^{*(P)}_{ml'}}{|\delta(k,w)|^{2}}{\mathcal P}(k,w)e(l,l',m)\nonumber \\
\end{eqnarray}

\section{Gauge transformations}
\label{gaugetran}

We assume that we are working with theories derived from generally covariant actions. Consequently, we may use the four dimensional
diffeomorphism invariance to impose restrictions on the perturbational quantities in order to simplify calculations. 
Consider an infinitesimal vector field $\xi^{\mu}$, which will be used to generate diffeomorphisms. In this sense infinitesimal
means that all components of $\xi^{\mu}$ are to be regarded as first order in smallness. We may define its components as follows:

\begin{eqnarray}
\xi_{\mu}dx^{\mu} = a^{2}\sum_{k,m}(\xi_{t}{\mathcal E}dt +\xi_{z}{\mathcal E}dz+\xi_{P}\nabla_{a}{\mathcal E}dx^{a}+\xi_{A}\bar{\nabla}_{a}{\mathcal E} dx^{a})
\end{eqnarray}
Under an infinitesimal diffeomorphism the metric changes at fixed coordinate location changes as follows:

\begin{eqnarray}
g_{\mu\nu} \rightarrow g_{\mu\nu} + 2 D_{(\mu}\xi_{\nu)}
\end{eqnarray}
where $D_{\mu}$ is the derivate operator compatible with the background spacetime metric. We now decompose the tensor $D_{(\mu}\xi_{\nu)}$ into polar and axial parts which are, respectively

\begin{eqnarray}
\label{poldif}
\nonumber(D_{(\mu}\xi_{\nu)})^{(P)}dx^{\mu}dx^{\nu} &=& a^{2}\sum_{k,m}( ({\mathcal H}\xi_{t}+\xi'_{t}){\mathcal E} dt^{2}+\frac{1}{2}(\xi'_{P}+\xi_{t})\nabla_{b}{\mathcal E}dtdx^{b}\\
  \nonumber                &&  + \frac{1}{2}(\xi'_{z}+\partial_{z}\xi_{t}){\mathcal E}dtdz+\frac{1}{2}(\partial_{z}\xi_{P}+\xi_{z})\nabla_{b}{\mathcal E}dzdx^{b}\\
  \nonumber                &&  +\xi_{P}\nabla_{a}\nabla_{b}{\mathcal E} dx^{a}dx^{b} - {\mathcal H}\xi_{t} \gamma_{ab}{\mathcal E}dx^{a}dx^{b}\\
                  && +(\partial_{z}\xi_{z}-{\mathcal H}\xi_{t}){\mathcal E}dzdz)
\end{eqnarray}

\begin{eqnarray}
\label{axdif}
\nonumber(D_{(\mu}\xi_{\nu)})^{(A)}dx^{\mu}dx^{\nu} &=&a^{2}\sum_{k,m} (\frac{1}{2}\partial_{z}\xi_{A} \bar{\nabla}_{b} {\mathcal E}dzdx^{b} +\frac{1}{2}\xi'_{A}\bar{\nabla}_{b}{\mathcal E}dtdx^{b} \\
  && +\xi_{A}\nabla_{(a}\bar{\nabla}_{b)}{\mathcal E}dx^{a}dx^{b})
\end{eqnarray}
Recall that the perturbed metric has the following polar and axial components:

\begin{eqnarray}
\delta g^{(P)}_{\mu\nu}dx^{\mu}dx^{\nu}&=& a^{2}\sum_{k,m}(V(z,t){\mathcal E}dt^{2}+E(z,t){\mathcal E}dzdt+F(z,t){\mathcal E}dz^{2}\nonumber\\
                                        && +B(z,t)\nabla_{d}{\mathcal E} dzdx^{d}+ C(z,t)\nabla_{d}{\mathcal E} dtdx^{d}\nonumber \\
                                        && +U(z,t)\gamma_{ab}{\mathcal E}dx^{a}dx^{b}+X(z,t)\nabla_{a}\nabla_{b}{\mathcal E} dx^{a}dx^{b}) \label{polmetra}
\end{eqnarray}
and 

\begin{eqnarray}
\delta g^{(A)}_{\mu\nu}dx^{\mu}dx^{\nu}  &=&  a^{2}\sum_{k,m}(R(z,t) \bar{\nabla}_{a}{\mathcal E} dtdx^{a}+S(z,t)\bar{\nabla}_{a}{\mathcal E}dzdx^{a}\nonumber\\
&&+2Q(z,t)
\nabla_{(a}\bar{\nabla}_{b)}{\mathcal E}dx^{a}dx^{b}) \label{axmetra}
\end{eqnarray}
It is clear from the form of (\ref{poldif}) and (\ref{axdif}) that none of the individual components in (\ref{polmetra}) and (\ref{axmetra}) are unalterable
under a gauge transformation. However, we may construct combinations of perturbations that are gauge invariant. Combinations in the polar case are:

\begin{eqnarray}
\Xi^{(P)}_{1} &=& iw B - \frac{F}{2}+\frac{U}{2}+\frac{w^{2}X}{2}\\
\Xi^{(P)}_{2} &=& \frac{V}{2}+\frac{U}{2}-C'+\frac{X''}{2}\\
\Xi^{(P)}_{3} &=& {\mathcal H}E-{\mathcal H}B'+iw{\mathcal H}C+iwU \\
\Xi^{(P)}_{4} &=& B'-\frac{iwX'}{2}-E-\frac{iwU}{2{\mathcal H}}
\end{eqnarray}
Combinations in the axial case are:

\begin{eqnarray}
\Xi^{(A)}_{1} &=& iwQ- S \\
 \Xi^{(A)}_{2} &=& iwR-S' 
 \end{eqnarray}

\section{Comparison to FRW scalar-tensor-vector decomposition in flat limit}
\label{compfrw}

The shearless class of spacetimes described by (\ref{back}) allow only spatially flat FRW spacetimes as a limiting ($K\rightarrow 0$) case. In a general gauge we may write a perturbed spatially flat FRW spacetime as follows:

\begin{eqnarray}
ds^{2} &=& a^{2}(-(1+2\Psi)dt^{2}+(\nabla_{i}\alpha+\alpha_{i})dtdx^{i}+(\delta_{ij}+\delta g_{ij})dx^{i}dx^{j}) \label{pfrw} \\
\delta g_{ij} &\equiv & -2\Phi \delta_{ij} +\nabla_{i}\nabla_{j}\beta+ \nabla_{(i}\beta_{j)}+ \tilde{h}_{ij} 
\end{eqnarray}
where $\nabla_{i}\alpha^{i}= \nabla_{i}\beta^{i}= \tilde{h}^{i}_{\phantom{i}i}= \nabla_{i}\tilde{h}^{i}_{\phantom{j}j}=0$. Indices 
$i$ and $j$ are three-dimensional, indices are raised and lowered with the background co-moving spatial metric $\delta_{ij}$, and 
$\nabla_{i}=\partial_{i}$ is the metric compatible derivative operator associated with $\delta_{ij}$. The perturbations
$\Psi,\Phi,\alpha,\beta$ are termed scalar perturbations, the divergenceless fields $\beta^{i}$ and $\alpha^{i}$ are 
termed vector modes each with two polarizations, whilst the field $\tilde{h}_{ij}$ carries two polarizations.

We now write $\alpha_{i}$ and $\beta_{i}$ in terms of the 2+1 polar and axial decomposition:

\begin{eqnarray}
\alpha_{z} &=& \sum_{k,m}\alpha_{z}{\mathcal E}  \\
\alpha_{a} &=& \sum_{k,m}( \zeta \nabla_{a}{\mathcal E}+\bar{\zeta}\bar{\nabla}_{a}{\mathcal E})\\
\beta_{z} &=&  \sum_{k,m}\beta_{z}{\mathcal E}  \\
\beta_{a} &=& \sum_{k,m}( \eta \nabla_{a}{\mathcal E}+\bar{\eta}\bar{\nabla}_{a}{\mathcal E})
\end{eqnarray}
where recall that in the flat case ${\mathcal E}$ is a Bessel function $J_{m}(kr)$. The vanishing divergence of $\alpha_{i}$ and $\beta_{i}$ then yields the following conditions:

\begin{eqnarray}
0 &=& \nabla_{z}\alpha_{z}-k^{2}\zeta \\
0 &=& \nabla_{z}\beta_{z}-k^{2}\eta
\end{eqnarray}
Similarly we can decompose the field $\tilde{h}_{ij}$ as follows:

\begin{eqnarray}
\tilde{h}_{ij}dx^{i}dx^{j} &=& \left(\sum_{k,m}h_{zz}{\mathcal E}\right) dzdz+\tilde{h}_{za}dzdx^{a}+\tilde{h}_{ab}dx^{a}dx^{b} \\
\tilde{h}_{za} &=& \sum_{k,m}( h^{(V)} \nabla_{a}{\mathcal E}+\bar{h}^{(V)}\bar{\nabla}_{a}{\mathcal E})\\
\tilde{h}_{ab} &=& \sum_{k,m}(h^{(T)}\gamma_{ab}+h^{(D)}\nabla_{a}\nabla_{b}{\mathcal E}+\bar{h}^{(D)}\nabla_{(a}\bar{\nabla}_{b)}{\mathcal E})
\end{eqnarray}
The condition $\tilde{h}^{i}_{\phantom{i}i}=0$ then implies that:

\begin{eqnarray}
0=h_{zz}+2h^{(T)}- k^{2} h^{(D)}
\end{eqnarray}
whilst the transverse condition $\partial_{i}\tilde{h}^{i}_{\phantom{i}j}=0$ implies that:

\begin{eqnarray}
0 &=& \partial_{z}h^{z}_{\phantom{z}z}-k^{2} h^{(V)} \\
0 &=& (\partial_{z}h^{(V)}-k^{2}h^{(D)})\nabla_{a}{\mathcal E} \\
0 &=& (\partial_{z}\bar{h}^{(V)}-\frac{k^{2}}{2}\bar{h}^{(D)})\bar{\nabla}_{a}{\mathcal E}
\end{eqnarray}
We would like to see how the polar and axial perturbations are related to tensor, vector, and scalar perturbations in the limit of spatial flatness. Comparison
between (\ref{pfrw}) and  (\ref{polmetra}) and (\ref{axmetra})  yields, in terms of Fourier modes:

\begin{eqnarray}
V &=&  -2\Psi \\
E&=& iw\alpha+\alpha_{z} \\
F  &=&  -2\Phi-w^{2}\beta+iw\beta_{z}+h_{zz} \\
B &=& iw\beta+\frac{1}{2}(iw\eta+\beta_{z})+h^{(V)} \\
C &=& \alpha+\zeta \\
U &=& h^{(T)}-2\Phi\\
X &=& \beta+ \eta+h^{(D)}
\end{eqnarray}
and

\begin{eqnarray}
R &=& \bar{\zeta} \\
S &=& \frac{1}{2}iw\bar{\eta}+\bar{h}^{(V)}\\
2Q &=& \bar{\eta}+ \bar{h}^{(D)}
\end{eqnarray}
We see then that in the flat limit, axial components consist entirely of tensor and vector perturbations whilst polar components are themselves sums of scalar, vector, and tensor perturbations. Consider the gauge invariant combinations of the previous section. 
We see that the following two are composed entirely of tensor perturbations in the flat limit:

\begin{eqnarray}
\Xi^{(P)}_{1} &=& iw h^{(V)} - \frac{h_{zz}}{2}+\frac{h^{(T)}}{2}+w^{2}\frac{h^{(D)}}{2}
\end{eqnarray}
\begin{eqnarray}
\Xi^{(A)}_{1} &=& iw\frac{\bar{h}^{(D)}}{2}-\bar{h}^{(V)}
\end{eqnarray}

\section{Table of selected notation}
\label{tableof}

\begin{tabular}{|c|r|}
	\hline
Symbol & Definition \\
	\hline
$\gamma_{ab}$ & Co-moving 2-surface background metric \\
$\epsilon_{ab}$ &  Co-moving 2-surface background volume form \\
$d^{2}S$ & Background volume element on the co-moving 2-surface \\
$R_{abf}^{\phantom{abf}d}$ & Background Riemann curvature tensor on the co-moving 2-surface \\
$h_{ij}$ & Background 3-dimensional spatial metric \\
$h^{(C)}_{ij}$ & Co-moving background 3-dimensional spatial metric \\
$|K|$ &  Curvature scalar of the co-moving 2-surface background metric \\
$t$  & Conformal time \\
$z$ & Co-moving coordinate in the z-direction \\
$x^{a}$ & Comoving coordinate on the co-moving 2-surface\\
$a(t)$ & The background expansion rate of the spatial surface orthogonal to $(\partial_{z})^{\mu}$\\
$b(t)$ & Background expansion rate along the z-direction \\
$D_{\mu}$ & Derivative operator compatible with background spacetime metric \\
$\nabla_{a}$   & Derivate operator compatible with co-moving 2-surface background metric  \\
$\nabla^{2}$   & Laplace-Beltrami on the background co-moving 2-surface   \\
$\delta g^{(P)}_{\mu\nu}$ & Component of polar perturbation to the spacetime metric \\
$\delta g^{(A)}_{\mu\nu}$ & Component of axial perturbation to the spacetime metric \\
$\delta G^{(P)\mu}_{\phantom{(P)\mu}\nu}$ & Component of the polar perturbation to the Einstein tensor \\
$\delta G^{(A)\mu}_{\phantom{(A)\mu}\nu}$ & Component of the polar perturbation to the Einstein tensor \\
${\mathcal E}^{k}_{m}$ &  Eigenfunction of $\nabla^{2}$ corresponding to eigenvalue $-kk^{*}$ for generic curvature\\
$w$  & Wavenumber in the z-direction in the shearless case \\
$\bar{\nabla}_{a}$  & Defined as $\epsilon_{a}^{\phantom{a}b}\nabla_{b}$ \\
$\phi(z,x^{a},t)$ & Scalar field supporting the shear-free anisotropic curvature \\
$\xi^{\mu}$ & Infinitesimal diffeomorphism generating vector field \\
$f^{\Psi}$ & Distribution function for species of massless particles $\Psi$ \\
$f^{T}$ & Distribution function for species of massive particles  \\
$P^{\mu}$ & Four momentum vector of particle (massless or massive) \\
$p^{2}$ & The norm-squared of the particle spatial momentum \\
$\alpha$ & The projection of the co-moving spatial momentum along $(\partial_{z})^{\mu}$ \\
$\hat{p}^{a}$ & Components of momenta in the 2-surface with normalization $\gamma_{ab}\hat{p}^{a}\hat{p}^{b}=1$ \\
$T(t)$ & The background photon temperature \\
$\Theta^{(P)}(x^{i},\hat{P}^{i},t)$ & The dimensionless polar temperature perturbation\\
$\Theta^{(A)}(x^{i},\hat{P}^{i},t)$ & The dimensionless axial temperature perturbation\\
${\mathcal B}^{(P)}_{NA} $  &  Basis function for polar mode-moment expansion  \\
${\mathcal B}^{(A)}_{NA} $  &  Basis function for axial mode-moment expansion  \\
$\Theta^{(P)}_{NA}(k,w,t)$   & Polar temperature perturbation moment  \\
$\Theta^{(A)}_{NA}(k,w,t)$ & Axial temperature perturbation moment \\ 
$c_{NA}^{(P)}$ & Normalization function for polar temperature moment \\
$c_{NA}^{(A)}$ & Normalization function for axial temperature moment \\
	\hline
\end{tabular}

\begin{tabular}{|c|r|}
	\hline
Symbol & Definition \\
	\hline
	
$C^{ab..M}$ & Polar Chebyshev tensor (M indices) \\
${\mathcal G}^{ab..M}$ & Axial Chebyshev tensor (M indices) \\
$\kappa_{N}^{2}$ & Function appearing in the Boltzmann equation\\ 
${\mathcal M}^{ij}_{(NA)}$ & Matrix relating directional derivatives of ${\mathcal B}_{NA}$ to sums of such functions\\
${\mathcal O}^{(P)}$ & Polar projection operator \\
${\mathcal O}^{(A)}$ & Axial projection operator \\
$(\textbf{A}^{+})^{a}$ & 2-vector defined via equation (\ref{ap}) \\
$(\textbf{A}^{-})^{a}$ & 2-vector defined via equation (\ref{am}) \\
$W^{ab..M}_{\phantom{ab..M}cd..M}$ & Tensor defined via equation (\ref{wdef}) \\
$Z^{ab..M}_{\phantom{ab..M}cd..M}$ & Tensor defined via equation (\ref{zdef}) \\
$\beta$ & Cosine of angle between z direction and 3-d wave-vector in the spatially flat limit\\
$j_{A}(x)$ & Spherical Bessel function \\
$\tau$ & Particle proper time \\
$Y^{l}_{m}$ & Spherical harmonic \\
$\tilde{A}^{(+)a}$ &   The equivalent of equation (\ref{ap}) but instead defined with respect to $\tilde{\gamma}_{cd}$\\
$\tilde{A}^{(-)a}$ & The equivalent of equation (\ref{am}) but instead defined with respect to
$\tilde{\gamma}_{cd}$	 \\
$T^{(\eta)}_{mn}$ & Matrix element for a representation $\eta$ of a group\\
${\mathcal P}(k,w,t)$ & Dark matter power spectrum \\
${\mathcal P}^{(l)}_{mn}$ & Function defined via equation (\ref{plmn}) \\
$P^{m}_{l}$ & Associated Legendre function\\
${\mathcal P}_{A}$ & Legendre polynomial of order A \\
${\mathcal D}_{+}$ & Differential raising operator (see for instance equation (\ref{dpl})) \\
${\mathcal D}_{-}$ & Differential lowering operator (see for instance equation (\ref{dmi})) \\
$Z^{k}_{m}$ & Eigenfunction of the co-moving 2-surface Laplace-Beltrami in the open case \\
${\mathcal Q}^{k}_{ab}$ & Function defined via equation (\ref{qdef})  \\
$J_{n-m}(x)$ & Bessel function of the first kind \\
${\mathcal J}^{k}_{n}$ & Function defined via equation (\ref{calj}) \\
$a_{lm}$ & Polar temperature perturbation in spherical harmonic space \\
$\Xi^{(P)}$ & Gauge-invariant polar perturbation \\
$\Xi^{(A)}$ & Gauge-invariant axial perturbation \\

	\hline
\end{tabular}

\end{document}